\documentclass[aps,pra,english,twocolumn,floatfix,showpacs,10pt]{revtex4-1}
\usepackage{amsfonts}
\usepackage{amsmath}
\usepackage{graphicx}
\usepackage{hyperref}
\usepackage{natbib}

\begin{document}

\title[Time-of-flight expansion of trapped dipolar Fermi gases]{Time-of-flight expansion of trapped dipolar Fermi gases:\\
From the collisionless to the hydrodynamic regime}

\author{Vladimir Velji\'{c}}
\email{vladimir.veljic@ipb.ac.rs}
\affiliation{Scientific Computing Laboratory, Center for the Study of Complex Systems, Institute of Physics Belgrade, University of Belgrade, Pregrevica 118, 11080 Belgrade, Serbia}
\author{Antun Bala\v{z}}
\email{antun.balaz@ipb.ac.rs}
\affiliation{Scientific Computing Laboratory, Center for the Study of Complex Systems, Institute of Physics Belgrade, University of Belgrade, Pregrevica 118, 11080 Belgrade, Serbia}
\author{Axel Pelster}
\email{axel.pelster@physik.uni-kl.de}
\affiliation{Physics Department and Research Center Optimas, Technical University of Kaiserslautern, Erwin-Schr\"odinger Strasse 46, 67663 Kaiserslautern, Germany}

\begin{abstract}
A recent time-of-flight (TOF) expansion experiment with polarized fermionic erbium atoms measured a Fermi surface deformation from a sphere to an ellipsoid due to dipole-dipole interaction, thus confirming previous theoretical predictions.
Here we perform a systematic study of the ground-state properties and TOF dynamics for trapped dipolar Fermi gases from the collisionless to the hydrodynamic regime at zero temperature.
To this end we solve analytically the underlying Boltzmann-Vlasov equation within the relaxation-time approximation in the vicinity of equilibrium by using a suitable rescaling of the equilibrium distribution.
The resulting ordinary differential equations for the respective scaling parameters are then solved numerically for experimentally realistic parameters and relaxation times that correspond to the collisionless, collisional, and hydrodynamic regime.
The equations for the collisional regime are first solved in the approximation of a fixed relaxation time, and then this approach is extended to include a self-consistent determination of the relaxation time.
The presented analytical and numerical results are relevant for a detailed quantitative understanding of ongoing experiments and the design of future experiments with ultracold fermionic dipolar atoms and molecules. In particular, the obtained results are relevant for systems with strong dipole-dipole interaction, which turn out to affect significantly the aspect ratios during the TOF expansion.
\end{abstract}


\maketitle

\section{Introduction}
\label{sec:intro}

Atomic and molecular ultracold gases offer many advantages for studying quantum phenomena, especially within the realm of many-body physics, due to the high degree of tunability of inter-atomic interactions \cite{TheBECBook, TheBECBook2}. In particular, dipolar quantum gases of atoms and molecules have received much attention in recent years, as the anisotropic and long-range nature of the magnetic or electric dipole-dipole interaction (DDI) gives rise to a rich spectrum of novel properties in such systems \cite{Santos, Glaum1, Glaum2, Baranov, PfauRep, Carr, Krumnow, Block, Alex1, Branko, Hamid, Radha, Ghabour}.
Such systems include those made up of ultracold atoms, as well as those consisting of heteronuclear molecules with large dipole-dipole interactions. 
Furthermore, in the recent theoretical and experimental research~\cite{Er2} a novel kind of strongly dipolar quantum gas was introduced. 
These are weakly bound polar molecules produced from atoms with large magnetic dipole moments, such as erbium and other lanthanides. These molecules can have a very large magnetic moment, which amounts to twice that of its individual atoms \cite{Gadway}.

In 2005 an anisotropic deformation of the expanding dipolar bosonic chromium  condensate due to DDI was observed \cite{Pfau}.
In the recent experiment \cite{PfauNature}, also the Rosensweig instability was detected in a $^{164}$Dy Bose-Einstein condensate, which represents a quantum ferrofluid due to the large atomic magnetic dipole moments. 
Namely, after a sudden decrease of the scattering length, the dipolar quantum gas creates self-ordered surface structures in form of droplet crystals, which can only be understood by taking into account DDI \cite{GP1, GP2, GP3, GP4, GP5, GP6} and the corresponding quantum fluctuations \cite{Lima3, Lima4, Blakie1, PfauPRL, Blakie2, Xi, Santos1, Santos2, Blakie3, Blakie4, Francesca2, Pfau2}.
 
For dipolar Fermi gases it was predicted that the long-range and anisotropic DDI leads in equilibrium to an anisotropic deformation of the Fermi surface from a sphere to an ellipsoid \cite{Sogo}.
A recent time-of-flight (TOF) expansion experiment has now unambiguously detected such an ellipsoidal Fermi surface (FS) deformation in a dipolar quantum gas of fermionic erbium atoms, which turns out to be of the order of a few percent \cite{Francesca}.
Within the Hartree-Fock mean-field theory for a many-body system, first-order contributions of DDI to the total energy of the system taken into account are in terms of both the
Hartree direct interaction and the Fock exchange interaction \cite{Sogo, Sogo2, Sogo2c, Zhang, Baillie, Lima1, Lima2, Falk}. 
In the case of a Fermi gas with isotropic interaction, the Hartree and the Fock interactions cancel out \cite{Sogo}, thus leading to a spherically symmetric FS. But in the case of a Fermi gas with anisotropic DDI the Hartree term gives rise to a distortion in real space \cite{Abad}, whereas
the Fock term gives rise to a distortion in momentum space, i.e., to an ellipsoidal deformation of the Fermi sphere.
Note in this context that the Fock exchange term in dipolar Fermi gases is the consequence of a combined effect of the DDI and the Pauli exclusion principle. 
In the current experimentally relevant range of dipolar interactions the theory beyond Hartree-Fock, where the total energy is determined up to second-order in the DDI, yields only
small differences, which cannot yet be resolved experimentally. Thus, the Hartree-Fock mean-field approximation yields already quantitatively accurate results for present-day experiments \cite{Liu-Yin,Kopietz1,Kopietz2}.

The investigations of collective oscillations and TOF dynamics of dipolar Fermi gases have so far focused on either the
collisionless (CL) regime \cite{Sogo2, Sogo2c, Zhang}, where collisions can be neglected, or on the hydrodynamic (HD) regime \cite{Lima1, Lima2},
where collisions occur so often that local equilibrium can be assumed. The recent paper of W\"achtler \textit{et al.} \cite{Falk} studied even the behavior of collective oscillations when the system undergoes a crossover from one regime to the other. 

Motivated by the experimental observation of the ellipsoidal FS deformation in the TOF experiment \cite{Francesca}, we continue here the analytical analysis along the lines of reference~\cite{Falk} and investigate in detail the expansion dynamics for the collisional regime, which represents the transition zone between the limiting CL and HD regimes. We also extend previous approaches based on the relaxation-time approximation by introducing a self-consistently determined relaxation time, and study how this quantitatively affects the TOF dynamics.

The paper is structured as follows. In Sec.~\ref{sec:amdfg} we introduce our notation and summarize recent experiments on atomic and molecular dipolar Fermi gases. In Sec.~\ref{sec:ge} we analyze the global equilibrium of the system by minimizing the Hartree-Fock total energy in order to obtain the Thomas-Fermi radii and momenta. Afterwards, in order to study the dynamics, in Sec.~\ref{sec:bve} we follow reference~\cite{Falk} and introduce the Boltzmann-Vlasov equation for dipolar Fermi gases as well as an approximative solution, which is based on a suitable rescaling ansatz for the equilibrium Wigner function. In Sec.~\ref{sec:tof} we study in detail the TOF expansion dynamics of an initially trapped Fermi gas.
To this end we present our analytical and numerical results of the TOF analysis all the way from the collisionless to the hydrodynamic regime and reveal how the expanding cloud bears the signature of the underlying DDI. 
Finally, Sec.~\ref{sec:con} gathers our concluding remarks and gives an outlook for future research.

\section{Atomic and molecular dipolar Fermi gases}
\label{sec:amdfg}

We consider a trapped ultracold quantum degenerate dipolar gas of single-component fermions of mass $M$ and magnetic dipole moment $\bf m$ or electric dipole moment $\bf d$ at zero temperature. 
The system is then described by the second-quantized Hamiltonian
\begin{eqnarray}
&&\hat{H}=\int d\mathbf{r}\, \hat{\Psi}^\dagger(\mathbf{r})\left[-\frac{\hbar^2}{2M} \nabla^2 +V(\mathbf{r})\right]\hat{\Psi} (\mathbf{r})\nonumber\\
&&+\frac{1}{2} \iint d\mathbf{r}\, d\mathbf{r'}\, \hat{\Psi}^\dagger(\mathbf{r'})\hat{\Psi}^\dagger(\mathbf{r}) V_\mathrm{int}({\bf r}-{\bf r'})\hat{\Psi} (\mathbf{r})\hat{\Psi} (\mathbf{r'})\, .
\label{eq:hamiltonian}
 \end{eqnarray}
Since the Pauli exclusion principle inhibits contact interaction, the long-range DDI between the polarized fermionic point dipoles is dominant. It is described by
\begin{equation}
 V_\mathrm{int}({\bf r})=\frac{C_\mathrm{dd}}{4\pi \left|{\bf r}\right|^3}\left(1-3 \cos^2 \vartheta \right)\, ,
\label{ddi}
\end{equation}
where $\bf r$ denotes the relative position between the dipoles, $\vartheta$ stands for the angle between $\bf r$ and the polarization axis of the dipoles, and
$C_\mathrm{dd}$ represents the dipolar interaction strength, which depends on the nature of the dipoles. Namely, for electric dipoles it is defined as $C^{e}_\mathrm{dd}=d^2 / \varepsilon_0$, where $\varepsilon_0$ is the vacuum permittivity,
while for magnetic dipoles one has $C^{m}_\mathrm{dd}=\mu_0 m^2$, where $\mu_0$ is the vacuum permeability. 
Magnetic dipolar moments are usually measured in units of Bohr magneton ($\mu_\mathrm{B}=9.27401 \times 10^{-24}\,\mathrm{JT}^{-1}$), and electric dipolar moments in units of Debye ($\mathrm{D}=3.33564 \times 10^{-30}\,\mathrm{Cm}$). Note that the
DDI of polar molecules is about $10^4$ times stronger than that of dipolar atoms, as $C^{e}_\mathrm{dd}/C^{m}_\mathrm{dd} \sim \alpha^{-2}_S$, with $\alpha_S=7.297 \cdot 10^{-3}$ being the Sommerfeld fine-structure constant. 

\begin{figure}[!t]
\centering
\includegraphics[width=5.5cm]{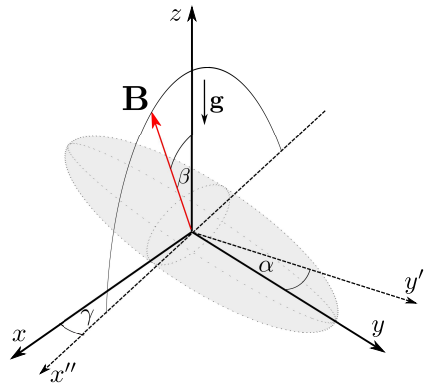}
\caption{Schematic illustration of the geometry in the Innsbruck experiment with $^{167}$Er \cite{Francesca}. Axes $x$, $y$, $z$ correspond to trap axes, while the orientation of the magnetic field ${\textbf B}$ and the atomic dipoles is determined by spherical coordinates $\beta$ and $\gamma$. Earth's gravitational field is parallel to $z$ axis, while imaging axis is denoted by $y'$, lies in the $xy$ plane, and forms an angle $\alpha$ with $y$ axis.}
\label{fig:fig1}
\end{figure}

\begin{table*}[!t]
\begin{tabular}{ccccccc}
\hline\hline
    gas &  $^{53}$Cr \cite{Cr} & $^{167}$Er \cite{Er}  & $^{161}$Dy \cite{Dy1}  & $^{167}$Er$^{168}$Er \cite{Er2}   &  $^{23}$Na$^{40}$K \cite{NaK1} & $^{40}$K$^{87}$Rb \cite{KRb}\\ \hline
    $m/d$ & $6\,\mu_\mathrm{B}$ & $7\,\mu_\mathrm{B}$ & $10\,\mu_\mathrm{B}$   & $14\,\mu_\mathrm{B}$ & $0.8\,\mathrm{D}$ & $0.566\,\mathrm{D}$ \\
    $\epsilon_\mathrm{dd}$ & $0.02$ & $0.15$ &  $0.30$& $1.76$& $5.44$ & $7.77$\\
\hline\hline
  \end{tabular}
\caption{Maximal values of dipole moments ($m$ for species with a magnetic dipole and $d$ for species with an electric dipole) and relative interaction strengths of fermionic atoms and molecules currently used in ultracold experiments, calculated according to Eq.~(\ref{ris}) using the trap parameters and particle number given in the text. Note that the electric dipole moments $d$ of molecular species $^{23}$Na$^{40}$K and $^{40}$K$^{87}$Rb can be tuned to smaller values by using an external electric field.}
\label{tab:tab1}
\end{table*}

Due to the anisotropy in the dipolar interaction potential~(\ref{ddi}), dipolar Fermi gases tend to be stretched along the polarization direction, since this leads to a lower total energy.
Here we consider the parameters of a recent Innsbruck experiment \cite{Francesca} performed with the fermionic erbium atoms $^{167}$Er in the collisionless regime, which are confined into a three-dimensional optical dipole trap with frequencies $(\omega_x, \omega_y, \omega_z)=(579,\, 91,\, 611) \times 2\pi\,\mathrm{Hz}$. It contained $N=7 \cdot 10^4$ atoms at a temperature $T$ of $0.18\, T_\mathrm{F}$, 
with the Fermi temperature being $T_\mathrm{F} = 1.1\,\mu\mathrm{K}$. The underlying geometry of the experiment is depicted in Fig.~\ref{fig:fig1}.
Gravity is oriented along the $z$ direction. The atomic cloud is imaged along the $y'$ axis, which forms an angle $\alpha=28^\circ$ with respect to the $y$ axis. 
The magnetic field $\textbf{B}$ forms an angle $\beta$ with the $z$ axis and lies in the $x''z$ plane, which is rotated for an angle $\gamma=14^\circ$ with respect to the $xz$ plane.
In the following we restrict ourselves to the general  geometry of the anisotropic trap, where the dipoles are oriented in the direction of one of the trap axis, which reflects the experimental
situation at the two limiting cases $\beta=0^\circ$ and $\beta=90^\circ$.

Previous theoretical works have predicted that the degree of deformation of the FS depends on the Fermi energy and the strength of the dipole moment \cite{Sogo, Lima1,Lima2, Falk}, therefore we use a relative interaction strength of the DDI when comparing its effect on different species of ultracold Fermi gases.
The relative interaction strength is given by
\begin{equation}
\epsilon_\mathrm{dd}= \frac{C_\mathrm{dd}}{4 \pi}\sqrt{\frac{M^3 \bar{\omega}}{\hbar^5}}N^{1/6}\, ,
\label{ris}
\end{equation}
where $\bar{\omega}=(\omega_x \omega_y \omega_z)^{1/3}$  denotes the geometric mean of the trap frequencies. 

The available dipolar Fermi gases in current ultracold experiments are listed in Table~\ref{tab:tab1}, together with the maximal values of their dipole moments and relative interaction strengths, considering the trap parameters and particle number given above.
The first quantum degenerate Fermi gas of the strongly magnetic atoms of dysprosium was produced in 2012 \cite{Dy1}. Later on, a degenerate Fermi gas of erbium atoms \cite{Er} and molecules \cite{Er2} was also realized. A dipolar Fermi sea of degenerate $^{53}$Cr, together with a BEC of $^{52}$Cr was produced in 2015 \cite{Cr}.
Few years ago a molecular dipolar gas of $^{40}$K$^{87}$Rb was realized using a single step of STIRAP \cite{Bergmann} (stimulated Raman adiabatic passage) with two-frequency laser irradiation \cite{KRb}, and with the same technique a new ultracold dipolar gas of fermionic molecules of $^{23}$Na$^{40}$K was created in a recent experiment \cite{NaK1, NaK2}. The same technique can be also used for thermal \cite{Kuznetsova} and ultracold dipolar Bose gases \cite{Gregory} of heteronuclear molecules.

We will consider the experimentally available range of relative strengths of the DDI and atom or molecule species given in Table I in the following sections. Therefore, the presented results are directly applicable to current and future experiments.

\section{Global equilibrium}
\label{sec:ge}

A quantum many-body system can be described in terms of a Wigner function  $\nu=\nu({\bf{r,k}},t)$, as it represents the Wigner-Weyl transform of the density matrix of the system and is equivalent to a quantum-mechanical wave function \cite{SchleichBook}.  
The Wigner function is a quasiprobability distribution function, and integrating it over the space or the momentum variables leads to the respective probability distribution functions. 
The quantum-mechanical expectation values of observables \cite{STE1,STE2,STE3,STE4,STE5} can be obtained as their phase-space averages, weighted by the Wigner function.

Considering a trapped ultracold dipolar Fermi gas, the equilibrium distribution function in the phase space will rapidly decrease to zero outside a certain closed surface, due to a combined effect of the Pauli exclusion principle, which is responsible for a formation of the FS in the momentum space, and the trapping in real space. 
Therefore, in order to model the global equilibrium distribution of the dipolar Fermi gas we use an ansatz for the semiclassical Wigner function, which resembles the form of the Wigner-transformed Fermi-Dirac distribution of a noninteracting Fermi gas. Note that the temperature of the dipolar Fermi gas in the experiment \cite{Francesca} is that low that thermal fluctuations are expected to be of the order of $(T/T_\mathrm{F})^2\approx 3\%$ due to the Sommerfeld expansion. This justifies to use for the Wigner-transformed Fermi-Dirac distribution of a noninteracting Fermi gas the zero-temperature approximation:

\begin{equation}
\nu^0({\bf r},{\bf k})=\Theta \left(1-\sum_i \frac{r^2_i}{R^2_i}-\sum_i \frac{k^2_i}{K^2_i} \right)\, .
\end{equation} 
Here $\Theta$ is the Heaviside step function. The variational parameters $R_i$ and $K_i$ represent the Thomas-Fermi (TF) radius and the Fermi momentum in the $i$-th direction, respectively, and describe the extension of the equilibrium Fermi surface in both coordinate and momentum space. 
With this ansatz, the normalization of the distribution $\nu^0({\bf r},{\bf k})$ to $N$ fermions leads to the condition

\begin{equation}
N= \int d{\bf r} \int \frac{d{\bf k}}{(2\pi)^3} \; \nu^0({\bf r},{\bf k})=\frac{1}{48}\bar{R}^3\bar{K}^3\, ,
\label{eq:particlenumberconservation}
\end{equation}
where $\bar{R}=(R_xR_yR_z)^{1/3}$ and $\bar{K}=(K_xK_yK_z)^{1/3}$ denote the geometric means of the respective TF radii and momenta. The total energy of the system in the Hartree-Fock approximation for dipoles oriented along $z$ axis, i.e., $\beta=0^\circ$ is given by
\begin{widetext}
\begin{equation}
E= \frac{N}{8}\sum_j\frac{\hbar^2K^2_j}{2M}+\frac{N}{8}\frac{M}{2}\sum_j \omega_j^2R_j^2
-\frac{48N^2 c_0}{8\bar{R}^3} f\left( \frac{R_x}{R_z},\frac{R_y}{R_z}\right)+\frac{48N^2 c_0}{8\bar{R}^3} f\left( \frac{K_z}{K_x},\frac{K_z}{K_y}\right),
\label{eq:totalenergy}
\end{equation}
\end{widetext}
where $c_0=\frac{2^{10}C_\mathrm{dd}}{3^4 \cdot 5 \cdot 7 \pi^3}$,  
while the function $f$ and its derivatives with respect to the first and second argument, $f_1$ and $f_2$, respectively, are anisotropy functions defined in references~\cite{Falk, Lima1,Lima2}. Note that the corresponding expression for the Hartree-Fock energy of the second considered case of dipoles oriented along $x$ axis, i.e., $\beta=90^\circ$ is obtained by a simple cyclic permutation of indices $x\rightarrow y\rightarrow z\rightarrow x$ in Eq.~(\ref{eq:totalenergy}). The same applies to all other equations throughout the paper.

The TF radii and momenta $R_i$ and $K_i$ are determined by minimizing the energy (\ref{eq:totalenergy}) with respect to them, which leads to the following set of algebraic equations:
\begin{widetext}
 \begin{eqnarray}
&&\frac{\hbar^2 K_x^2}{2M}-\frac{1}{3}\sum_j\frac{\hbar^2K_j^2}{2M}-\frac{48Nc_0}{\bar{R}^3}\frac{K_z}{K_x}  f_1\left( \frac{K_z}{K_x},\frac{K_z}{K_y}\right)=0\, ,\label{1} \\
&&\frac{\hbar^2 K_y^2}{2M}-\frac{1}{3}\sum_j\frac{\hbar^2K_j^2}{2M}-\frac{48Nc_0}{\bar{R}^3}\frac{K_z}{K_y}  f_2\left( \frac{K_z}{K_x},\frac{K_z}{K_y}\right)=0\, ,\label{2}\\
&&\frac{\hbar^2 K_z^2}{2M}-\frac{1}{3}\sum_j\frac{\hbar^2K_j^2}{2M}+\frac{48Nc_0}{\bar{R}^3}\left[\frac{K_z}{K_x}  f_1\left( \frac{K_z}{K_x},\frac{K_z}{K_y}\right)+\frac{K_z}{K_y}  f_2\left( \frac{K_z}{K_x},\frac{K_z}{K_y}\right)\right]=0\, , \label{3}\\
&&\omega_x^2R_x^2-\frac{1}{3}\sum_j\frac{\hbar^2K_j^2}{M^2}-\frac{48Nc_0}{M\bar{R}^3}\left[f\left( \frac{K_z}{K_x},\frac{K_z}{K_y} \right)
- f\left( \frac{R_x}{R_z},\frac{R_y}{R_z}\right)+\frac{R_x}{R_z} f_1\left(\frac{R_x}{R_z},\frac{R_y}{R_z}\right)  \right]=0\, , \label{VSTRx}\\
&&\omega_y^2R_y^2-\frac{1}{3}\sum_j\frac{\hbar^2K_j^2}{M^2}-\frac{48Nc_0}{M\bar{R}^3}\left[ f\left( \frac{K_z}{K_x},\frac{K_z}{K_y} \right)
- f\left( \frac{R_x}{R_z},\frac{R_y}{R_z}\right)+\frac{R_y}{R_z} f_2\left(\frac{R_x}{R_z},\frac{R_y}{R_z}\right)  \right]=0\, , \label{VSTRy}\\
&&\omega_z^2R_z^2-\frac{1}{3}\sum_j\frac{\hbar^2K_j^2}{M^2} -\frac{48Nc_0}{M\bar{R}^3}\left[ f\left( \frac{K_z}{K_x},\frac{K_z}{K_y} \right)
- f\left( \frac{R_x}{R_z},\frac{R_y}{R_z}\right)-\frac{R_x}{R_z} f_1\left(\frac{R_x}{R_z},\frac{R_y}{R_z}\right)-\frac{R_y}{R_z} f_2\left(\frac{R_x}{R_z},\frac{R_y}{R_z}\right)  \right]=0\, . \label{VSTRz}
 \end{eqnarray}
 \end{widetext}
 Note that Eqs.~(\ref{1})--(\ref{3}) are linearly dependent, and due to the symmetry of the anisotropy function $f(x,y)=f(y,x)$ can be reduced to two independent equations \cite{Lima1, Lima2, Falk},
 \begin{eqnarray}
\label{k1}
&&\hspace*{-4mm}K_x=K_y\, , \\
&&\hspace*{-4mm}K_z^2- K_x^2=
\frac{144 M N c_0}{\hbar^2\bar{R}^3}\Bigg[ 1-\frac{( 2 K_x^2+ K_z^2 ) f_s\Big( \frac{K_z}{K_x} \Big)}{2(K_x^2- K_z^2)} \Bigg],\label{k2}
 \end{eqnarray}
where $f_s(x) \equiv f(x,x)$ denotes the diagonal part of the anisotropy function. This implies that the momentum distribution of a dipolar
Fermi gas in global equilibrium remains cylindrically symmetric despite a general triaxial harmonic confinement \cite{Lima1, Lima2}. Due to the anisotropy of the dipolar interaction potential, dipolar quantum gases tend to be stretched along the polarization direction, i.e., the direction of an external magnetic or electric field, since this leads to a lower total energy. We note that this is valid not only for fermions, but for bosons as well \cite{Pfau,Lima3,Lima4}.
Equations~(\ref{VSTRx})--(\ref{k2}), together with Eq.~(\ref{eq:particlenumberconservation}), represent a closed set of six algebraic equations, which fix all variational parameters $R_i$ and $K_i$ in global equilibrium.

\begin{figure*}[!t]
\centering
\includegraphics[width=4.8cm]{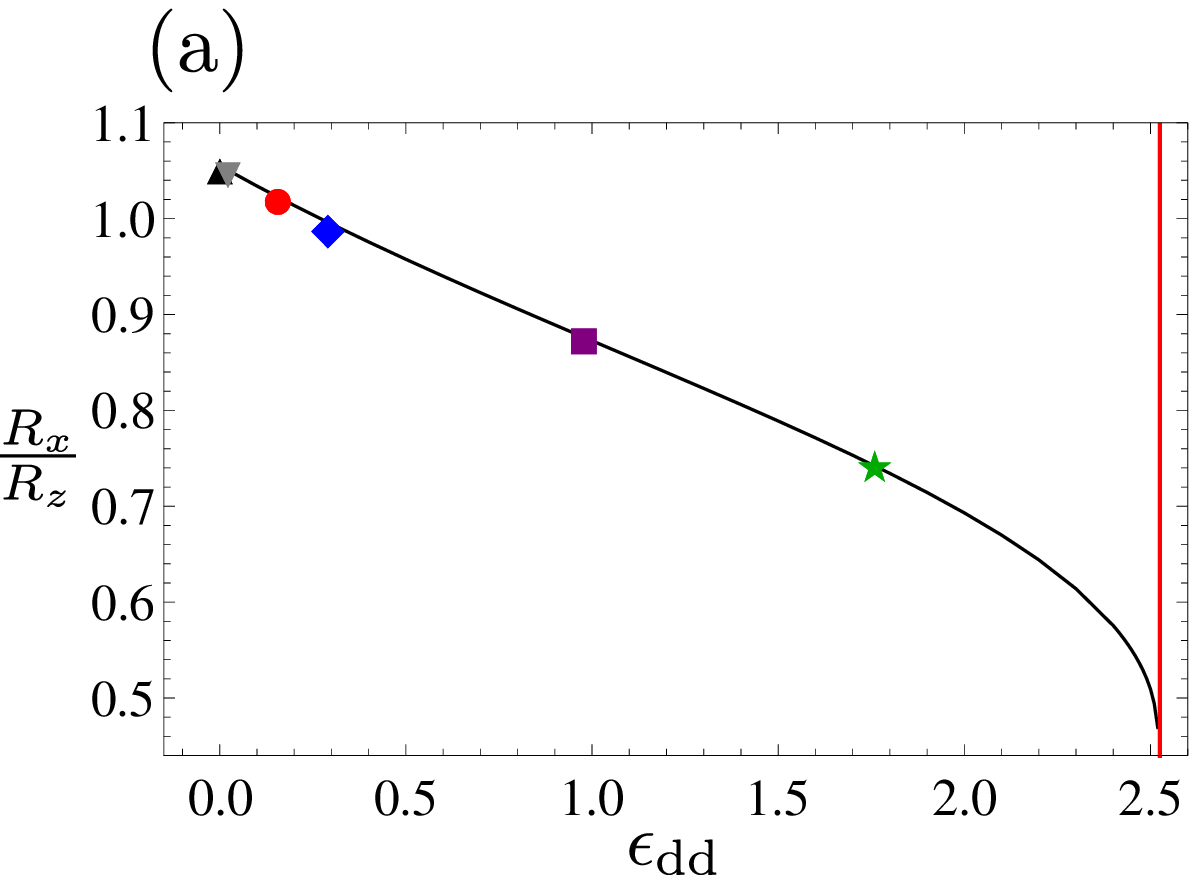} \hspace{0.3cm}
\includegraphics[width=4.8cm]{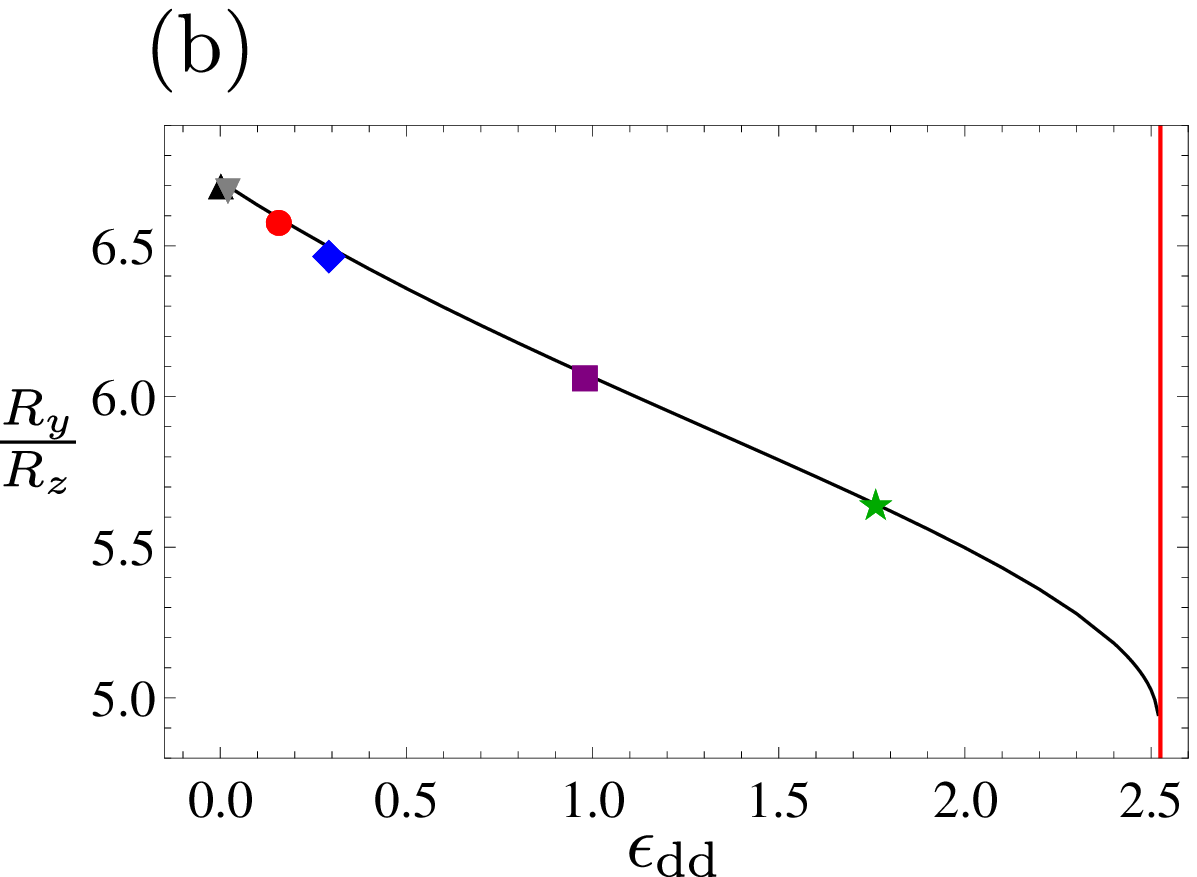} \hspace{0.3cm}
\includegraphics[width=4.8cm]{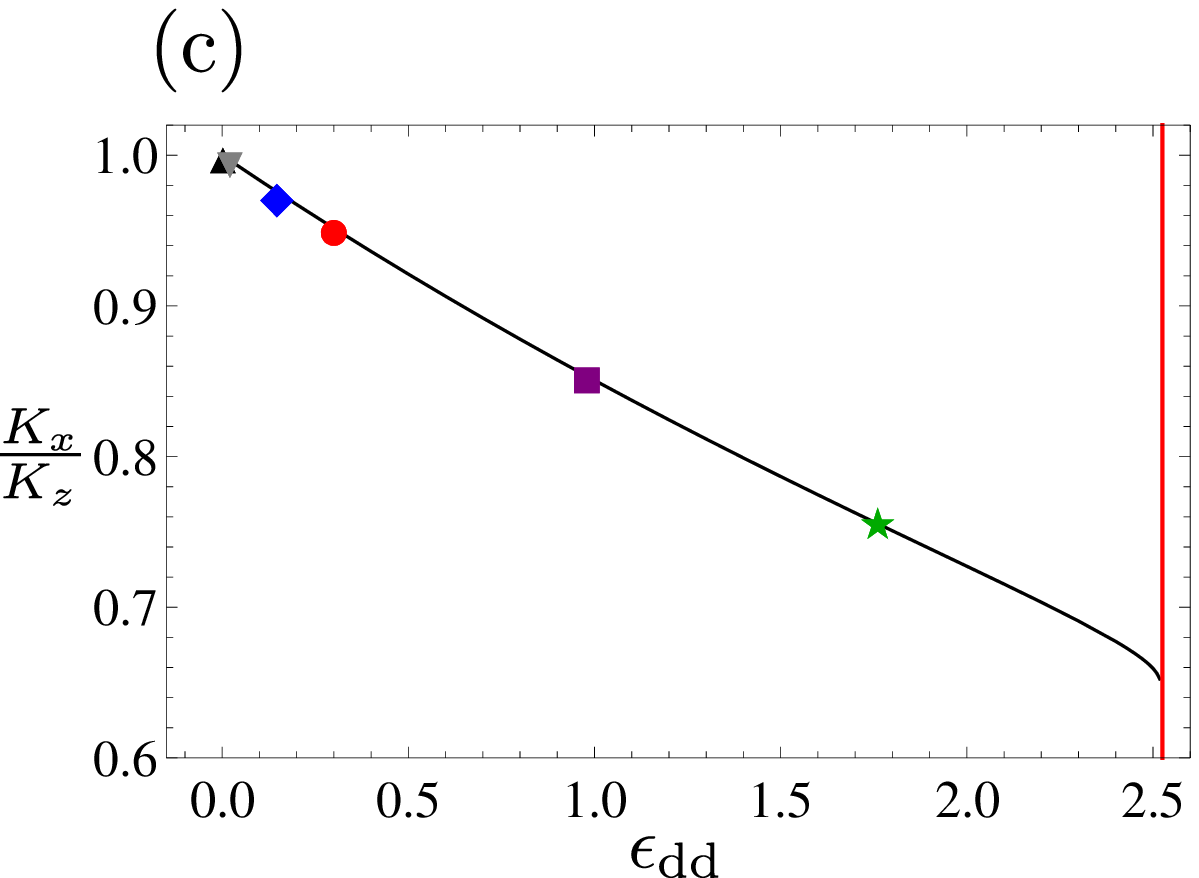}
\caption{Aspect ratios in real and momentum space as functions of relative dipolar interaction strength $\epsilon_\mathrm{dd}$ for Fermi gases in global equilibrium for considered trap geometry with dipoles parallel to $z$ axis: (a) ${R_x}/{R_z}$, (b) ${R_y}/{R_z}$, and (c) ${K_x}/{K_z}$. 
Black up-pointing triangles represent aspect ratios for the limiting case of a noninteracting Fermi gas: in real space ${R_x}/{R_z} = \omega_z/\omega_x$ and ${R_y}/{R_z} = \omega_z/\omega_y$, while in momentum space ${K_x}/{K_z} = 1$ (Fermi sphere).
Other symbols represent aspect ratios for dipolar atoms and molecules from Table~\ref{tab:tab2}: $^{53}$Cr (gray down-pointing triangles), $^{167}$Er (red circles), $^{161}$Dy (blue diamonds), $^{40}$K$^{87}$Rb (purple squares), $^{167}$Er$^{168}$Er (green stars).
Red vertical line corresponds to a critical value of the relative dipolar interaction strength $\epsilon_\mathrm{dd}^\mathrm{crit}\approx 2.52$ for considered trap geometry; for $\epsilon_\mathrm{dd} > \epsilon_\mathrm{dd}^\mathrm{crit}$ no stable stationary solution exists for a system of Eqs.~(\ref{VSTRx})--(\ref{k2}), together with Eq.~(\ref{eq:particlenumberconservation}).}
\label{fig:fig2}
\end{figure*}

For weak enough interactions, a local minimum might exist to which the system would return after a small perturbation. 
The regions of system parameters satisfying this property are called stable and the mathematical criterion behind this classification is given by positive eigenvalues of the Hessian matrix of the energy functional~\cite{Lima2}.
Figure~\ref{fig:fig2} depicts aspect ratios of stable solutions, i.e., the deformation of the Fermi surface in real and momentum space in global equilibrium for the dipolar Fermi gases given in Table~\ref{tab:tab2}. 
These results are obtained for the dipoles oriented in the direction of the $z$ axis, i.e., for the angle $\beta=0^\circ$ (see Fig.~\ref{fig:fig1}). 
For the limiting case of a noninteracting Fermi gas we know that the aspect ratios in real space are ${R_x}/{R_z} = \omega_z/\omega_x$ and ${R_y}/{R_z} = \omega_z/\omega_y$, while in momentum space the Fermi surface becomes the Fermi sphere and therefore we have ${K_x}/{K_z} = 1$.

\begin{table*}[!t]
\begin{tabular}{ccccccc}
\hline\hline
    gas &  $^{53}$Cr \cite{Cr} & $^{167}$Er \cite{Er}  & $^{161}$Dy \cite{Dy1} & $^{40}$K$^{87}$Rb \cite{KRb}  & $^{167}$Er$^{168}$Er \cite{Er2} \\ \hline
    $m/d$ & $6\,\mu_\mathrm{B}$ & $7\,\mu_\mathrm{B}$ & $10\,\mu_\mathrm{B}$  & $0.2\,\mathrm{D}$ & $14\,\mu_\mathrm{B}$\\
    $\epsilon_\mathrm{dd}$ & $0.02$ & $0.15$ &  $0.30$& $0.97$& $1.76$\\
\hline\hline
  \end{tabular}
\caption{Dipole moments ($m$ for species with a magnetic dipole and $d$ for species with an electric dipole) and relative interaction strengths of fermionic atoms and molecules to be used throughout the paper, calculated using the trap parameters and particle number given in the text.}
\label{tab:tab2}
\end{table*}

Red vertical lines in Fig.~\ref{fig:fig2} represent a critical value of the relative interaction strength $\epsilon_\mathrm{dd}^\mathrm{crit} \approx 2.52$ for the considered trap geometry. Namely, for $\epsilon_\mathrm{dd} > \epsilon_\mathrm{dd}^\mathrm{crit}$
stable stationary solutions for Eqs.~(\ref{VSTRx})--(\ref{k2}), together with Eq.~(\ref{eq:particlenumberconservation}), do not exist \cite{Lima2, Blakie} for system parameters from the Innsbruck experiment \cite{Francesca}.
Note that the value of $\epsilon_\mathrm{dd}^\mathrm{crit}$ does not depend on the mass of the species and is universal for a given trap geometry, as can be shown by rewriting Eqs.~(\ref{VSTRx})--(\ref{k2}) in the dimensionless form.

From Table~\ref{tab:tab1} we see that electric dipolar molecules $^{23}$Na$^{40}$K and $^{40}$K$^{87}$Rb with the largest values of relative dipolar interaction strength $\epsilon_\mathrm{dd}$ are unstable for the considered system parameters if their maximal values of electric dipole moments are used, since in both cases  $\epsilon_\mathrm{dd} > \epsilon_\mathrm{dd}^\mathrm{crit}$.
However, by using an external electric field, their dipole moments can be tuned to smaller values, and therefore we will consider the case of $^{40}$K$^{87}$Rb with the value of electric dipole moment tuned down to $d = 0.2\,\mathrm{D}$ \cite{sigma}, for which one obtains $\epsilon_\mathrm{dd}=0.97 < \epsilon_\mathrm{dd}^\mathrm{crit}$. Table~\ref{tab:tab2} gives the corresponding parameters of the five atomic and molecular dipolar species we will consider in the rest of this paper.

In Fig.~\ref{fig:fig2} corresponding aspect ratios for the noninteracting case are shown as black up-pointing triangles in comparison with aspect ratios for interacting Fermi gases. 
For atomic gases of $^{53}$Cr, $^{167}$Er, and $^{161}$Dy the DDI is not that strong, and their aspect ratios in momentum space deviate less than $5\%$ from unity, see Fig.~\ref{fig:fig2}(c). 
Actually, for  $^{53}$Cr (gray down-pointing triangles) the aspect ratio in momentum space is just $1\%$ smaller than 1, which is quite challenging to be observable in an experiment. 
Nevertheless, for $^{167}$Er (red circles) the aspect ratio in momentum space turns out to be about $3\%$ less than 1 and has already been experimentally observed in reference~\cite{Francesca}, meaning that the $5\%$ deformation for $^{161}$Dy (blue diamonds) should also be observable. For the considered parameters for $^{40}$K$^{87}$Rb (purple squares) with $\epsilon_\mathrm{dd}=0.97$ we obtain even larger value of the FS deformation of about $15 \%$. Furthermore, a molecule of $^{168}$Er$^{167}$Er (green stars) with $\epsilon_\mathrm{dd}=1.76$ would yield a FS deformation of nearly $25\%$.

\begin{figure}[!b]
\centering
\includegraphics[width=5.7cm]{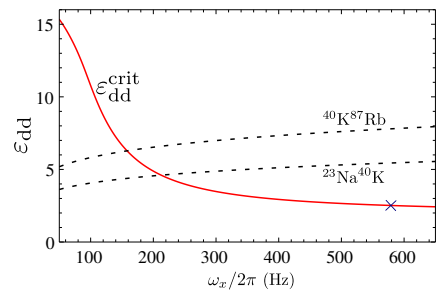}
\caption{Critical value of relative dipolar interaction strength $\epsilon_\mathrm{dd}^\mathrm{crit}$ (red solid line) as function of trap frequency $\omega_x$ for fixed values $(\omega_y, \omega_z)= (91, 611)\times 2\pi\,\mathrm{Hz}$ and particle number $N=7\times 10^4$. Blue cross corresponds to experimental value of frequency $\omega_x=579\times 2\pi\,\mathrm{Hz}$ from the Innsbruck experiment \cite{Francesca}, for which $\epsilon_\mathrm{dd}^\mathrm{crit} \approx 2.52$. Black dashed lines depict relative dipolar interaction strength $\epsilon_\mathrm{dd}$ for dipolar molecular species $^{23}$Na$^{40}$K and $^{40}$K$^{87}$Rb according to Eq.~(\ref{ris}), for the same parameters and for maximal values of their electric dipole moments from Table~\ref{tab:tab1}.}
\label{fig:fig3}
\end{figure}

Note that the critical value $\epsilon_\mathrm{dd}^\mathrm{crit}$ strongly depends on the trap geometry, as can be seen in Fig.~\ref{fig:fig3}, where we show its dependence on the frequency $\omega_x$ for fixed values $(\omega_y, \omega_z)= (91, 611)\times 2\pi\,\mathrm{Hz}$ from the Innsbruck experiment \cite{Francesca}. For the corresponding experimental value $\omega_x=579\times 2\pi\,\mathrm{Hz}$ we obtain $\epsilon_\mathrm{dd}^\mathrm{crit} \approx 2.52$ (blue cross), the same value that can be deduced from Fig.~\ref{fig:fig2}. In Fig.~\ref{fig:fig3} we also show relative dipolar interaction strength $\epsilon_\mathrm{dd}$ for molecular species  $^{23}$Na$^{40}$K and $^{40}$K$^{87}$Rb for maximal values of their electric dipole moments from Table~\ref{tab:tab1}. Note that the relative interaction strengths also depend on the trap geometry according to Eq.~(\ref{ris}). As already pointed out, for the cigar-shaped trap geometry of the Innsbruck experiment \cite{Francesca} both molecular species turn out to be unstable. However, for the pancake-shaped trap with sufficiently small value of the frequency $\omega_x$, i.e., $\omega_x<210\times 2\pi\,\mathrm{Hz}$ for $^{23}$Na$^{40}$K and $\omega_x<155\times 2\pi\,\mathrm{Hz}$ for $^{40}$K$^{87}$Rb, both species can be made stable even if their maximal electric dipole moments are used.

\section{Scaling ansatz for the Boltzmann-Vlasov equation}
\label{sec:bve}

The dynamics of a trapped ultracold dipolar degenerate Fermi gas can be described in terms of the Botzmann-Vlasov (BV) equation, which was previously prominently used in the realm of nuclear \cite{Baran, Alex2} and plasma \cite{ZLjP1, ZLjP2} physics. It was already used to study the TOF dynamics of ultracold fermions with the contact interaction \cite{Dusling, Urban}, as well as their collective modes \cite{Urban2, Urban3}. The BV equation determines the time evolution of the Wigner function $\nu$ \cite{Falk} and for a dipolar Fermi gas it reads:
 \begin{eqnarray}
 &&\frac{\partial \nu({\bf{r,k}},t)}{\partial t} + \frac{\hbar\bf{k}}{M} \nabla_{\bf{r}} \nu+ \frac{1}{\hbar}\nabla_{\bf{k}} U({\bf{r,k}},t) \nabla_{\bf{r}} \nu({\bf{r,k}},t) \nonumber\\
&&-\frac{1}{\hbar}\nabla_{\bf{r}} U({\bf{r,k}},t) \nabla_{\bf{k}} \nu({\bf{r,k}},t) =I_\mathrm{coll}[\nu]({\bf{r,k}},t)\, .
\label{eq:bve}
\end{eqnarray}
Here $U({\bf{r,k}},t)=U_\mathrm{ext}({\bf r})+\int d{\bf r}' V_\mathrm{int}({\bf r}-{\bf r}')n({\bf r}',t)-\int \frac{d {\bf k}'}{(2\pi \hbar)^3} \tilde{V}_\mathrm{int}({\bf k}-{\bf k}')\nu({\bf r}, {\bf k}',t)$ denotes the mean-field potential, which
includes external trap potential, as well as the respective Hartree and Fock terms, where $V_\mathrm{int}({\bf r})$ represents the DDI potential (\ref{ddi}) and $\tilde{V}_\mathrm{int}({\bf k})$ its Fourier transform.
Note that this Hartree-Fock dynamic mean-field description is self-consistent and is of the first order in the interaction potential. 
On the right-hand side of Eq.~(\ref{eq:bve}) we have the collision integral $I_\mathrm{coll}[\nu]({\bf{r,k}},t)$ which is of second order of the interaction potential and describes collisions between two particles \cite{Baym}. 
Instead of using a full expression for the collision integral, which would require a detailed modeling of scattering processes between atoms or molecules, we apply here the relaxation-time approximation \cite{TheBECBook, Stringari-ansatz} in the form
\begin{equation}
I_\mathrm{coll}[\nu({\bf r, k}, t)]=-\frac{\nu({\bf r, k}, t)-\nu^\mathrm{le}({\bf r, k})}{\tau}\, .
\label{eq:rta}
\end{equation}
Here $\tau$ denotes the relaxation time, which is related to the average time between collisions, and $\nu^\mathrm{le}$ stands for the distribution function corresponding to local equilibrium. The physical idea is that the particles interact via collisions and exchange energy and momentum, which eventually leads to a relaxation of the system into a local equilibrium state in which the collisions will no longer change the distribution function.
In contrast to that, the local velocity field or the density can still be spatially dependent.
The local thermodynamical equilibrium of a dipolar Fermi gas is defined by $I_\mathrm{coll}[\nu^\mathrm{le}]=0$.
If the time-dependent distribution function $\nu({\bf r},{\bf k},t)$ is close to the global equilibrium $\nu^0({\bf r},{\bf k})$, it can be approximately expressed by a suitable rescaling of the equilibrium distribution \cite{Stringari-ansatz}:
\begin{equation}
 \nu({\bf r},{\bf k},t) \rightarrow  \Gamma(t) \nu^0(\boldsymbol{\mathcal{R}}({\bf r},t), \boldsymbol{\mathcal{K}}({\bf r},{\bf k},t))\, ,
\end{equation}
with the rescaled variables defined by 
\begin{equation}
\mathcal{R}_i({\bf r},t)=\frac{r_i}{b_i(t)}\, , \label{eq:r}
\end{equation}
and
\begin{equation}
\mathcal{K}_i({\bf r},{\bf k},t)=\frac{1}{\sqrt{\theta_i(t)}}\left[ k_i-\frac{M \dot{b}_i(t) r_i}{\hbar  b_i(t)}\right]\, , \label{eq:k}
\end{equation}
\noindent
where $b_i(t)$ and $\theta_i(t)$ are time-dependent dimensionless scaling parameters. The normalization factor  $\Gamma(t)$ is given by \cite{Stringari-ansatz}
\begin{equation}
 \Gamma(t)^{-1} = \prod_i b_i(t) \sqrt{\theta_i(t)}\, . \label{eq:normalizationfactor}
\end{equation}
The second term in the bracket of Eq.~(\ref{eq:k}) is proportional to the local velocity. Namely, taking the derivative with respect to time in Eq.~(\ref{eq:r}) we get $\dot{\mathcal{R}}_i({\bf r},t)\sim k_i-M \dot{b}_i(t) r_i/(\hbar  b_i(t))$ with $k_i=M \dot{r}_i/\hbar$ \cite{Zhang, Castin}.
Subtracting the drift velocity $\dot{b}_i(t)r_i/b_i(t)$ in the ansatz (\ref{eq:k}) it is ensured that the momentum $\boldsymbol{\mathcal{K}}({\bf r},{\bf k},t)$ is not affected by the time dependence of the ansatz for $\boldsymbol{\mathcal{R}}({\bf r},t)$.

The time dependence of the distribution function is governed by the scaling parameters $b_i(t)$ and $\theta_i(t)$, which denote the time-dependent deformations of the spatial and momentum variables, respectively. 
Inserting the above ansatz into the Boltzmann-Vlasov Eq.~(\ref{eq:bve}) one obtains coupled ordinary differential equations of motion for the respective scaling parameters \cite{Falk}:
\begin{widetext}
\begin{eqnarray}
&& \ddot{b}_i+\omega_i^2b_i-\frac{\hbar^2 K_i^2  \theta_i}{M^2b_i R_i^2 }
 + \frac{48N c_0}{ Mb_i R_i^2 \prod_j b_j R_j}\left[ f\left( \frac{b_xR_x}{b_zR_z},\frac{b_yR_y}{b_zR_z}\right)-b_iR_i\frac{\partial}{\partial b_i R_i}f\left( \frac{b_xR_x}{b_zR_z},\frac{b_yR_y}{b_zR_z}\right)\right] \nonumber \\ 
 &&-\frac{48N c_0}{ Mb_i R_i^2 \prod_j b_j R_j}\left[ f\left(\frac{\theta_z^{\frac{1}{2}}K_z}{\theta_x^{\frac{1}{2}}K_x}, \frac{\theta_z^{\frac{1}{2}}K_z}{\theta_y^{\frac{1}{2}}K_y}\right)+\theta_i^{\frac{1}{2}}K_i \frac{\partial}{\partial \theta_i^{\frac{1}{2}}K_i} f\left(\frac{\theta_z^{\frac{1}{2}}K_z}{\theta_x^{\frac{1}{2}}K_x}, \frac{\theta_z^{\frac{1}{2}}K_z}{\theta_y^{\frac{1}{2}}K_y}\right)\right]=0 \, , \label{eq:bi}\\ 
 &&\dot{\theta_i}+2\frac{\dot{b_i}}{b_i}\theta_i +\frac{1}{\tau}(\theta_i-\theta^\mathrm{le}_i)=0\, \label{eq:theta_i}.
\end{eqnarray}
\end{widetext}
Note that in the case of the global equilibrium the three Eqs.~(\ref{eq:bi}) with the initial conditions $b_i(0)=\theta_i(0)=1$ and $\dot{b}_i(0)=\dot{\theta}_i(0)=0$ at $t=0$ reduce to Eqs.~(\ref{VSTRx})--(\ref{VSTRz}), as expected.
Also, we remark that the initial conditions correspond to $\nu({\bf r}, {\bf k}, t=0) \equiv \nu^0({\bf r},{\bf k})$.

\section{Time-of-flight expansion}
\label{sec:tof}

The most ubiquitous method to study the physics of trapped ultracold gases is their absorption imaging after the release of the atomic or molecular cloud from the trap.
Turning off the trap potential allows the ultracold gas cloud to expand for tens of milliseconds and an absorption image is taken afterwards, when the cloud is large enough for the image to be recorded by a CCD camera.
This technique, known as the TOF imaging, is one of the most important probes of ultracold quantum systems and 
TOF expansion experiments are a key diagnostic tool to study their properties.
From the size of the expanded cloud and the known time of flight one can directly obtain, for instance, the Fermi energy for a non-interacting degenerate Fermi gas.
In the case of free ballistic expansion, which is generically applied to theoretically model TOF, the ellipsoidal FS deformation due to DDI is taken into account before TOF, while all interactions between atoms during TOF are neglected. 
In contrast to that, a nonballistic expansion model takes into account interactions for calculating both global equilibrium before TOF and the subsequent expansion. 
In this section we show how quantitative information about the ellipsoidal FS deformation relevant for the current experiments can be determined from solving the BV equation for a TOF expansion of the dipolar Fermi gas.

Bearing in mind that the trap potential is turned off during TOF, Eqs.~(\ref{eq:bi}) and (\ref{eq:theta_i}) can be used to describe the TOF dynamics if we remove the terms $\omega_i^2b_i$ which stem from the harmonic trap potential.
Within this formalism, the average sizes of the Fermi gas cloud in real space are given by (see Appendix~\ref{sec:AR} for more details)
\begin{equation}
\langle r^2_i\rangle=\frac{1}{N}\int \frac{d{\bf k}}{(2 \pi)^3}\int d{\bf r}\, \nu({\bf r},{\bf k},t) r^2_i=\frac{1}{8}R^2_ib_i^2(t)\, . \label{eq:r2mean}
\end{equation}
The deformation of the cloud shape is described in terms of the cloud aspect ratio $A_R(t)$, which is defined by the ratio of the root mean square of the transverse and longitudinal cloud radii, i.e., the average sizes of the cloud in vertical $\sqrt{\langle r^2_\mathrm{v}\rangle}$ and horizontal $\sqrt{\langle r^2_\mathrm{h}\rangle}$ direction in the imaging plane.  
Since the imaging axis in the Innsbruck experiment~\cite{Francesca} has an angle of $\alpha=28^\circ$ with respect to the $y$ axis, according to Eq.~(\ref{eq:ARa}) from Appendix~\ref{sec:AR} this leads to 
\begin{equation}
\label{eq:AR}
A_R(t)=\frac{R_z b_z(t)}{\sqrt{R_x^2b^2_x(t)\cos^2\alpha+R_y^2b^2_y(t)\sin^2\alpha}}\, .
\end{equation}
This aspect ratio in real space $A_R(t)$ represents a directly measurable quantity in the TOF dynamics experiments. In order to describe the influence of DDI on the FS we also use a corresponding aspect ratio in momentum space. In analogy to $A_R(t)$,
the average sizes of the Fermi gas cloud in momentum space read (see Appendix~\ref{sec:AK} for more details)
\begin{eqnarray}
 \langle k_i^2 \rangle&=&\frac{1}{N} \int d{\bf r}\int \frac{d{\bf k}}{(2 \pi)^3}  \nu({\bf r},{\bf k},t) k^2_i\nonumber\\
 &=&\frac{1}{8}\left( K_i^2\theta_i(t)+\frac{M^2R_i^2 \dot{b}^2_i(t)}{\hbar^2}\right)\, , \label{eq:k2}
\end{eqnarray}
and the corresponding aspect ratio in momentum space is given by
\begin{equation}
\label{eq:AK}
A_K(t)=\sqrt{\frac{\langle k^2_z\rangle}{\langle k^2_x\rangle\cos^2\alpha+\langle k^2_y\rangle\sin^2\alpha}}\, .
\end{equation}

The relaxation time $\tau$ in Eq.~(\ref{eq:theta_i}) determines the regime of the dipolar Fermi gas and, therefore, by solving the appropriate equations for varying values of $\tau$, we are able to describe dynamic properties of the Fermi gas all the way from the collisionless ($\bar{\omega} \tau \gg 1$) to the hydrodynamic ($\bar{\omega} \tau \ll 1$) regime. Here, as before, $\bar{\omega}$ represents the geometric mean of the trap frequencies.
In Sec.~\ref{subsec:cl} we will study the collisionless regime, in Sec.~\ref{subsec:hy} the hydrodynamic regime, while in Sec.~\ref{subsec:cr} we will investigate the system behavior in the intermediate, collisional regime.
In Sec.~\ref{subsec:crsc} we will improve the relaxation-time approximation in the collisional regime even further by determining the relaxation time in a self-consistent way.

\subsection{Collisionless regime}
\label{subsec:cl}

The value of the relaxation time $\tau$ determines the regime of the Fermi gas during the expansion. 
In the low-density or collisionless regime, which is determined by the condition $\bar{\omega} \tau \gg 1$, the relaxation time $\tau$ can be taken to be infinite. In the limit $\tau \rightarrow \infty$ 
the differential Eqs.~(\ref{eq:theta_i}) for the scaling parameters $\theta_i$ decouple and the dynamic behavior in each direction is independent from the others.
Due to this, Eqs.~(\ref{eq:theta_i}) can be solved analytically. With the respective initial conditions $b_i(0)=\theta_i(0)=1$ and $\dot{b}_i(0)=\dot{\theta}_i(0)=0$ we obtain $\theta_i(t)=b_i(t)^{-2}$. 
Inserting this solution in Eqs.~(\ref{eq:bi}) for the scaling parameters $b_i(t)$ yields the equations of motion in the collisionless regime \cite{Sogo2, Sogo2c, Zhang}.
We numerically solve them for a general system geometry, where the trap frequencies in the three directions are different and correspond to the values of the Innsbruck experiment \cite{Francesca}, and the magnetic field is oriented either in $z$ direction ($\beta=0^\circ$) or in $x$ direction ($\beta=90^\circ$).
Although at $\beta=90^\circ$ the dipoles' orientation forms an angle of $\gamma=14^\circ$ (see Fig.~\ref{fig:fig1}) with respect to the $x$ axis, we assume for simplicity in our calculations that the dipoles are parallel to the $x$ axis.

\begin{figure*}[!t]
\centering
\includegraphics[width=6.5cm]{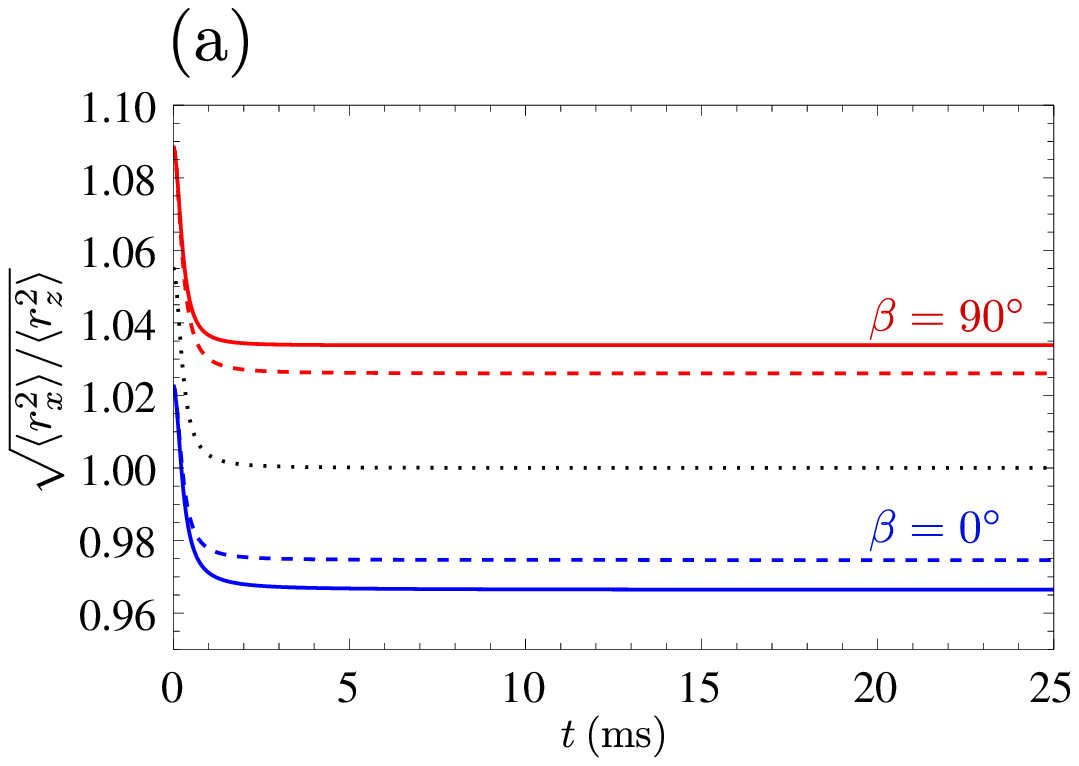}\hspace*{10mm}
\includegraphics[width=6.5cm]{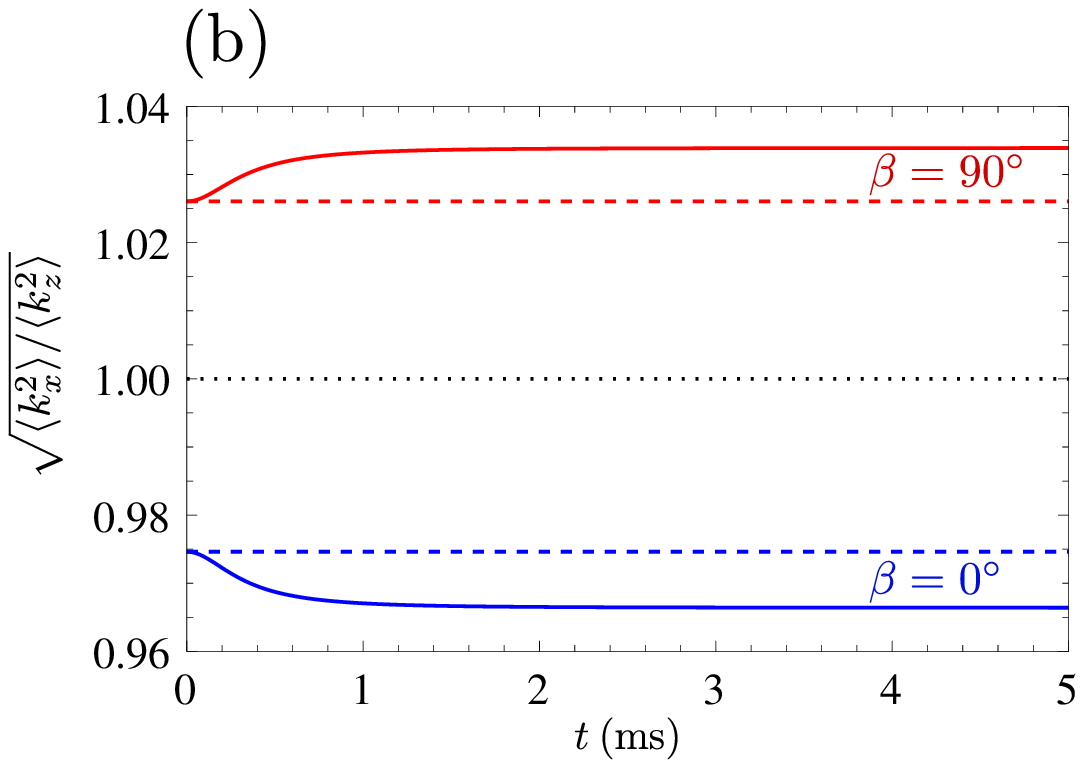}
\includegraphics[width=6.5cm]{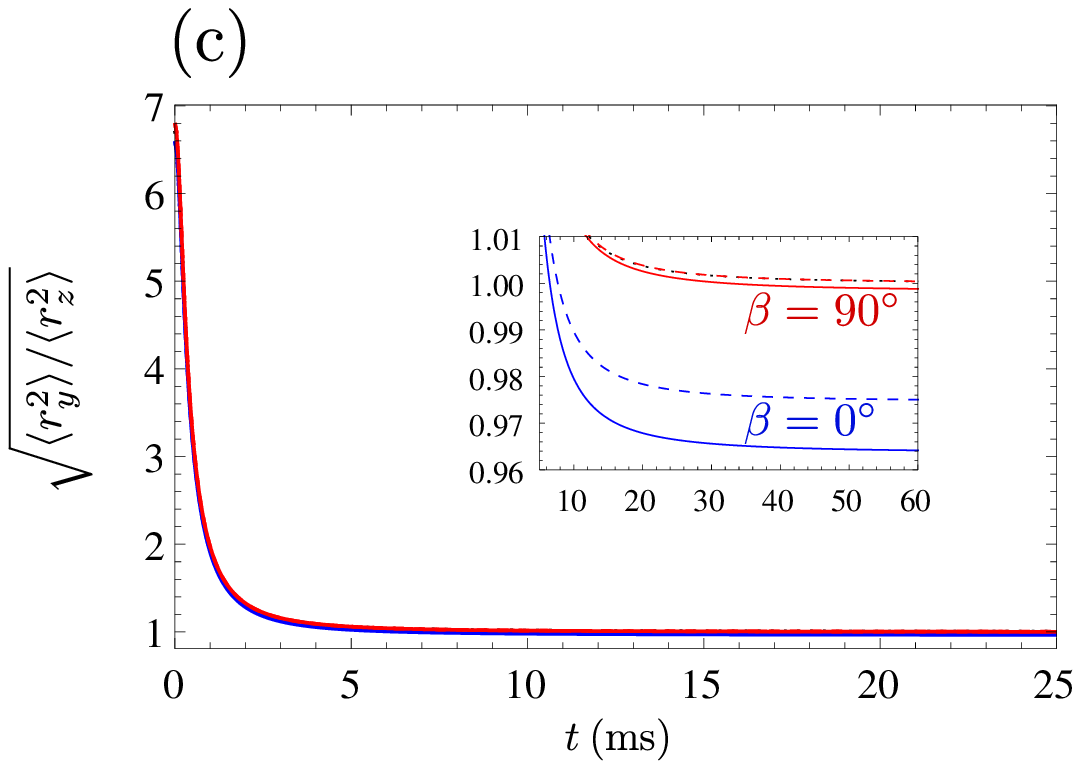}\hspace*{10mm}
\includegraphics[width=6.5cm]{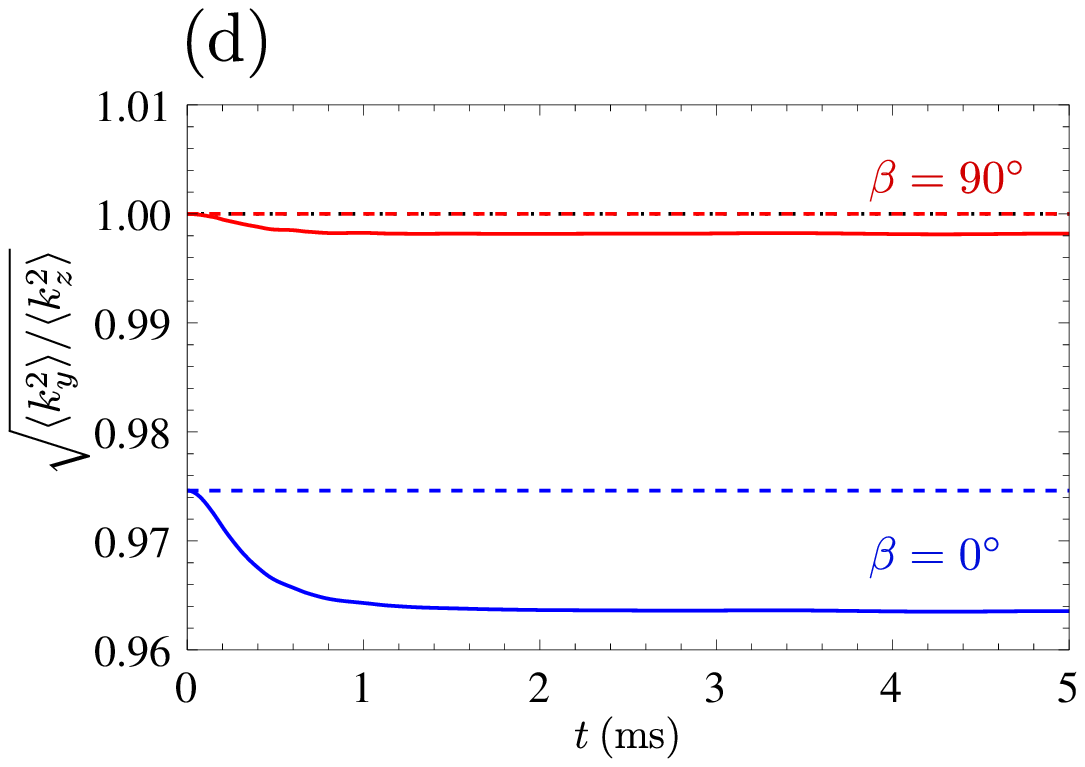}
\includegraphics[width=6.5cm]{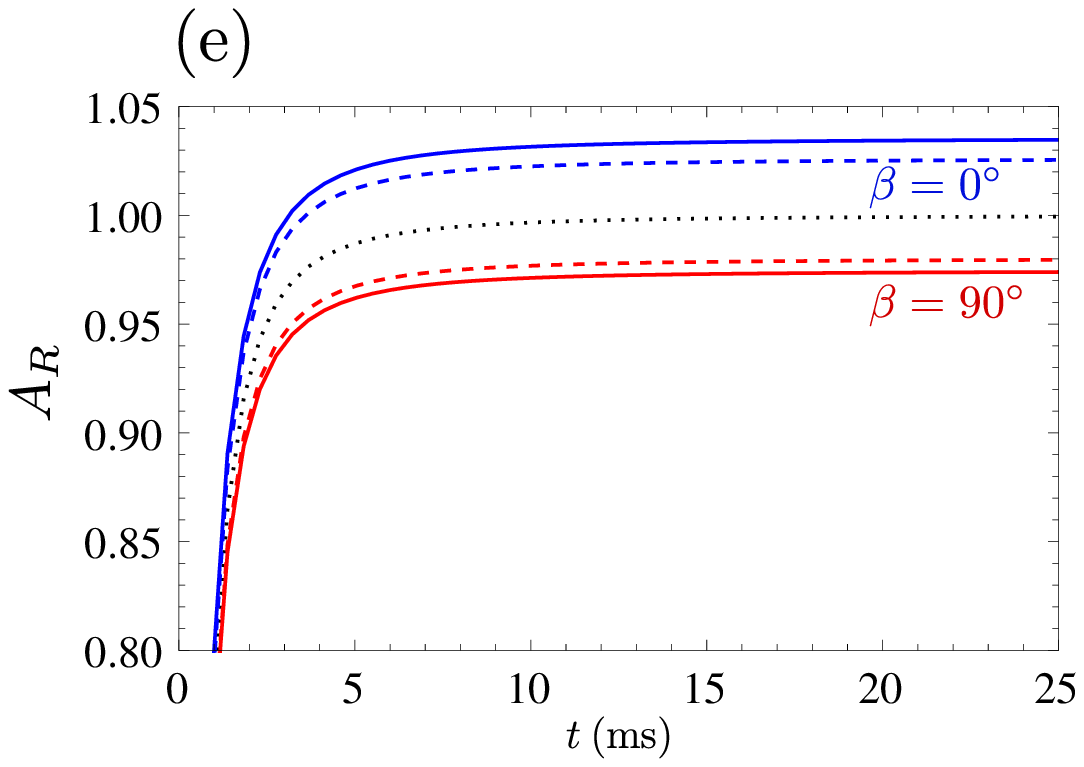}\hspace*{10mm}
\includegraphics[width=6.5cm]{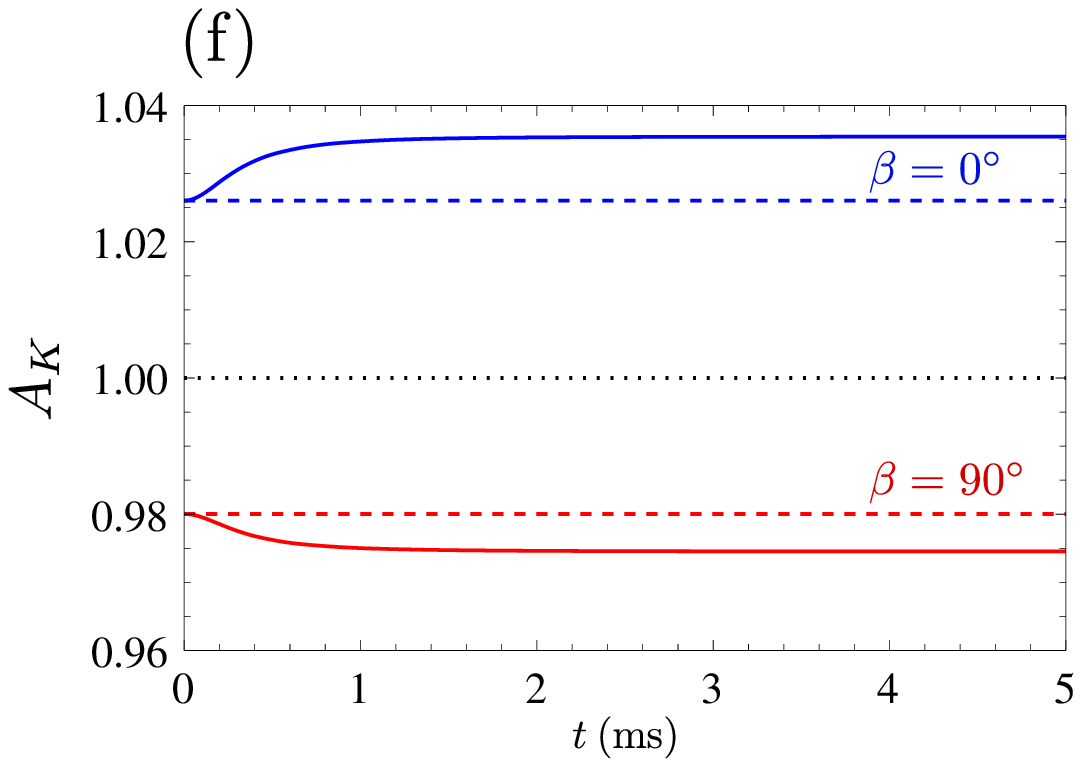}
\caption{Aspect ratios in real and momentum space in the collisionless regime during TOF expansion of ultracold gas of $^{167}$Er:
(a) $\sqrt{\langle r^2_x\rangle/\langle r^2_z\rangle}$,
(b) $\sqrt{\langle r^2_y\rangle/\langle r^2_z\rangle}$,
(c) $A_R$,
(d) $\sqrt{\langle k^2_x\rangle/\langle k^2_z\rangle}$, 
(e) $\sqrt{\langle k^2_y\rangle/\langle k^2_z\rangle}$, 
(f) $A_K$. Black dotted lines represent aspect ratios for noninteracting case, dashed lines represent ballistic expansion, and solid lines represent nonballistic expansion. As indicated in the graphs (a)-(d), two lower blue solid and dashed lines correspond to $\beta=0^\circ$, and two upper red solid and dashed lines correspond to $\beta=90^\circ$, while in graphs (e) and (f) the position of lines is reversed: two upper blue solid and dashed lines are for $\beta=0^\circ$, and two lower red solid and dashed lines are for $\beta=90^\circ$.}
\label{fig:fig4}
\end{figure*}

Graphs in the left-hand side column of Fig.~\ref{fig:fig4} show aspect ratios $\sqrt{\langle r^2_x\rangle/\langle r^2_z\rangle}$,  $\sqrt{\langle r^2_y\rangle/\langle r^2_z\rangle}$, as well the cloud aspect ratio $A_\mathrm{R}$ in real space during TOF in the collisionless regime. The black dotted line in the middle corresponds to the case of a noninteracting Fermi gas, i.e., $c_0=0$, where the differential equations for the scaling parameters $b_i(t)$ can be solved analytically, yielding
\begin{equation}
\label{eq:bip}
 b^{(0)}_i(t)=\sqrt{1+\left(\frac{\hbar K^{(0)}_i}{M R^{(0)}_i}\right)^2t^2}\, ,
\end{equation} 
with $R^{(0)}_i$ and $K^{(0)}_i$ denoting the global equilibrium radius and momentum in the $i$-th direction, respectively. These scaling parameters are solutions of Eqs.~(\ref{VSTRx})--(\ref{k2}), together with Eq.~(\ref{eq:particlenumberconservation}), for the case of a noninteracting Fermi gas with
\begin{equation}
R^{(0)}_i=\sqrt{\frac{2E_\mathrm{F}}{M \omega_i^2}}\, , \quad \quad K^{(0)}_i=\sqrt{\frac{2ME_\mathrm{F}}{\hbar^2}}\, ,
\end{equation}
where $E_\mathrm{F}=(6N)^{1/3}\hbar \bar{\omega}$ denotes the Fermi energy.
All aspect ratios for a noninteracting Fermi gas in real space asymptotically approach one in the long TOF limit. This shows that a cloud of noninteracting fermions becomes spherical after a long enough expansion, reflecting its isotropic momentum distribution even in the triaxial harmonic trap. As DDI is absent here, the orientation of the magnetic or the electric field, i.e., of the dipole moments of atoms or molecules, has no influence on the FS deformation \cite{Stringari2}. Graphs in the right-hand side column of Fig.~\ref{fig:fig4} show the corresponding aspect ratios in momentum space. As expected, the black dotted line is constant and equal to one, as for the noninteracting fermions the FS is not deformed.

\begin{figure*}[!t]
\centering
\includegraphics[width=6.5cm]{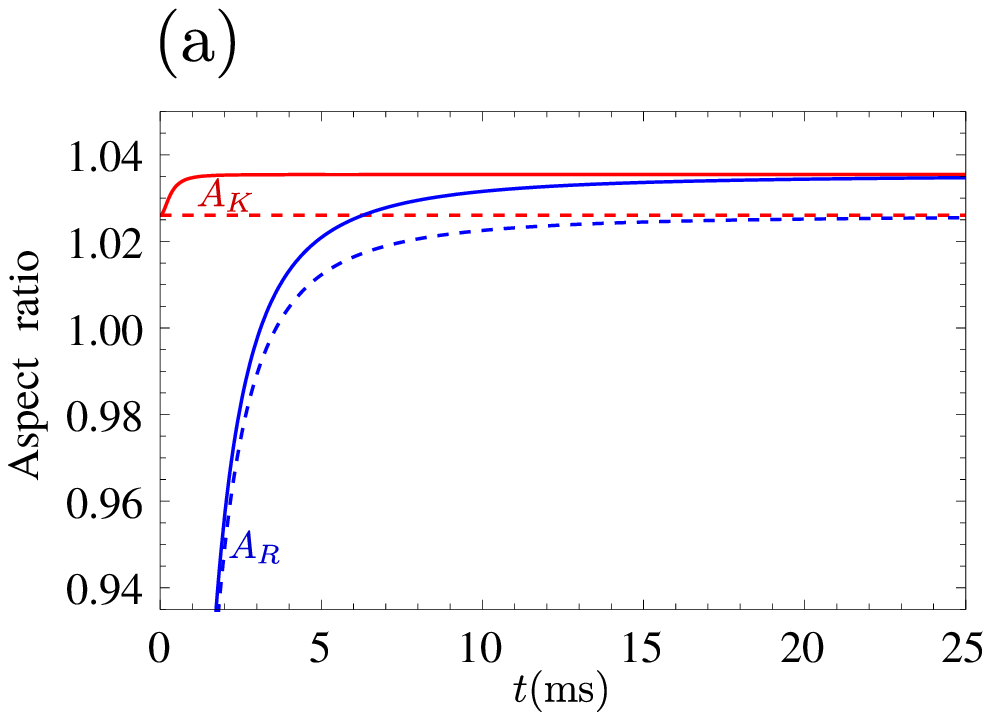}\hspace*{10mm}
\includegraphics[width=6.5cm]{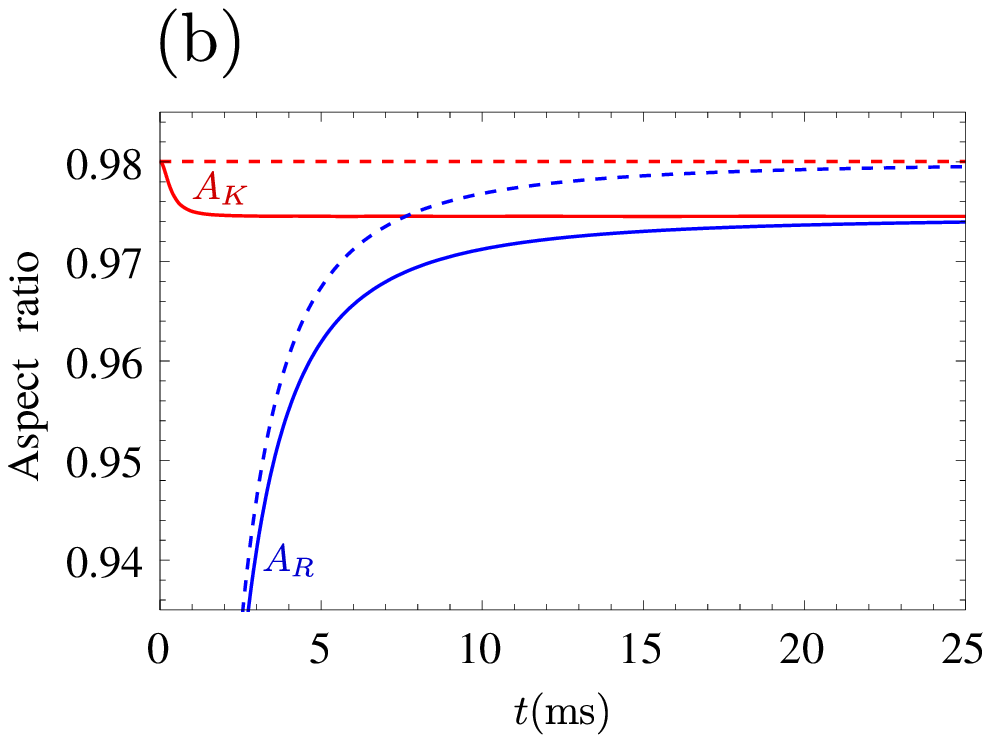}
\caption{Aspect ratios in real and momentum space in the collisionless regime converge to the same asymptotic values during TOF expansion of ultracold gas of $^{167}$Er: (a) $\beta=0^\circ$, (b) $\beta=90^\circ$. Solid (dashed) lines represent aspect ratios for nonballistic (ballistic) expansion of $^{167}$Er. The initially lower branch of blue lines corresponds to real space aspect ratios $A_R$, while the initially upper branch of red lines corresponds to momentum space aspect ratios $A_K$.}
\label{fig:fig5}
\end{figure*}

Furthermore, Fig.~\ref{fig:fig4} also depicts the time dependence of the aspect ratios when the DDI is taken into account. Dashed lines correspond to the ballistic expansion, when DDI is assumed to affect the initial ground state, but not later during the expansion. Mathematically, this means that the ballistic expansion is also determined by Eq.~(\ref{eq:bip}), but now with the parameters $R_{i}$ and $K_{i}$ instead of $R^{(0)}_i$ and $K^{(0)}_i$, respectively:
\begin{equation}
\label{eq:bipbal}
 b^\mathrm{bal}_i(t)=\sqrt{1+\left(\frac{\hbar K_i}{M R_i}\right)^2t^2}\, .
\end{equation} 

Solid lines in Fig.~\ref{fig:fig4} represent results for the nonballistic expansion, when we take DDI into account for calculating both the initial ground state and the subsequent expansion. To obtain these results, one has to solve numerically the coupled differential Eqs.~(\ref{eq:bi}) together with $\theta_i(t)=b_i(t)^{-2}$.
In Figs.~\ref{fig:fig4}(a) to \ref{fig:fig4}(d), top red solid and dashed lines correspond to the orientation of dipoles in the $x$ direction, and bottom blue solid and dashed lines correspond to dipoles' orientation in the $z$ direction, while in Figs.~\ref{fig:fig4}(e) and \ref{fig:fig4}(f) the position of lines turns out to be reversed: top blue lines give results for dipoles in the $z$ direction, and bottom red lines for dipoles in the $x$ direction.

From the graphs in the right-hand column of Fig.~\ref{fig:fig4} we read off that the aspect ratios in momentum space are constant if ballistic expansion approximation is used (all dashed lines). This is not surprising, since here DDI is neglected during the expansion. This can also be shown mathematically if we insert the solution for $b^\mathrm{bal}_i(t)$ from Eq.~(\ref{eq:bipbal}) into expression (\ref{eq:k2}) for $\langle k_i^2 \rangle$, using $\theta_i(t)=b_i(t)^{-2}$, which is valid for the collisionless regime. With this we obtain $\langle k_i^2 \rangle^\mathrm{bal}=K_i^2$, thus the momentum space aspect ratios $\sqrt{\langle k_i^2 \rangle^\mathrm{bal}/\langle k_j^2 \rangle^\mathrm{bal}}=K_i/K_j$ for the ballistic expansion are clearly time-independent and are therefore determined by the initial ground state distribution.

From Fig.~\ref{fig:fig4} we see that the cloud aspect ratios in real space reach their corresponding plateaus after several tens of milliseconds. 
The asymptotic value of $A_R$ for $\beta=0^\circ$ for ballistic expansion is $1.025$, whereas for nonballistic expansion it is $1.035$, thus resulting in a $1\%$ difference due to DDI.
For $\beta=90^\circ$ the asymptotic value of $A_R$ for ballistic expansion is $0.98$, while for nonballistic expansion it is $0.97$, representing again a $1\%$ difference. We also note that for $\beta=0^\circ$ the usual inversion of the cloud shape occurs, while for $\beta=90^\circ$ this is not the case. All these results are in excellent quantitative agreement with the experimental values reported in reference~\cite{Francesca}.

Aspect ratios in momentum space behave similarly, and again a difference of around $0.5-1\%$ between their asymptotic values in a ballistic and nonballistic expansion are observed. But one important difference is that here they are reached much faster, already after several milliseconds. A more detailed analysis reveals that the two terms in Eq.~(\ref{eq:k2}) compete with each other during TOF, but the second term becomes dominant quite fast. Although the corresponding term in Eq.~(\ref{eq:r2mean}) has the same asymptotic behavior, the initial value of $A_K$ is much closer to its asymptotic value than in the case of $A_R$ and, as a consequence, all aspect ratios in momentum space converge faster.

Note that the aspect ratio in momentum space at the initial time $t=0$ coincides with the asymptotic aspect ratio in real space for ballistic expansion:
\begin{equation}
\label{eq:AK0}
A^\mathrm{bal}_K(0)=\lim_{t\to\infty}A^\mathrm{bal}_R(t)\, .
\end{equation}
The ballistic expansion aspect ratio in momentum space at $t=0$ can be calculated from Eq.~(\ref{eq:AK}) by using the initial conditions for the scaling parameters to yield
\begin{equation}
A^\mathrm{bal}_K(0)=\frac{K_z}{\sqrt{K^2_x\cos^2\alpha+K^2_y\sin^2\alpha}}\, .
\label{eq:AK0bal}
\end{equation}
On the other hand, the asymptotic value of the ballistic expansion aspect ratio in real space can be obtained if we insert the approximate expressions for the long-time behavior of the scaling parameters $b^\mathrm{bal}_i(t)\approx\hbar K_i t/(M R_i)$ from Eq.~(\ref{eq:bipbal}) into Eq.~(\ref{eq:AR}), which yields the same value as $A^\mathrm{bal}_K(0)$ in Eq.~(\ref{eq:AK0bal}).
This fact was used in reference~\cite{Francesca} in order to observe the ellipsoidal deformation of the FS, as the real space aspect ratios can be readily measured during TOF. However, this is only correct within the ballistic approximation, as for the truly nonballistic expansion such a relationship is no longer valid.

But from Fig.~\ref{fig:fig5} we read off that both for ballistic (dashed lines) and nonballistic (solid lines) expansion another relationship seems to hold. Namely the aspect ratios in momentum space and the corresponding aspect ratios in real space turn out to have the same asymptotic values:
\begin{equation}
\label{eq:AKeqAR}
\lim_{t\to\infty}A_K(t)=\lim_{t\to\infty}A_R(t)\, .
\end{equation}
The above is true for both considered orientations of dipoles, i.e., $\beta=0^\circ$ and $\beta=90^\circ$. A similar conclusion was reached in reference~\cite{Sogo} for a dipolar Fermi gas that was initially in a cylindrically symmetric harmonic trap, but we see here that this is true even for a fully anisotropic harmonic trapping potential.
Note that this finding cannot be directly used to determine the aspect ratio in momentum space at $t=0$ and the corresponding initial deformation of the FS, as for the ballistic expansion according to Eq.~(\ref{eq:AK0}). But this observation still allows to theoretically extract information on the momentum space distribution from experimental data. However, this requires that the corresponding equations are propagated backwards in time, so that the initial distribution in momentum space is calculated starting from the experimentally measured distribution in real space. Here the numerical challenge is that this backward propagation has to be calculated for an infinitely long expansion time.
 
\subsection{Hydrodynamic regime}
\label{subsec:hy}

In contrast to the previously considered collisionless regime, where collisions are completely neglected, we now turn to the hydrodynamic regime, where the system is supposed to have such a high density and, therefore, such a high collision rate, that it is always in local equilibrium. Although realistic systems, even if initially in the hydrodynamic regime, eventually become collisionless during the expansion, we follow references~\cite{Lima1, Lima2} and consider this theoretical limiting case for the sake of completeness.

In the hydrodynamic regime, the scaling parameters $\theta_i^\mathrm{hd}$ always coincide with the local equilibrium values, i.e., we have $\theta_i^\mathrm{hd}=\theta_i^\mathrm{le}$. However, since the limit $\tau\to 0$ holds, the last term in the left-hand side of Eqs.~(\ref{eq:theta_i}) is undetermined. Therefore, instead of  Eqs.~(\ref{eq:theta_i}), the hydrodynamic regime is defined via the condition \cite{Stringari-ansatz}
\begin{equation}
 \Gamma^\mathrm{hd}(t)^{-1}=\prod_i b_i^\mathrm{hd}(t)\sqrt{\theta_i^\mathrm{hd}(t)}=1\, . \label{eq:hydef}
\end{equation}
Using this condition, minimizing the Hartree-Fock energy (\ref{eq:totalenergy}) in the local equilibrium leads to the equations \cite{Lima1, Lima2}
 \begin{eqnarray}
&&\hspace*{-5mm}\theta_x^\mathrm{hd}=\theta_y^\mathrm{hd}\, ,\label{eq:th_hy}\\
&&\hspace*{-5mm}\frac{\hbar^2 \theta_z^\mathrm{hd}K_z^2}{2 M}-\frac{\hbar^2 \theta_x^\mathrm{hd}K_x^2}{2 M}=\nonumber\\
&&\hspace*{-5mm}    \frac{72 N c_0}{\prod_j b^\mathrm{hd}_j R_j}\left[1+ \frac{\left(2\theta_x^\mathrm{hd}K_x^2+\theta_z^\mathrm{hd}K_z^2\right)f_s \left( \frac{\sqrt{\theta_z^\mathrm{hd}}K_z}{\sqrt{\theta_x^\mathrm{hd}}K_x}\right)}{2\left(\theta_z^\mathrm{hd}K_z^2- \theta_x^\mathrm{hd}K_x^2\right)} \right] .
\label{eq:hy2}
\end{eqnarray}
Equations (\ref{eq:bi}), together with the identifications $b_i(t)=b^\mathrm{hd}_i(t)$ and $\theta_i(t)=\theta^\mathrm{hd}_i(t)$, with Eqs.~(\ref{eq:th_hy}) and (\ref{eq:hy2}), and the normalization condition (\ref{eq:hydef}) represent a closed set of six equations for the six respective scaling parameters in the hydrodynamic regime. These equations are solved numerically during the nonballistic TOF expansion. For comparison, we have also solved the corresponding equations for the ballistic expansion, although the hydrodynamic regime implies that DDI cannot be neglected at any point.

\begin{figure*}[!t]
\centering
\includegraphics[width=6.5cm]{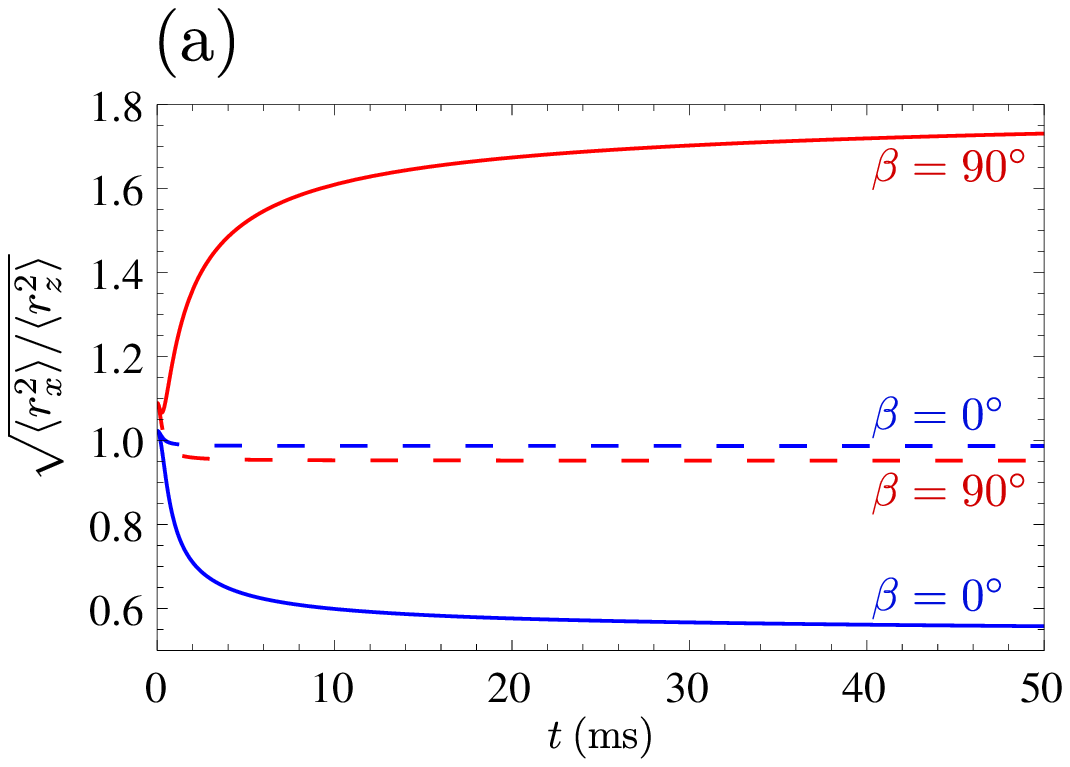}\hspace*{10mm}
\includegraphics[width=6.5cm]{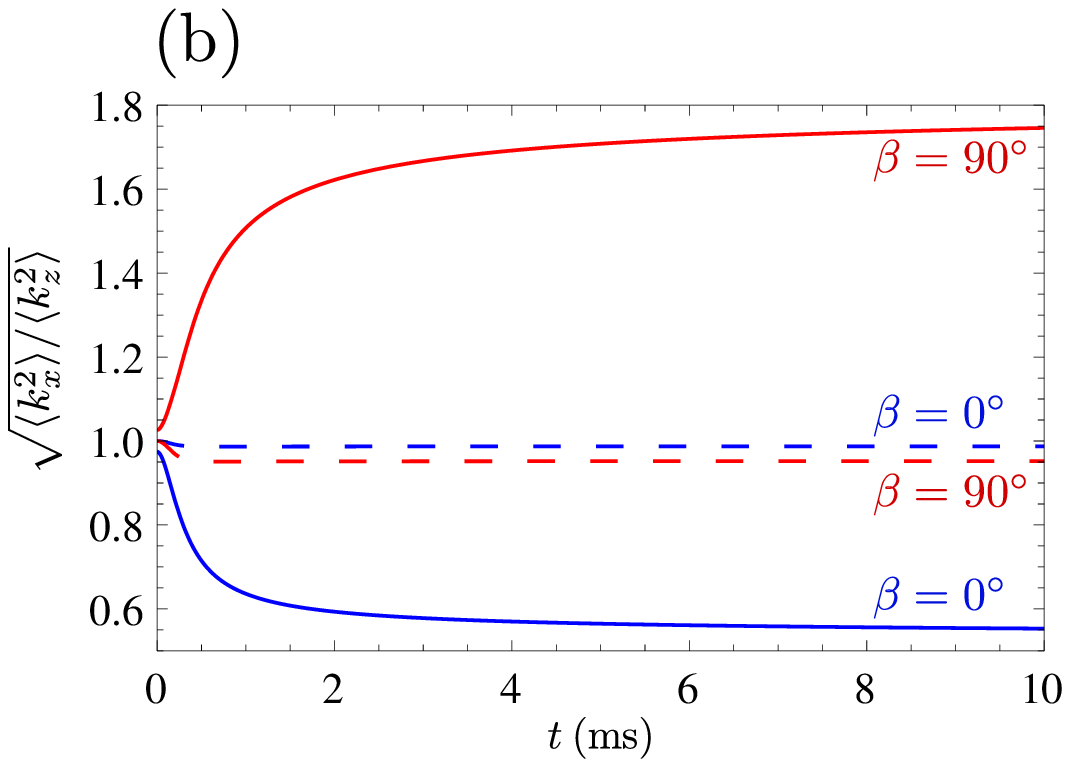}
\includegraphics[width=6.5cm]{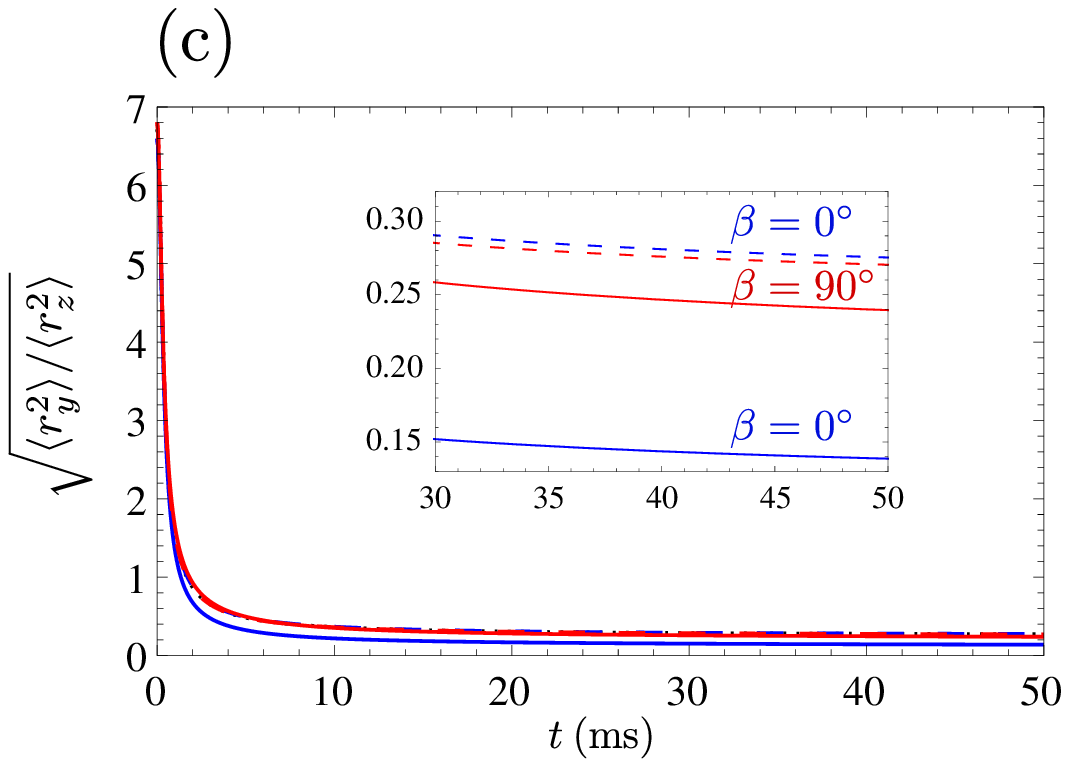}\hspace*{10mm}
\includegraphics[width=6.5cm]{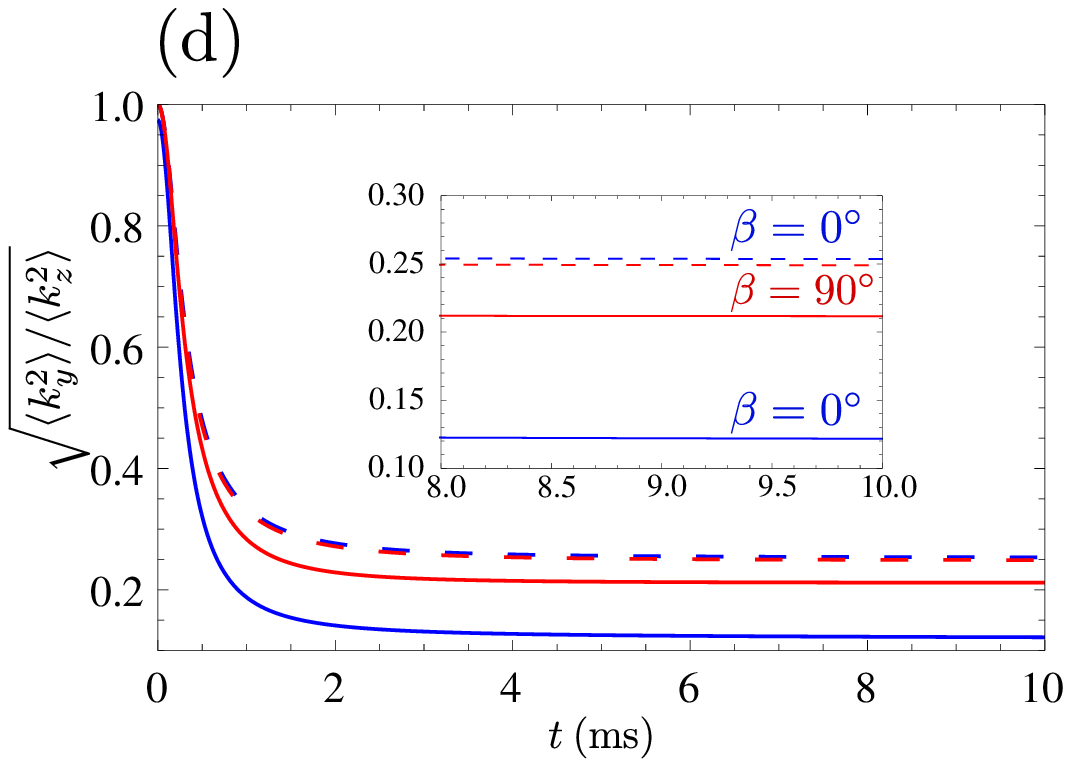}
\includegraphics[width=6.5cm]{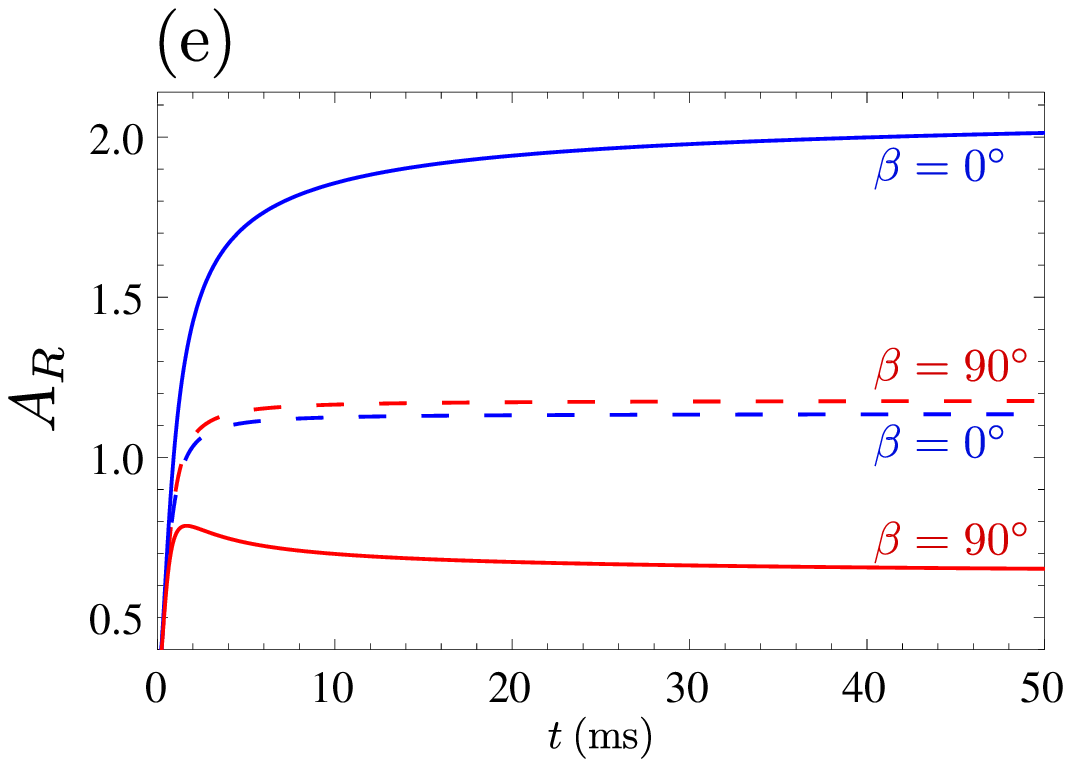}\hspace*{10mm}
\includegraphics[width=6.5cm]{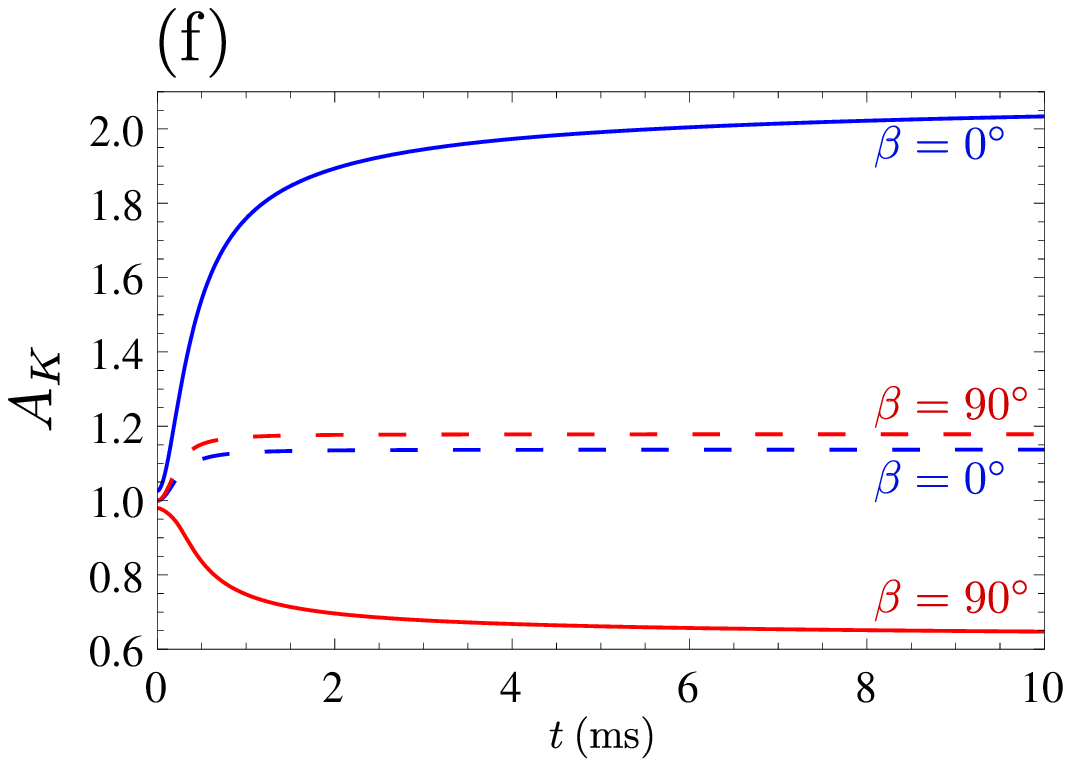}
\caption{Aspect ratios in real and momentum space in the hydrodynamic regime during TOF expansion of ultracold gas of $^{167}$Er:
(a) $\sqrt{\langle r^2_x\rangle/\langle r^2_z\rangle}$,
(b) $\sqrt{\langle r^2_y\rangle/\langle r^2_z\rangle}$,
(c) $A_R$,
(d) $\sqrt{\langle k^2_x\rangle/\langle k^2_z\rangle}$, 
(e) $\sqrt{\langle k^2_y\rangle/\langle k^2_z\rangle}$, 
(f) $A_K$. Dashed lines represent ballistic expansion and solid lines represent nonballistic expansion. As indicated in graphs (a)-(d), the lower blue solid and the upper blue dashed line correspond to $\beta=0^\circ$, while the upper red solid and the lower red dashed line correspond to $\beta=90^\circ$. In graphs (e) and (f) the position of lines is reversed: the upper blue solid and the lower blue dashed line are for $\beta=0^\circ$; the lower red solid and the upper red dashed line are for $\beta=90^\circ$.}
\label{fig:fig6}
\end{figure*}

Figure~\ref{fig:fig6} shows the corresponding aspect ratios in real and momentum space for $^{167}$Er. As expected, we see that there is a significant difference between the ballistic and the nonballistic expansion, in contrast to the collisionless regime in Fig.~\ref{fig:fig4}. From the graphs in the left column of Fig.~\ref{fig:fig6} we observe that the real space aspect ratios for $\beta=0^\circ$ behave generally similar to those in the collisionless regime, including the cloud shape inversion, although the asymptotic values differ more from their initial values for nonballistic expansion. On the other hand, for $\beta=90^\circ$ we see a qualitatively different behavior in Fig.~\ref{fig:fig6}(a), where the aspect ratio $\sqrt{\langle r^2_x\rangle/\langle r^2_z\rangle}$ increases, while in Fig.~\ref{fig:fig4}(a) it decreases. In Fig.~\ref{fig:fig6}(c) for $\beta=90^\circ$ we read off that the aspect ratio $A_R$ even behaves nonmonotonously, with a local maximum at around 1~ms, while in the collisionless regime it only increases monotonously until it reaches its asymptotic value. However, again the inversion of the cloud shape is not present for $\beta=90^\circ$, unlike for $\beta=0^\circ$. We also note that the positions of ballistic expansion curves are reversed in all graphs compared to the collisionless regime, including those for momentum space aspect ratios.

\begin{figure*}[!t]
\centering
\includegraphics[width=6.5cm]{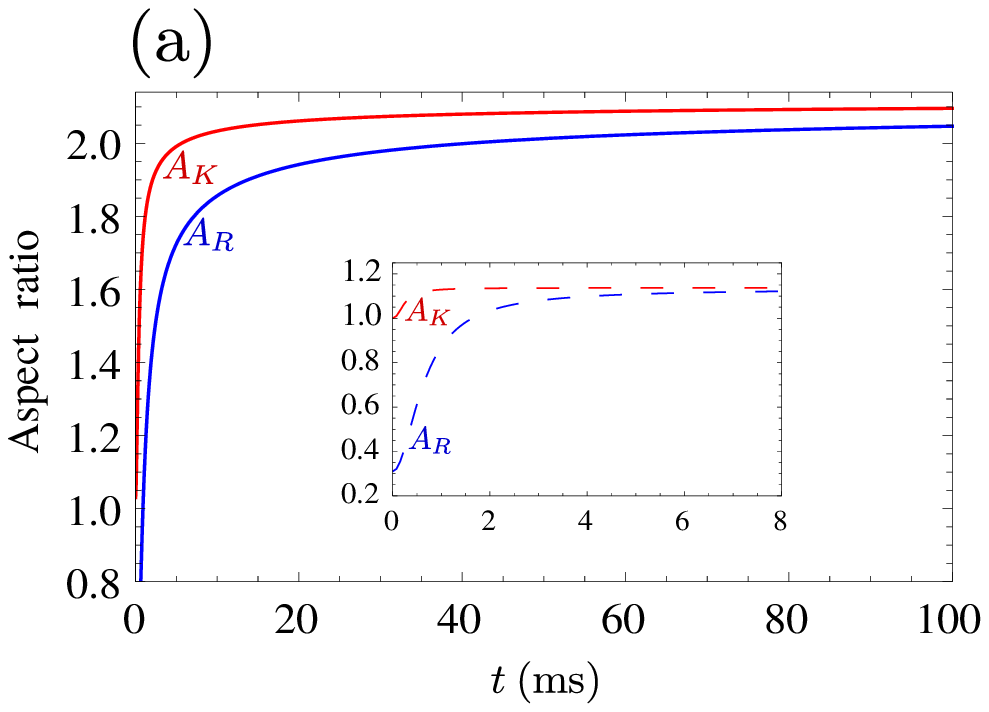}\hspace*{10mm}
\includegraphics[width=6.5cm]{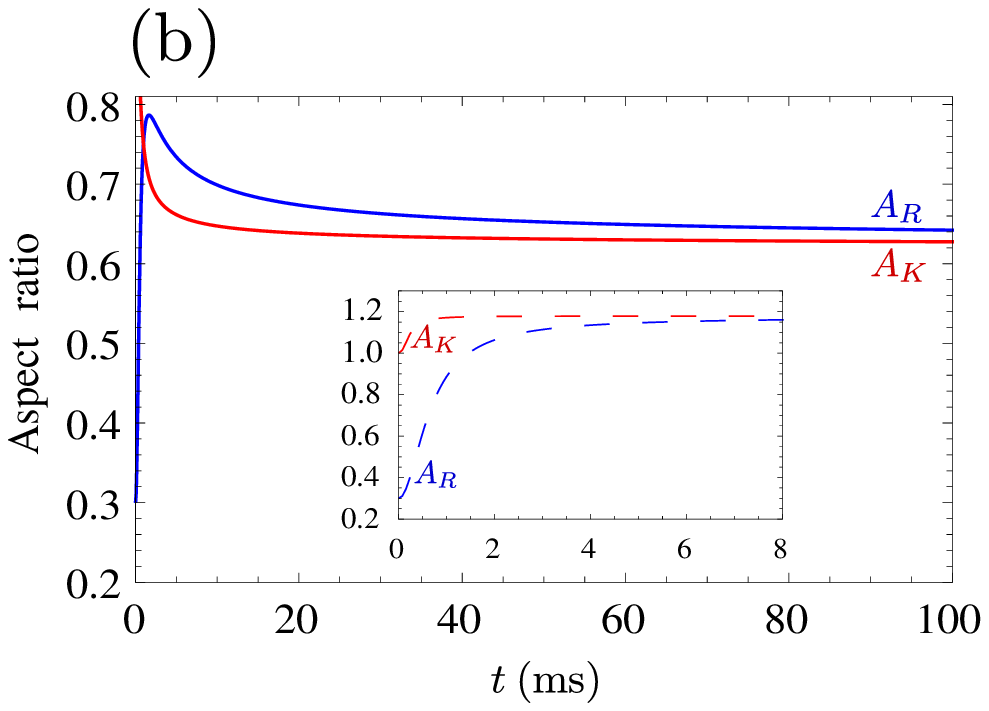}
\caption{Aspect ratios in real and momentum space in the hydrodynamic regime converge to the same asymptotic values during TOF expansion of ultracold gas of $^{167}$Er:
(a) $\beta=0^\circ$, (b) $\beta=90^\circ$. Solid lines give aspect ratios for nonballistic expansion, while dashed lines in the insets show the corresponding ballistic results. Blue lower lines in (a) and blue upper lines in (b) correspond to $A_R$, while red upper lines in (a) and red lower lines in (b) correspond to $A_K$.}
\label{fig:fig7}
\end{figure*}

The behavior of momentum space aspect ratios in the right column of Fig.~\ref{fig:fig6} is generally the same as in Fig.~\ref{fig:fig4} for the collisionless regime, just with larger differences between initial and asymptotic values, for both cases $\beta=0^\circ$ and $\beta=90^\circ$.

The final cloud aspect ratio in real space for nonballistic expansion is twice as large as the corresponding collisionless value for $\beta=0^\circ$, while for $\beta=90^\circ$ the asymptotic value of the aspect ratio is around 0.65, which amounts to a decrease of around $35\%$ compared to the collisionless value. For the ballistic expansion, which we know to be unrealistic in the hydrodynamic regime, the corresponding increase and decrease amounts to both around $12\%$. Similar numbers are also obtained for the momentum space aspect ratio $A_K$, as can be seen from the graphs on the right-hand side in Fig.~\ref{fig:fig6}. Since the corresponding values in the collisionless regime are all close to one, the above percentages also apply here, and represent the results for the ellipsoidal deformation of the FS in the hydrodynamic regime.

The same conclusion can be also obtained from Fig.~\ref{fig:fig7}, where we compare aspect ratios in real and momentum space. Furthermore, with these graphs we confirm that the asymptotic values of the aspect ratios $A_R$ and $A_K$ also coincide in the hydrodynamic regime for both cases $\beta=0^\circ$ and $\beta=90^\circ$, as stated by Eq.~(\ref{eq:AKeqAR}) for the collisionless regime. If we compare the convergence of aspect ratios to their asymptotic values in Figs.~\ref{fig:fig5} and \ref{fig:fig7}, we see that in the hydrodynamic regime typical times to reach the plateau are similar in real and in momentum space, and have the value of several tens of milliseconds. This coincides with the corresponding convergence times for real space aspect ratios in the collisionless regime, where also a significant difference between the initial and the asymptotic value of aspect ratios occurred. Only in the case of momentum space aspect ratios in the collisionless regime, where the deformation of the FS is small during the whole expansion, asymptotic values can be reached faster, namely in just few milliseconds.
 
However, as already emphasized, even if initially in the hydrodynamic regime, the dipolar Fermi gas becomes more and more dilute during the TOF expansion, and the hydrodynamic regime continuously goes over into the collisional regime, and, finally, into the collisionless regime. Therefore, we model the collisional regime in the remainder of this section, since it is relevant for experiments where the density of the Fermi gas is high enough so that we can assume it is initially in the collisional or in the hydrodynamic regime.

\subsection{Collisional regime with constant relaxation time}
\label{subsec:cr}

Here we start considering the collisional regime and assume that the relaxation-time approximation (\ref{eq:rta}) can be applied. Furthermore, in this section we presume that the relaxation time $\tau$ remains constant during the TOF. The latter assumption is only valid for short times of flight, before the density of the gas decreases significantly. We will improve upon this approximation in Sec.~\ref{subsec:crsc}, where the relaxation time is determined self-consistently.

However, provided that the relaxation time is constant, the TOF dynamics can be obtained by directly solving Eqs.~(\ref{eq:bi}) and (\ref{eq:theta_i}) for a given value of $\tau$. Note that the values of the scaling parameters $\theta_i^\mathrm{le}$ in local equilibrium are obtained according to Sec.~\ref{subsec:hy}, i.e., they represent the solutions of the equations for the hydrodynamic regime $\theta^\mathrm{hd}_i$.

The physical meaning of Eqs.~(\ref{eq:theta_i}) is that dissipation occurs when the system is outside of a local equilibrium as long as there are collisions, i.e., as long as the relaxation time $\tau$ remains finite. Effects of collisions are therefore described through Eqs.~(\ref{eq:theta_i}), whereas Eqs.~(\ref{eq:bi}) for the scaling parameters $b_i$ do not directly contain such terms. However, effects of collisions enter indirectly into Eqs.~(\ref{eq:bi}) through the scaling parameters $\theta_i$.

Here we numerically solve the coupled system of Eqs.~(\ref{eq:bi}) and (\ref{eq:theta_i}) during the nonballistic expansion for a fixed value of the relaxation time $\tau$.
Varying the value of the relaxation time we are able to describe all regimes, from the collisionless, obtained in the limit $\tau \to \infty$, to the hydrodynamic, obtained in the limit $\tau \to 0$. In particular, although the approximation of a fixed relaxation time is not realistic for longer expansion times, it allows us to understand and describe in more detail different collisional regimes, for finite values of $\tau$, when the system undergoes a crossover from one limiting regime to the other.

\begin{figure}[!b]
\centering
\includegraphics[width=6.2cm]{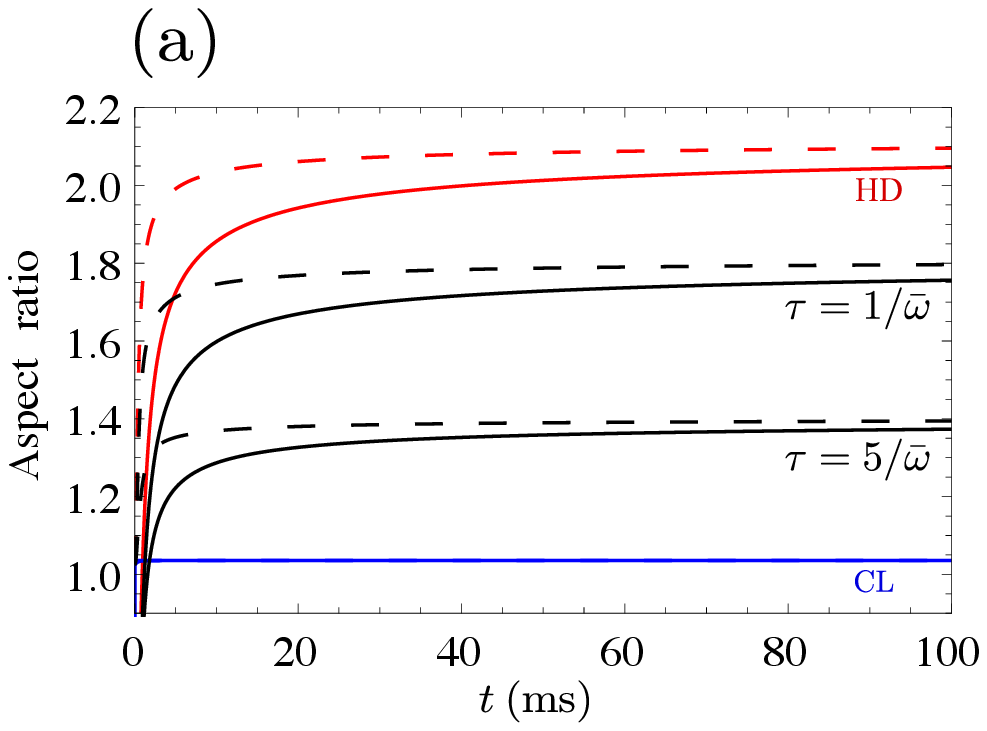}
\includegraphics[width=6.2cm]{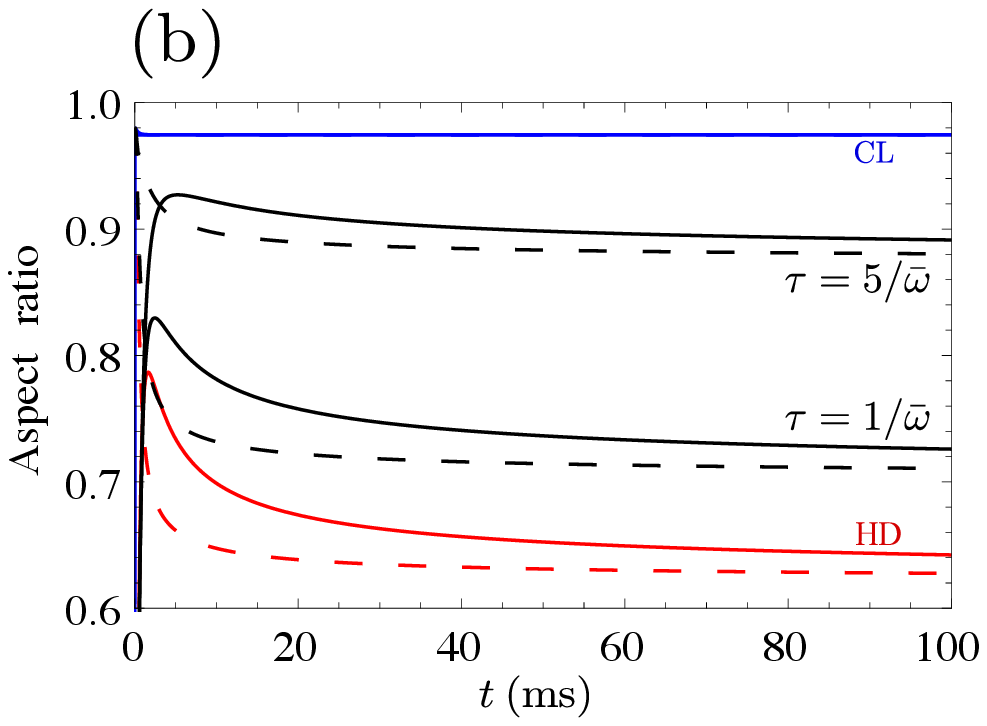}
\caption{Aspect ratios in real (solid lines) and momentum space (dashed lines) in the collisional regime during TOF expansion of ultracold gas of $^{167}$Er:
(a) $\beta=0^\circ$, (b) $\beta=90^\circ$. The pairs of curves in (a) from top to bottom and in (b) from bottom to top correspond to: hydrodynamic regime (HD, red), collisional regime (black) for fixed relaxation times $\tau=1/\bar{\omega}$ and $\tau=5/\bar{\omega}$, and collisionless regime (CL, blue).}
\label{fig:fig8}
\end{figure}

Figure~\ref{fig:fig8} shows the obtained aspect ratios for $^{167}$Er in real and momentum space for the two limiting cases considered previously, the collisionless and the hydrodynamic regime, as well as for the collisional regime with the fixed relaxation times $\tau=1/\bar{\omega}$ and $\tau=5/\bar{\omega}$, where $\bar{\omega}$ represents a geometric mean of the trap frequencies. Depending on the respective geometry, asymptotic values of aspect ratios either decrease with increasing relaxation time, see Figs.~\ref{fig:fig8}(a) for $\beta=0^\circ$, or vice versa, see Fig.~\ref{fig:fig8}(b) for $\beta=90^\circ$. We also read off from these figures that the corresponding asymptotic values of aspect ratios in real and momentum space in the collisional regime are again equal according to relation (\ref{eq:AKeqAR}).

\begin{figure}[!b]
\centering
\includegraphics[width=6.5cm]{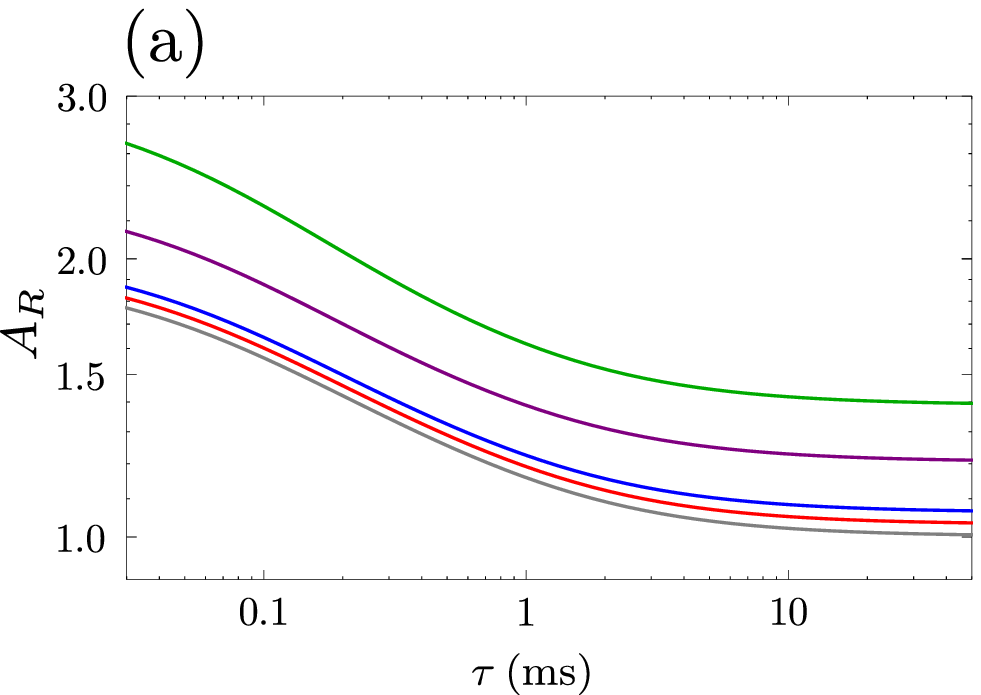}
\includegraphics[width=6.5cm]{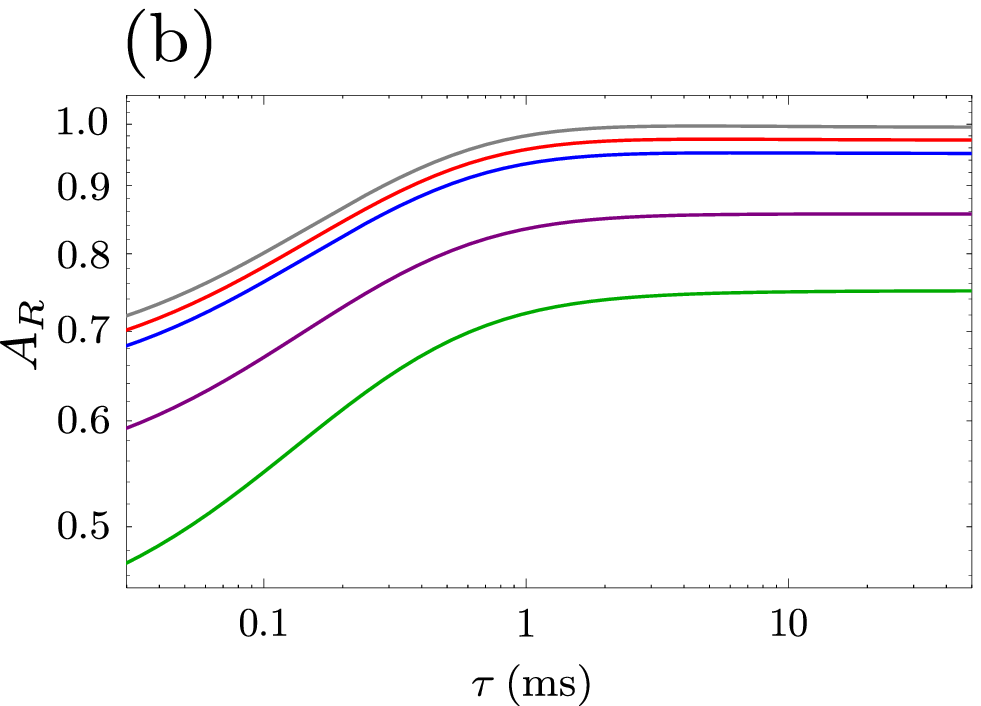}
\caption{Aspect ratios in real space after $t=10$~ms TOF as function of fixed relaxation time $\tau$ for different ultracold Fermi gases: (a) $\beta=0^\circ$, (b) $\beta=90^\circ$. The curves in (a) from bottom to top and in (b) from top to bottom correspond to: $^{53}$Cr (gray), $^{167}$Er (red), $^{161}$Dy (blue), $^{40}$K$^{87}$Rb (purple), and $^{167}$Er$^{168}$Er (green).}
\label{fig:fig9}
\end{figure}

Motivated by the experiment reported in reference~\cite{Francesca}, in Fig.~\ref{fig:fig9} we plot the aspect ratio in real space $A_R$ obtained after $t=10$~ms TOF as a function of a fixed relaxation time $\tau$ for two different orientations of dipoles for the respective ultracold Fermi gases given in Table~\ref{tab:tab2}. If the dipoles are oriented along $z$ axis, i.e., Fig.~\ref{fig:fig9}(a), the corresponding aspect ratios for any fixed value of the relaxation time $\tau$ increase monotonously with the relative dipolar interaction strength $\epsilon_\mathrm{dd}$, while for the dipoles along $x$ axis, i.e., Fig.~\ref{fig:fig9}(b), the situation is opposite, as expected. Note that the corresponding curves for the noninteracting case $\epsilon_\mathrm{dd}=0$ would be quite close to those for $^{53}$Cr, as can already be expected according to Fig.~\ref{fig:fig2}.

Plots like those in Fig.~\ref{fig:fig9} represent powerful diagnostic tools for estimating the relaxation time $\tau$ from experimentally measured values of aspect ratios $A_R$ for sufficiently short TOF, when the fixed relaxation-time approximation is still applicable. Furthermore, these graphs can be used for estimating the time scale $t$ to approach the asymptotic values of the aspect ratios from experimentally available TOF expansion data. Provided that it turns out for a TOF $t$ that the corresponding relaxation time $\tau$ satisfies the condition $\bar{\omega}\tau\gg 1$, one has already reached the collisionless regime. This means that for longer times $t$ no further change of the aspect ratio is expected as one is already quite close to its asymptotic value.
 
\subsection{Collisional regime with self-consistently determined relaxation time}
\label{subsec:crsc}

Whereas in Sec.~\ref{subsec:cr} we assumed that the relaxation time is constant, now we model the TOF expansion of ultracold dipolar Fermi gases more realistically and take into account that the relaxation time is also time-dependent. Namely, during TOF the gas rapidly expands, the distance between atoms grows, and as a consequence the relaxation time increases, thus eventually leading the system into the collisionless regime, even if initially it was in the hydrodynamic or in the collisional regime.

\begin{figure*}[!t]
\centering
\includegraphics[width=6.5cm]{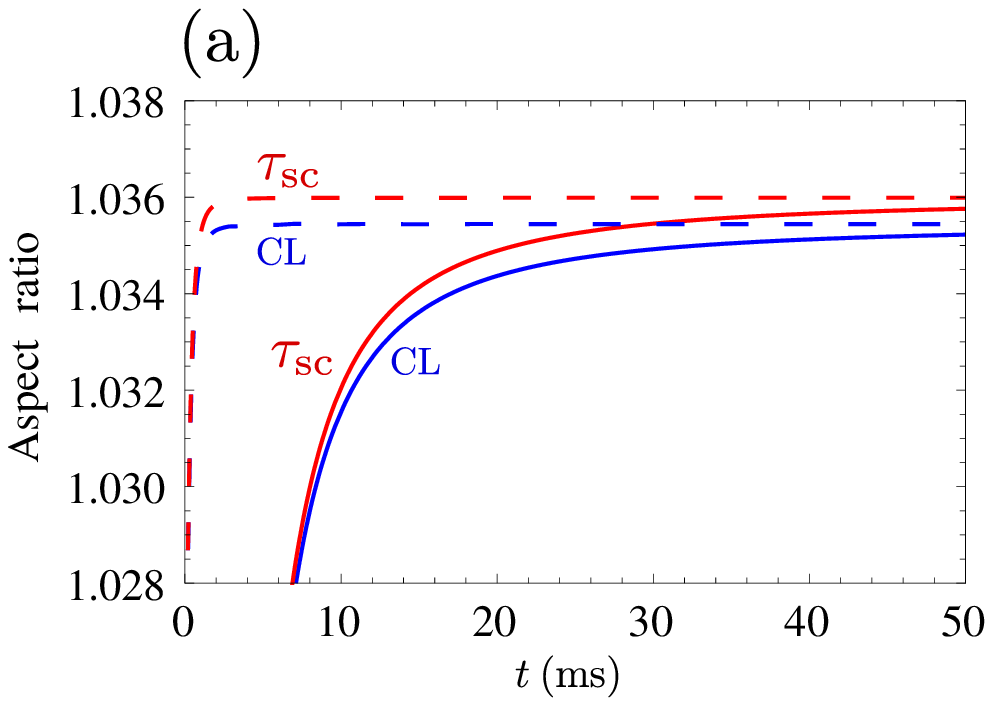}\hspace*{10mm}
\includegraphics[width=6.5cm]{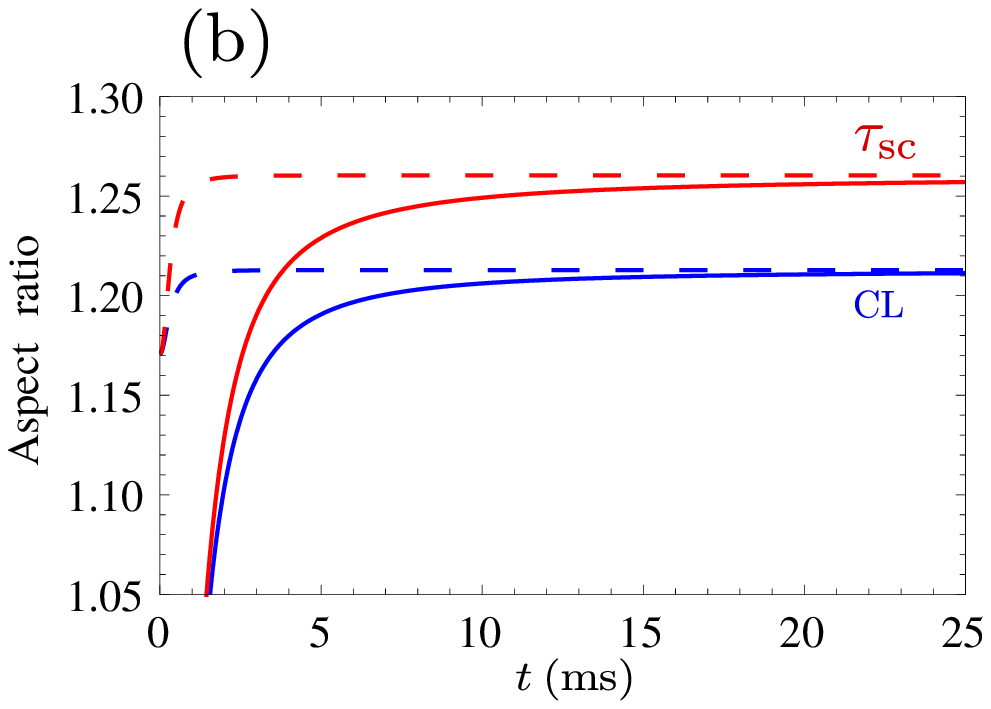}
\caption{Aspect ratios in real (solid lines) and momentum space (dashed lines) in the collisional regime during TOF expansion for $\beta=0^\circ$: (a) $^{167}$Er, (b) $^{40}$K$^{87}$Rb. Red upper solid and dashed line correspond to expansion dynamics with self-consistently determined relaxation time $\tau_\mathrm{SC}$. For comparison, blue lower solid and dashed line give the corresponding aspect ratios for the collisionless regime (CL).}
\label{fig:fig10}
\end{figure*}

In order to quantify this physical notion, one would have to calculate the collision integral on the right-hand side of Eq.~(\ref{eq:bve}), which requires a detailed modeling of scattering processes in the system, i.e., the elastic collisions of fermionic atoms or molecules that arise purely from universal dipolar scattering. The standard approach for the case of a system close to local equilibrium is to use the relaxation-time approximation \cite{TheBECBook, Stringari-ansatz}, which is given by Eq.~(\ref{eq:rta}). In Ref.~\cite{Bohn-Jin} it was derived that the characteristic relaxation time for a classical gas can be expressed as
\begin{equation}
\tau=\frac{\alpha_\mathrm{coll}}{ \bar{n}\sigma_\mathrm{el}v}\, ,
\label{eq:taucl}
\end{equation}
where the parameter $\alpha_\mathrm{coll}$ denotes a geometry-dependent average number of collisions which is necessary to rethermalize the system after a collision, $\bar{n}$ represents the mean number density, $\sigma_\mathrm{el}$ is the total elastic cross section, and $v$ is the mean relative velocity. In Ref.~\cite{tau} it was heuristically argued and experimentally confirmed that for quantum degenerate dipolar fermionic systems at low temperatures and parameter regimes considered here, the relaxation time can be modeled by a modified expression
\begin{equation}
\tau_\mathrm{SC}=\frac{\alpha_\mathrm{coll}}{\eta \bar{n}\sigma_\mathrm{el}v}\, ,
\label{eq:tau}
\end{equation}
which allows us to calculate it self-consistently, hence the subscript SC. In the above equation, $\eta$ stands for a Pauli suppression factor, which represents the reduction of the rethermalization rate in a degenerate Fermi gas due to Pauli blocking, and amounts to $\eta=1$ for non-degenerate gases. The Pauli suppression factor depends on the degeneration level of fermions and is usually expressed as a function of the dimensionless temperature $T/T_\mathrm{F}$, where $T$ denotes the temperature and $T_\mathrm{F}$ is the Fermi temperature for the corresponding system.

In the considered case, the mean number density is given by $\bar{n}=N/V(t)$, where the volume $V(t)$ of the Fermi gas increases during the TOF expansion according to
\begin{equation}
 V(t)=\frac{4\pi}{3}\prod_i R_i b_i(t)\, .
\label{eq:volume}
\end{equation}
The total elastic cross section $\sigma_\mathrm{el}$ is universally related to the dipole moment of fermions \cite{sigma} according to
\begin{equation}
 \sigma_\mathrm{el}=\frac{32\pi}{15}a_\mathrm{dd}^2\, ,
\label{eq:sigma}
\end{equation}
where $a_\mathrm{dd}=C_\mathrm{dd}M/8\pi\hbar^2$ represents a characteristic dipole length. The mean relative velocity $v$ is given by
\begin{equation}
\label{eq:vmean}
v=\sqrt{\frac{16k_\mathrm{B}T}{\pi M}}\, .
\end{equation}

For the parameters of the experiment \cite{Francesca} with atomic $^{167}$Er used throughout this paper, the universal dipolar scattering theory \cite{tau} predicts the total elastic cross section value $\sigma_\mathrm{el}=1.8 \times 10^{-12}\, \mathrm{cm}^2$, which agrees with the value measured in reference~\cite{Er2}. The temperature of the system was set to $T/T_\mathrm{F}=0.18$, with $T_\mathrm{F}=1.1\, \mu\mathrm{K}$, which yields the Pauli suppression factor $\eta=0.3$ \cite{tau}, as well as the mean relative velocity $v$ according to Eq.~(\ref{eq:vmean}). To completely fix all parameters which are necessary for a self-consistent determination of the relaxation time with Eq.~(\ref{eq:tau}), we still need to take the appropriate value of the number of collisions $\alpha_\mathrm{coll}$ for the given geometry, i.e., for the given angle $\beta$ from reference~\cite{tau}.

Figure~\ref{fig:fig10}(a) shows the corresponding aspect ratios in real and momentum space for $^{167}$Er during TOF expansion for $\beta=0^\circ$, for which the average number of collisions to rethermalize is $\alpha_\mathrm{coll}=3.2$ \cite{tau}. The red upper solid and dashed line in Fig.~\ref{fig:fig10}(a) are obtained by numerically solving Eqs.~(\ref{eq:bi}) and (\ref{eq:theta_i}), with the relaxation time determined self-consistently through Eq.~(\ref{eq:tau}). In the same plot we also see for the sake of comparison the results for the collisionless regime in terms of the blue lower solid and dashed line. The difference between the corresponding lines is less than $0.1\%$, which is certainly within the experimental error bars, and confirms that the system is indeed very close to the collisionless regime, as assumed in reference~\cite{Francesca}.

However, systems with a stronger DDI can easily reach the collisional regime, where a finite value for the relaxation time has to be taken into account. In order to demonstrate this, we analyze the TOF expansion of a $^{40}$K$^{87}$Rb dipolar Fermi gas \cite{KRb}, whose relative dipolar interaction strength is $\epsilon_\mathrm{dd}=0.97$, compared to $\epsilon_\mathrm{dd}=0.15$ for $^{167}$Er (see Table~\ref{tab:tab2}). Polar molecules have generically stronger electric dipole moments in comparison with the magnetic dipole moments of atoms, which is expected to yield a sensible difference in the respective aspect ratios.

In Fig.~\ref{fig:fig10}(b) we show the TOF expansion dynamics for $^{40}$K$^{87}$Rb for the same number of fermions and trap frequencies as in reference~\cite{Francesca}. The temperature of the system is assumed to be $T=350\, \mathrm{nK}=0.3\, T_\mathrm{F}$, as in reference~\cite{KRb}, which yields the Pauli suppression factor $\eta=0.5$ \cite{tau}. The total elastic cross section according to Eq.~(\ref{eq:sigma}) in this case is $\sigma_\mathrm{el}=9.6 \times 10^{-11}\, \mathrm{cm}^2$, in agreement with the results of reference \cite{sigma}. The average number of collisions to rethermalize is again taken to be $\alpha_\mathrm{coll}=3.2$ for $\beta=0^\circ$ \cite{tau}. The difference between the aspect ratios calculated using the self-consistently determined relaxation time and those calculated assuming that the system is in the collisionless regime are here around $10\%$, which could be clearly observed in future experiments. Furthermore, for polar molecules with a stronger DDI the differences are expected to be even more pronounced. Thus in experiments with such systems the relaxation time must be taken into account, for instance through the self-consistent approach presented here. We also note that the asymptotic values of aspect ratios in real and momentum space turn out to be again the same, as was already the case in the collisionless and in the hydrodynamic regime.

\begin{figure}[!t]
\centering
\includegraphics[width=6.5cm]{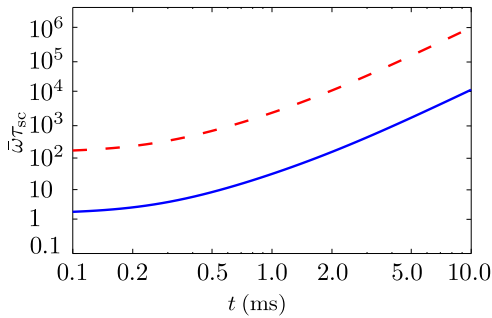}
\caption{Self-consistently determined relaxation time (\ref{eq:tau}) as function of TOF $t$ for ultracold Fermi gas of $^{167}$Er (red dashed line) and $^{40}$K$^{87}$Rb (blue solid line) for $\beta=0^\circ$. The collisionless regime is achieved for $\bar{\omega}\tau_\mathrm{SC}\gg 1$.}
\label{fig:fig11}
\end{figure}

Figure~\ref{fig:fig11} shows the resulting time dependence of the self-consistently determined relaxation time during TOF expansion for both analyzed species, i.e., $^{167}$Er (red dashed line) and $^{40}$K$^{87}$Rb (blue solid line). As we can see, for an atomic gas of $^{167}$Er the relaxation time satisfies the condition $\bar{\omega}\tau_\mathrm{SC}\gg 1$ right from the beginning, which further justifies the previous conclusion that the system is always in the collisionless regime \cite{Francesca}. For a molecular gas of $^{40}$K$^{87}$Rb, however, this condition is satisfied only after 1-2~ms, so initially the system is in the collisional regime. Furthermore, we recognize that the relaxation time increases quite fast, namely faster than exponential, as we can see from the log-log plot of Fig.~\ref{fig:fig11}. Thus, the approximation of Sec.~\ref{subsec:cr} with a fixed relaxation time would clearly not be suitable, and a self-consistent approach as presented here is indispensable.

\begin{figure}[!b]
\centering
\includegraphics[width=6.5cm]{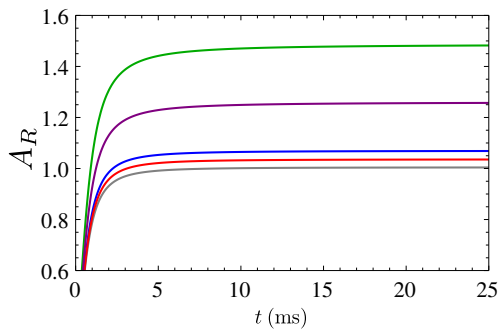}
\caption{Aspect ratios in real space during TOF expansion in the collisional regime with self-consistently determined relaxation time for different ultracold Fermi gases for $\beta=0^\circ$. The curves from bottom to top correspond to: $^{53}$Cr (gray), $^{167}$Er (red), $^{161}$Dy (blue), $^{40}$K$^{87}$Rb (purple), and $^{167}$Er$^{168}$Er (green).}
\label{fig:fig12}
\end{figure}

To summarize our results for the aspect ratios during the TOF expansion in the collisional regime with self-consistently determined relaxation time, in Fig.~\ref{fig:fig12} we combine our results for the time dependence of aspect ratios in real space $A_R$ for $\beta=0^\circ$ for $^{167}$Er from Fig.~\ref{fig:fig10}(a) and $^{40}$K$^{87}$Rb from Fig.~\ref{fig:fig10}(b) with the results for three other considered dipolar fermionic species from Table~\ref{tab:tab2}. We see that increasing relative dipolar interaction strength leads to increasing aspect ratios after long TOF. While for $^{53}$Cr, $^{167}$Er, and $^{161}$Dy asymptotic values of $A_R$ are just few percent over 1, for $^{40}$K$^{87}$Rb we obtain a value of about 1.26, and for $^{167}$Er$^{168}$Er about 1.48.

\section{Conclusions}
\label{sec:con}

In conclusion, we have explored the properties of trapped dipolar Fermi gases at zero temperature in global equilibrium, as well as their dynamics during TOF expansion by using the Boltzmann-Vlasov formalism in the relaxation-time approximation for the collision integral. We have studied the aspect ratios of the fermionic cloud in real and momentum space, including the deformation of the Fermi sphere due to the presence of the dipole-dipole interaction. In particular, we have extended the existing theoretical models such that we could describe all experimentally relevant regimes: collisionless, collisional, and hydrodynamic.

The obtained results for the global equilibrium and the TOF expansion aspect ratios in the collisionless regime are in excellent agreement with experimental results of reference~\cite{Francesca}. In the collisional regime we have introduced an approach for self-consistently determining the relaxation time, which allows a detailed modeling of the global equilibrium and the TOF expansion in cases when the collision integral cannot be neglected. We have also shown that a strong dipole-dipole interaction, available for some experimentally accessible ultracold Fermi species, could place the system into the collisional regime, which requires to use a self-consistent determination of the relaxation time presented here. Furthermore, we find that in the collisional regime the TOF dynamics can be accurately studied only if the nonballistic expansion is used, and the dipole-dipole interaction is properly taken into account not only to calculate the ground state, but also during the whole TOF. Therefore, the obtained theoretical results are relevant for designing future experiments with strongly dipolar Fermi gases, for identifying results of the corresponding TOF measurements, and for identifying effects of dipole-dipole interaction in general.

For future work, it would be of interest to go beyond Refs.~\cite{Bohn-Jin, tau} and derive more accurate results for the relaxation time from first principles. This would amount to linearizing the BV equation and treating the linearization with the rescaling technique introduced in Ref.~\cite{Stringari-ansatz}. Furthermore, the approach developed here, based on the relaxation-time approximation for the Boltzmann-Vlasov equation, can also be applied to other fields of physics. The examples include nuclear physics, such as a study of viscosity of the quark-gluon plasma \cite{qgp1, qgp2} and ultra-relativistic heavy-ion collisions \cite{urhic}, as well as plasma physics \cite{Bittencourt}, where, e.g., transient regimes of degenerate electrons can be studied using the relaxation-time approximation \cite{Brennan}.

\section*{Acknowledgements}
We acknowledge  L. Chomaz, F. Ferlaino, A. R. P. Lima, I. Vasi\' c, and F. W\"achtler for inspiring discussions.
This work was supported in part by the Ministry of Education, Science, and Technological Development of the Republic of Serbia under projects ON171017 and IBEC, by the German Academic and Exchange Service (DAAD) under project IBEC, by the German Research Foundation (DFG) via the Collaborative Research Centers SFB/TR49 and SFB/TR185. Numerical simulations were run on the PARADOX supercomputing facility at the Scientific Computing Laboratory of the Institute of Physics Belgrade.

\appendix

\section{Aspect ratio in real space} 
\label{sec:AR}

To calculate aspect ratios in real space, we use the same geometry as in reference~\cite{Francesca}, see Fig.~\ref{fig:fig1}. The imaging plane is $x'z$, i.e., the imaging is performed along the $y'$ axis, which is rotated counterclockwise for an angle $\alpha$ with respect to the $y$ axis. The TOF absorption images correspond to density profiles of the system, so we first calculate the particle density $n({\bf r},t)$ from the Wigner quasiprobability-distribution function,
\begin{widetext}
\begin{equation*}
 n({\bf r},t)=\int \frac{d{\bf k}}{(2\pi)^3} \nu({\bf r},{\bf k},t)
=\int \frac{d{\bf k}}{(2\pi)^3} \Gamma(t) \nu^0(\boldsymbol{\mathcal{R}}({\bf r}, t),\boldsymbol{\mathcal{K}}({\bf r}, {\bf k}, t))
=  \int \frac{d{\bf k}}{(2\pi)^3} \Gamma(t) \Theta\Bigg(1-\sum_i \frac{\mathcal{R}_i^2({\bf r}, t)}{R^2_i}-\sum_i \frac{\mathcal{K}_i^2({\bf r}, {\bf k}, t)}{K^2_i} \Bigg),
\end{equation*}
where expressions for $\mathcal{R}_i({\bf r}, t)$ and $\mathcal{K}_i({\bf r}, {\bf k}, t)$ are given by Eqs.~(\ref{eq:r}) and (\ref{eq:k}), respectively. Changing the momentum variables $k_i$ to
$u_i=\frac{1}{K_i \sqrt{\theta_i(t)}}\left(k_i- \frac{M \dot{b}_i(t)}{\hbar b_i(t)}r_i\right)$ and switching to spherical coordinates yields
\begin{equation}
 n({\bf r},t)= \frac{\prod_i K_i}{6\pi^2\prod_i b_i(t)}\, \Bigg( 1-\sum_i \frac{r^2_i}{R^2_ib_i^2(t)} \Bigg)^\frac{3}{2}
  \Theta\left(1-\sum_i \frac{r^2_i}{R^2_ib_i^2(t)} \right)\, .\label{eq:n}
\end{equation}
\end{widetext}
The expectation value of a quantity $Q({\bf r})$ in real space is given by
\begin{equation}
\langle Q \rangle=\frac{1}{N}\int d{\bf r}\, n({\bf r},t) Q({\bf r})\, , \label{eq:meanQreal}
\end{equation}
so we immediately obtain that the expectation values of the coordinates vanish: ${\langle r_i \rangle}=0$. Therefore, the size in the $i$-th direction of an atomic or molecular cloud in real space is described in terms of the root mean squares $\sqrt{\langle r^2_i \rangle}$. Using the expression (\ref{eq:n}) for the particle density, the corresponding expectation values are found to be
\begin{equation}
{\langle r^2_i \rangle}=\frac{1}{N}\int d{\bf r}\, n({\bf r},t) r_i^2=\frac{1}{8}R^2_i b_i(t)\, .
\end{equation}
Since the imaging is performed in the $x'z$ plane, the aspect ratio in real space is defined by
\begin{equation}
A_R(t)=\sqrt{\frac{\langle r^2_z\rangle}{\langle r'^2_x\rangle}}\, ,
\label{eq:AAR}
\end{equation}
so we also need to calculate the expectation value $\langle r'^2_x \rangle$, where $(r'_x,r'_y,r_z)=(r_x\cos\alpha+r_y\sin\alpha, r'_y\cos\alpha-r_x\sin\alpha,r_z)$. After a straightforward but tedious calculation we get
\begin{equation}
\langle r'^2_x \rangle=\frac{1}{8}\Big[R^2_x b^2_x(t)\cos^2\alpha+R^2_yb^2_y(t)\sin^2\alpha\Big]\, ,
\end{equation}
and finally the aspect ratio (\ref{eq:AAR}) is given by
\begin{equation}
\label{eq:ARa}
A_R(t)=\frac{R_z b_z(t)}{\sqrt{R_x^2b^2_x(t)\cos^2\alpha+R_y^2b^2_y(t)\sin^2\alpha}}\, .
\end{equation}
Note that in the Innsbruck experiment \cite{Francesca} the angle $\alpha$ had the value $\alpha=28^\circ$. 

\section{Aspect ratio in momentum space}
\label{sec:AK}

In order to describe effects of the DDI on the Fermi surface, we use an aspect ratio in momentum space, which is defined similarly as the aspect ratio in real space. First, we calculate the particle density in momentum space $n({\bf k},t)$ from the Wigner quasiprobability-distribution function,
\begin{widetext}
\begin{equation*}
n({\bf k},t)=\int d{\bf r}\, \nu({\bf r},{\bf k},t)=\int d{\bf r}\, \Gamma(t) \nu^0(\boldsymbol{\mathcal{R}}({\bf r}, t),\boldsymbol{\mathcal{K}}({\bf r}, {\bf k}, t))
=\int d{\bf r}\, \Gamma(t) \Theta\Bigg(1-\sum_i \frac{\mathcal{R}_i({\bf r}, t)^2}{R^2_i}-\sum_i \frac{\mathcal{K}_i({\bf r}, {\bf k}, t)^2}{K^2_i} \Bigg),
\end{equation*}
where again expressions for $\mathcal{R}_i({\bf r}, t)$ and $\mathcal{K}_i({\bf r}, {\bf k}, t)$ are given by Eqs.~(\ref{eq:r}) and (\ref{eq:k}), respectively. After a change of spatial variables $r_i$ according to
$u_i =  \frac{D_i(t) r_i}{R_ib_i(t)} - \frac{M R_i \dot{b}_i(t)k_i}{\hbar K^2_i \theta_i(t) D_i}$ with $D_i(t)= \sqrt{1+\frac{M^2R_i^2 \dot{b}^2_i(t)}{\hbar^2 K_i^2\theta_i(t)}}$, we switch to spherical coordinates and obtain
\begin{equation}
 n({\bf k},t)=
\frac{4\pi}{3}\frac{\prod_i R_i}{\prod_i \sqrt{\theta_i(t)}D_i(t)} \Bigg(1-\sum_i \frac{k^2_i}{K^2_i\theta_i(t)D_i^2(t)} \Bigg)^\frac{3}{2}\Theta\Bigg(1-\sum_i \frac{k^2_i}{K^2_i\theta_i(t)D_i^2(t)} \Bigg) .
\end{equation}
\end{widetext}
The expectation value of a variable $Q({\bf k})$ in momentum space is given by
\begin{equation}
\langle Q \rangle=\frac{1}{N}\int \frac{d{\bf k}}{(2\pi)^3}\, n({\bf k},t) Q({\bf k})\, , \label{eq:meanQmom}
\end{equation}
so we get $\langle k_i \rangle=0$ and the cloud sizes in momentum space are also defined by root mean squares $\sqrt{\langle k_i^2 \rangle}$. The corresponding expectation values can be explicitly calculated and yield
\begin{equation}
{\langle k^2_i \rangle}=\frac{1}{N}\int \frac{d{\bf k}}{(2\pi)^3}\, n({\bf k},t) k_i^2=
\frac{1}{8}\left( K_i^2\theta_i(t)+\frac{M^2R_i^2 \dot{b}^2_i(t)}{\hbar^2}\right),
\end{equation}
where we have used the same variable change as above, as well as Eq.~(\ref{eq:particlenumberconservation}). The aspect ratio in momentum space is defined as
\begin{equation}
A_K(t)=\sqrt{\frac{\langle k^2_z\rangle}{\langle k'^2_x\rangle}}\, ,
\label{eq:BAK}
\end{equation}
where $(k'_x,k'_y,k_z)=(k_x\cos\alpha+k_y\sin\alpha, k'_y\cos\alpha-k_x\sin\alpha,k_z)$. After a lengthy calculation we get
\begin{equation}
\langle k'^2_x \rangle=\frac{1}{8}\Big[D^2_xK^2_x \theta_x(t)\cos^2\alpha+D^2_y K^2_y \theta_y(t)\sin^2\alpha\Big]\, ,
\end{equation}
and finally the momentum space aspect ratio (\ref{eq:BAK}) reduces to
\begin{widetext}
\begin{equation}
A_K(t)= \sqrt{\frac{\hbar^2K_z^2\theta_z(t)+M^2R_z^2 \dot{b}^2_z(t)}{[\hbar^2K_x^2\theta_x(t)+M^2R_x^2 \dot{b}^2_x(t)]\cos^2\alpha+[\hbar^2K_y^2\theta_y(t)+
M^2R_y^2 \dot{b}^2_y(t)]\sin^2\alpha}}\, .
\end{equation}
\end{widetext}


\begin{thebibliography}{86}%
\makeatletter
\providecommand \@ifxundefined [1]{%
 \@ifx{#1\undefined}
}%
\providecommand \@ifnum [1]{%
 \ifnum #1\expandafter \@firstoftwo
 \else \expandafter \@secondoftwo
 \fi
}%
\providecommand \@ifx [1]{%
 \ifx #1\expandafter \@firstoftwo
 \else \expandafter \@secondoftwo
 \fi
}%
\providecommand \natexlab [1]{#1}%
\providecommand \enquote  [1]{``#1''}%
\providecommand \bibnamefont  [1]{#1}%
\providecommand \bibfnamefont [1]{#1}%
\providecommand \citenamefont [1]{#1}%
\providecommand \href@noop [0]{\@secondoftwo}%
\providecommand \href [0]{\begingroup \@sanitize@url \@href}%
\providecommand \@href[1]{\@@startlink{#1}\@@href}%
\providecommand \@@href[1]{\endgroup#1\@@endlink}%
\providecommand \@sanitize@url [0]{\catcode `\\12\catcode `\$12\catcode
  `\&12\catcode `\#12\catcode `\^12\catcode `\_12\catcode `\%12\relax}%
\providecommand \@@startlink[1]{}%
\providecommand \@@endlink[0]{}%
\providecommand \url  [0]{\begingroup\@sanitize@url \@url }%
\providecommand \@url [1]{\endgroup\@href {#1}{\urlprefix }}%
\providecommand \urlprefix  [0]{URL }%
\providecommand \Eprint [0]{\href }%
\providecommand \doibase [0]{http://dx.doi.org/}%
\providecommand \selectlanguage [0]{\@gobble}%
\providecommand \bibinfo  [0]{\@secondoftwo}%
\providecommand \bibfield  [0]{\@secondoftwo}%
\providecommand \translation [1]{[#1]}%
\providecommand \BibitemOpen [0]{}%
\providecommand \bibitemStop [0]{}%
\providecommand \bibitemNoStop [0]{.\EOS\space}%
\providecommand \EOS [0]{\spacefactor3000\relax}%
\providecommand \BibitemShut  [1]{\csname bibitem#1\endcsname}%
\let\auto@bib@innerbib\@empty
%</preamble>
\bibitem [{\citenamefont {Pethick}\ and\ \citenamefont
  {Smith}(2008)}]{TheBECBook}%
  \BibitemOpen
  \bibfield  {author} {\bibinfo {author} {\bibfnamefont {C.~J.}\ \bibnamefont
  {Pethick}}\ and\ \bibinfo {author} {\bibfnamefont {H.}~\bibnamefont
  {Smith}},\ }\href@noop {} {\emph {\bibinfo {title} {Bose-Einstein
  Condensation in Dilute Gases}}},\ \bibinfo {edition} {2nd}\ ed.\ (\bibinfo
  {publisher} {Cambridge University Press},\ \bibinfo {address} {Cambridge},\
  \bibinfo {year} {2008})\BibitemShut {NoStop}%
\bibitem [{\citenamefont {Pitaevskii}\ and\ \citenamefont
  {Stringari}(2016)}]{TheBECBook2}%
  \BibitemOpen
  \bibfield  {author} {\bibinfo {author} {\bibfnamefont {L.}~\bibnamefont
  {Pitaevskii}}\ and\ \bibinfo {author} {\bibfnamefont {S.}~\bibnamefont
  {Stringari}},\ }\href@noop {} {\emph {\bibinfo {title} {Bose-Einstein
  Condensation and Superfluidity}}},\ \bibinfo {edition} {2nd}\ ed.\ (\bibinfo
  {publisher} {Oxford University Press},\ \bibinfo {address} {Oxford},\
  \bibinfo {year} {2016})\BibitemShut {NoStop}%
\bibitem [{\citenamefont {Santos}\ \emph {et~al.}(2003)\citenamefont {Santos},
  \citenamefont {Shlyapnikov},\ and\ \citenamefont {Lewenstein}}]{Santos}%
  \BibitemOpen
  \bibfield  {author} {\bibinfo {author} {\bibfnamefont {L.}~\bibnamefont
  {Santos}}, \bibinfo {author} {\bibfnamefont {G.~V.}\ \bibnamefont
  {Shlyapnikov}}, \ and\ \bibinfo {author} {\bibfnamefont {M.}~\bibnamefont
  {Lewenstein}},\ }\href {\doibase 10.1103/PhysRevLett.90.250403} {\bibfield
  {journal} {\bibinfo  {journal} {Phys. Rev. Lett.}\ }\textbf {\bibinfo
  {volume} {90}},\ \bibinfo {pages} {250403} (\bibinfo {year}
  {2003})}\BibitemShut {NoStop}%
\bibitem [{\citenamefont {Glaum}\ \emph {et~al.}(2007)\citenamefont {Glaum},
  \citenamefont {Pelster}, \citenamefont {Kleinert},\ and\ \citenamefont
  {Pfau}}]{Glaum1}%
  \BibitemOpen
  \bibfield  {author} {\bibinfo {author} {\bibfnamefont {K.}~\bibnamefont
  {Glaum}}, \bibinfo {author} {\bibfnamefont {A.}~\bibnamefont {Pelster}},
  \bibinfo {author} {\bibfnamefont {H.}~\bibnamefont {Kleinert}}, \ and\
  \bibinfo {author} {\bibfnamefont {T.}~\bibnamefont {Pfau}},\ }\href {\doibase 10.1103/PhysRevLett.98.080407} {\bibfield  {journal} {\bibinfo  {journal}
  {Phys. Rev. Lett.}\ }\textbf {\bibinfo {volume} {98}},\ \bibinfo {pages}
  {080407} (\bibinfo {year} {2007})}\BibitemShut {NoStop}%
\bibitem [{\citenamefont {Glaum}\ and\ \citenamefont {Pelster}(2007)}]{Glaum2}%
  \BibitemOpen
  \bibfield  {author} {\bibinfo {author} {\bibfnamefont {K.}~\bibnamefont
  {Glaum}}\ and\ \bibinfo {author} {\bibfnamefont {A.}~\bibnamefont
  {Pelster}},\ }\href {\doibase 10.1103/PhysRevA.76.023604} {\bibfield
  {journal} {\bibinfo  {journal} {Phys. Rev. A}\ }\textbf {\bibinfo {volume}
  {76}},\ \bibinfo {pages} {023604} (\bibinfo {year} {2007})}\BibitemShut
  {NoStop}%
\bibitem [{\citenamefont {Baranov}(2008)}]{Baranov}%
  \BibitemOpen
  \bibfield  {author} {\bibinfo {author} {\bibfnamefont {M.}~\bibnamefont
  {Baranov}},\ }\href {\doibase 10.1016/j.physrep.2008.04.007} {\bibfield
  {journal} {\bibinfo  {journal} {Phys. Rep.}\ }\textbf {\bibinfo {volume}
  {464}},\ \bibinfo {pages} {71 } (\bibinfo {year} {2008})}\BibitemShut
  {NoStop}%
\bibitem [{\citenamefont {Lahaye}\ \emph {et~al.}(2009)\citenamefont {Lahaye},
  \citenamefont {Menotti}, \citenamefont {Santos}, \citenamefont {Lewenstein},\
  and\ \citenamefont {Pfau}}]{PfauRep}%
  \BibitemOpen
  \bibfield  {author} {\bibinfo {author} {\bibfnamefont {T.}~\bibnamefont
  {Lahaye}}, \bibinfo {author} {\bibfnamefont {C.}~\bibnamefont {Menotti}},
  \bibinfo {author} {\bibfnamefont {L.}~\bibnamefont {Santos}}, \bibinfo
  {author} {\bibfnamefont {M.}~\bibnamefont {Lewenstein}}, \ and\ \bibinfo
  {author} {\bibfnamefont {T.}~\bibnamefont {Pfau}},\ }\href {\doibase 10.1088/0034-4885/72/12/126401} {\bibfield  {journal} {\bibinfo  {journal}
  {Rep. Prog. Phys.}\ }\textbf {\bibinfo {volume} {72}},\ \bibinfo {pages}
  {126401} (\bibinfo {year} {2009})}\BibitemShut {NoStop}%
\bibitem [{\citenamefont {{Carr}}\ and\ \citenamefont {{Ye}}(2009)}]{Carr}%
  \BibitemOpen
  \bibfield  {author} {\bibinfo {author} {\bibfnamefont {L.~D.}\ \bibnamefont
  {{Carr}}}\ and\ \bibinfo {author} {\bibfnamefont {J.}~\bibnamefont {{Ye}}},\
  }\href {\doibase 10.1088/1367-2630/11/5/055009} {\bibfield  {journal}
  {\bibinfo  {journal} {New J. Phys.}\ }\textbf {\bibinfo {volume} {11}},\
  \bibinfo {eid} {055009} (\bibinfo {year} {2009})}\BibitemShut {NoStop}%
\bibitem [{\citenamefont {Krumnow}\ and\ \citenamefont
  {Pelster}(2011)}]{Krumnow}%
  \BibitemOpen
  \bibfield  {author} {\bibinfo {author} {\bibfnamefont {C.}~\bibnamefont
  {Krumnow}}\ and\ \bibinfo {author} {\bibfnamefont {A.}~\bibnamefont
  {Pelster}},\ }\href {\doibase 10.1103/PhysRevA.84.021608} {\bibfield
  {journal} {\bibinfo  {journal} {Phys. Rev. A}\ }\textbf {\bibinfo {volume}
  {84}},\ \bibinfo {pages} {021608(R)} (\bibinfo {year} {2011})}\BibitemShut
  {NoStop}%
\bibitem [{\citenamefont {Block}\ \emph {et~al.}(2012)\citenamefont {Block},
  \citenamefont {Zinner},\ and\ \citenamefont {Bruun}}]{Block}%
  \BibitemOpen
  \bibfield  {author} {\bibinfo {author} {\bibfnamefont {J.~K.}\ \bibnamefont
  {Block}}, \bibinfo {author} {\bibfnamefont {N.~T.}\ \bibnamefont {Zinner}}, \
  and\ \bibinfo {author} {\bibfnamefont {G.~M.}\ \bibnamefont {Bruun}},\ }\href
  {http://stacks.iop.org/1367-2630/14/i=10/a=105006} {\bibfield  {journal}
  {\bibinfo  {journal} {New J. Phys.}\ }\textbf {\bibinfo {volume} {14}},\
  \bibinfo {pages} {105006} (\bibinfo {year} {2012})}\BibitemShut {NoStop}%
\bibitem [{\citenamefont {Nicolin}(2013)}]{Alex1}%
  \BibitemOpen
  \bibfield  {author} {\bibinfo {author} {\bibfnamefont {A.~I.}\ \bibnamefont
  {Nicolin}},\ }\href
  {http://www.academiaromana.ro/sectii2002/proceedings/doc2013-1/06-Nicolin.pdf}
  {\bibfield  {journal} {\bibinfo  {journal} {Proc. Rom. Acad. Ser. A-Math.
  Phys.}\ }\textbf {\bibinfo {volume} {14}},\ \bibinfo {pages} {35 } (\bibinfo
  {year} {2013})}\BibitemShut {NoStop}%
\bibitem [{\citenamefont {Nikoli\'{c}}\ \emph {et~al.}(2013)\citenamefont
  {Nikoli\'{c}}, \citenamefont {Bala\v{z}},\ and\ \citenamefont
  {Pelster}}]{Branko}%
  \BibitemOpen
  \bibfield  {author} {\bibinfo {author} {\bibfnamefont {B.}~\bibnamefont
  {Nikoli\'{c}}}, \bibinfo {author} {\bibfnamefont {A.}~\bibnamefont
  {Bala\v{z}}}, \ and\ \bibinfo {author} {\bibfnamefont {A.}~\bibnamefont
  {Pelster}},\ }\href {\doibase 10.1103/PhysRevA.88.013624} {\bibfield
  {journal} {\bibinfo  {journal} {Phys. Rev. A}\ }\textbf {\bibinfo {volume}
  {88}},\ \bibinfo {pages} {013624} (\bibinfo {year} {2013})}\BibitemShut
  {NoStop}%
\bibitem [{\citenamefont {Al-Jibbouri}\ \emph {et~al.}(2013)\citenamefont
  {Al-Jibbouri}, \citenamefont {Vidanovi\'{c}}, \citenamefont {Bala\v{z}},\
  and\ \citenamefont {Pelster}}]{Hamid}%
  \BibitemOpen
  \bibfield  {author} {\bibinfo {author} {\bibfnamefont {H.}~\bibnamefont
  {Al-Jibbouri}}, \bibinfo {author} {\bibfnamefont {I.}~\bibnamefont
  {Vidanovi\'{c}}}, \bibinfo {author} {\bibfnamefont {A.}~\bibnamefont
  {Bala\v{z}}}, \ and\ \bibinfo {author} {\bibfnamefont {A.}~\bibnamefont
  {Pelster}},\ }\href {http://stacks.iop.org/0953-4075/46/i=6/a=065303}
  {\bibfield  {journal} {\bibinfo  {journal} {J. Phys. B: At. Mol. Opt. Phys.}\
  }\textbf {\bibinfo {volume} {46}},\ \bibinfo {pages} {065303} (\bibinfo
  {year} {2013})}\BibitemShut {NoStop}%
\bibitem [{\citenamefont {Bala\v{z}}\ \emph {et~al.}(2014)\citenamefont
  {Bala\v{z}}, \citenamefont {Paun}, \citenamefont {Nicolin}, \citenamefont
  {Balasubramanian},\ and\ \citenamefont {Ramaswamy}}]{Radha}%
  \BibitemOpen
  \bibfield  {author} {\bibinfo {author} {\bibfnamefont {A.}~\bibnamefont
  {Bala\v{z}}}, \bibinfo {author} {\bibfnamefont {R.}~\bibnamefont {Paun}},
  \bibinfo {author} {\bibfnamefont {A.~I.}\ \bibnamefont {Nicolin}}, \bibinfo
  {author} {\bibfnamefont {S.}~\bibnamefont {Balasubramanian}}, \ and\ \bibinfo
  {author} {\bibfnamefont {R.}~\bibnamefont {Ramaswamy}},\ }\href {\doibase 10.1103/PhysRevA.89.023609} {\bibfield  {journal} {\bibinfo  {journal} {Phys.
  Rev. A}\ }\textbf {\bibinfo {volume} {89}},\ \bibinfo {pages} {023609}
  (\bibinfo {year} {2014})}\BibitemShut {NoStop}%
\bibitem [{\citenamefont {Ghabour}\ and\ \citenamefont
  {Pelster}(2014)}]{Ghabour}%
  \BibitemOpen
  \bibfield  {author} {\bibinfo {author} {\bibfnamefont {M.}~\bibnamefont
  {Ghabour}}\ and\ \bibinfo {author} {\bibfnamefont {A.}~\bibnamefont
  {Pelster}},\ }\href {\doibase 10.1103/PhysRevA.90.063636} {\bibfield
  {journal} {\bibinfo  {journal} {Phys. Rev. A}\ }\textbf {\bibinfo {volume}
  {90}},\ \bibinfo {pages} {063636} (\bibinfo {year} {2014})};
  \bibfield  {author} {\bibinfo {author} {\bibfnamefont {S.~K.}\ \bibnamefont
  {Adhikari}},\ }\href {\doibase 10.1103/PhysRevA.88.043603} {\bibfield
  {journal} {\bibinfo  {journal} {Phys. Rev. A}\ }\textbf {\bibinfo {volume}
  {88}},\ \bibinfo {pages} {043603} (\bibinfo {year}
  {2013}{\natexlab{a}})};
  \bibfield  {author} {\bibinfo {author} {\bibfnamefont {S.~K.}\ \bibnamefont
  {Adhikari}},\ }\href {\doibase 10.1088/0953-4075/46/11/115301} {\bibfield
  {journal} {\bibinfo  {journal} {J. Phys. B: At. Mol. Opt. Phys.}\ }\textbf
  {\bibinfo {volume} {46}},\ \bibinfo {pages} {115301} (\bibinfo {year}
  {2013}{\natexlab{b}})};
  \bibfield  {author} {\bibinfo {author} {\bibfnamefont {S.~K.}\ \bibnamefont
  {Adhikari}},\ }\href {\doibase 10.1088/0953-4075/45/23/235303} {\bibfield
  {journal} {\bibinfo  {journal} {J. Phys. B: At. Mol. Opt. Phys.}\ }\textbf {\bibinfo {volume} {45}},\ \bibinfo {pages}
  {235303} (\bibinfo {year} {2012})}
  \BibitemShut {NoStop}%
\bibitem [{\citenamefont {Frisch}\ \emph {et~al.}(2015)\citenamefont {Frisch},
  \citenamefont {Mark}, \citenamefont {Aikawa}, \citenamefont {Baier},
  \citenamefont {Grimm}, \citenamefont {Petrov}, \citenamefont {Kotochigova},
  \citenamefont {Qu\'em\'ener}, \citenamefont {Lepers}, \citenamefont
  {Dulieu},\ and\ \citenamefont {Ferlaino}}]{Er2}%
  \BibitemOpen
  \bibfield  {author} {\bibinfo {author} {\bibfnamefont {A.}~\bibnamefont
  {Frisch}}, \bibinfo {author} {\bibfnamefont {M.}~\bibnamefont {Mark}},
  \bibinfo {author} {\bibfnamefont {K.}~\bibnamefont {Aikawa}}, \bibinfo
  {author} {\bibfnamefont {S.}~\bibnamefont {Baier}}, \bibinfo {author}
  {\bibfnamefont {R.}~\bibnamefont {Grimm}}, \bibinfo {author} {\bibfnamefont
  {A.}~\bibnamefont {Petrov}}, \bibinfo {author} {\bibfnamefont
  {S.}~\bibnamefont {Kotochigova}}, \bibinfo {author} {\bibfnamefont
  {G.}~\bibnamefont {Qu\'em\'ener}}, \bibinfo {author} {\bibfnamefont
  {M.}~\bibnamefont {Lepers}}, \bibinfo {author} {\bibfnamefont
  {O.}~\bibnamefont {Dulieu}}, \ and\ \bibinfo {author} {\bibfnamefont
  {F.}~\bibnamefont {Ferlaino}},\ }\href {\doibase 10.1103/PhysRevLett.115.203201} {\bibfield  {journal} {\bibinfo  {journal}
  {Phys. Rev. Lett.}\ }\textbf {\bibinfo {volume} {115}},\ \bibinfo {pages}
  {203201} (\bibinfo {year} {2015})}\BibitemShut {NoStop}%
\bibitem [{\citenamefont {Gadway}\ and\ \citenamefont {Yan}(2016)}]{Gadway}%
  \BibitemOpen
  \bibfield  {author} {\bibinfo {author} {\bibfnamefont {B.}~\bibnamefont
  {Gadway}}\ and\ \bibinfo {author} {\bibfnamefont {B.}~\bibnamefont {Yan}},\
  }\href {http://stacks.iop.org/0953-4075/49/i=15/a=152002} {\bibfield
  {journal} {\bibinfo  {journal} {J. Phys. B: At. Mol. Opt. Phys.}\ }\textbf
  {\bibinfo {volume} {49}},\ \bibinfo {pages} {152002} (\bibinfo {year}
  {2016})}\BibitemShut {NoStop}%
\bibitem [{\citenamefont {Stuhler}\ \emph {et~al.}(2005)\citenamefont
  {Stuhler}, \citenamefont {Griesmaier}, \citenamefont {Koch}, \citenamefont
  {Fattori}, \citenamefont {Pfau}, \citenamefont {Giovanazzi}, \citenamefont
  {Pedri},\ and\ \citenamefont {Santos}}]{Pfau}%
  \BibitemOpen
  \bibfield  {author} {\bibinfo {author} {\bibfnamefont {J.}~\bibnamefont
  {Stuhler}}, \bibinfo {author} {\bibfnamefont {A.}~\bibnamefont {Griesmaier}},
  \bibinfo {author} {\bibfnamefont {T.}~\bibnamefont {Koch}}, \bibinfo {author}
  {\bibfnamefont {M.}~\bibnamefont {Fattori}}, \bibinfo {author} {\bibfnamefont
  {T.}~\bibnamefont {Pfau}}, \bibinfo {author} {\bibfnamefont {S.}~\bibnamefont
  {Giovanazzi}}, \bibinfo {author} {\bibfnamefont {P.}~\bibnamefont {Pedri}}, \
  and\ \bibinfo {author} {\bibfnamefont {L.}~\bibnamefont {Santos}},\ }\href
  {\doibase 10.1103/PhysRevLett.95.150406} {\bibfield  {journal} {\bibinfo
  {journal} {Phys. Rev. Lett.}\ }\textbf {\bibinfo {volume} {95}},\ \bibinfo
  {pages} {150406} (\bibinfo {year} {2005})}\BibitemShut {NoStop}%
\bibitem [{\citenamefont {{Kadau}}\ \emph {et~al.}(2016)\citenamefont
  {{Kadau}}, \citenamefont {{Schmitt}}, \citenamefont {{Wenzel}}, \citenamefont
  {{Wink}}, \citenamefont {{Maier}}, \citenamefont {{Ferrier-Barbut}},\ and\
  \citenamefont {{Pfau}}}]{PfauNature}%
  \BibitemOpen
  \bibfield  {author} {\bibinfo {author} {\bibfnamefont {H.}~\bibnamefont
  {{Kadau}}}, \bibinfo {author} {\bibfnamefont {M.}~\bibnamefont {{Schmitt}}},
  \bibinfo {author} {\bibfnamefont {M.}~\bibnamefont {{Wenzel}}}, \bibinfo
  {author} {\bibfnamefont {C.}~\bibnamefont {{Wink}}}, \bibinfo {author}
  {\bibfnamefont {T.}~\bibnamefont {{Maier}}}, \bibinfo {author} {\bibfnamefont
  {I.}~\bibnamefont {{Ferrier-Barbut}}}, \ and\ \bibinfo {author}
  {\bibfnamefont {T.}~\bibnamefont {{Pfau}}},\ }\href {\doibase 10.1038/nature16485} {\bibfield  {journal} {\bibinfo  {journal} {\nat}\
  }\textbf {\bibinfo {volume} {530}},\ \bibinfo {pages} {194} (\bibinfo {year}
  {2016})}\BibitemShut {NoStop}%
\bibitem [{\citenamefont {Muruganandam}\ and\ \citenamefont
  {Adhikari}(2009)}]{GP1}%
  \BibitemOpen
  \bibfield  {author} {\bibinfo {author} {\bibfnamefont {P.}~\bibnamefont
  {Muruganandam}}\ and\ \bibinfo {author} {\bibfnamefont {S.}~\bibnamefont
  {Adhikari}},\ }\href {\doibase 10.1016/j.cpc.2009.04.015} {\bibfield
  {journal} {\bibinfo  {journal} {Comput. Phys. Commun.}\ }\textbf {\bibinfo
  {volume} {180}},\ \bibinfo {pages} {1888 } (\bibinfo {year}
  {2009})}\BibitemShut {NoStop}%
\bibitem [{\citenamefont {Vudragovi\'{c}}\ \emph {et~al.}(2012)\citenamefont
  {Vudragovi\'{c}}, \citenamefont {Vidanovi\'{c}}, \citenamefont {Bala\v{z}},
  \citenamefont {Muruganandam},\ and\ \citenamefont {Adhikari}}]{GP2}%
  \BibitemOpen
  \bibfield  {author} {\bibinfo {author} {\bibfnamefont {D.}~\bibnamefont
  {Vudragovi\'{c}}}, \bibinfo {author} {\bibfnamefont {I.}~\bibnamefont
  {Vidanovi\'{c}}}, \bibinfo {author} {\bibfnamefont {A.}~\bibnamefont
  {Bala\v{z}}}, \bibinfo {author} {\bibfnamefont {P.}~\bibnamefont
  {Muruganandam}}, \ and\ \bibinfo {author} {\bibfnamefont {S.~K.}\
  \bibnamefont {Adhikari}},\ }\href {\doibase 10.1016/j.cpc.2012.03.022}
  {\bibfield  {journal} {\bibinfo  {journal} {Comput. Phys. Commun.}\ }\textbf
  {\bibinfo {volume} {183}},\ \bibinfo {pages} {2021 } (\bibinfo {year}
  {2012})}\BibitemShut {NoStop}%
\bibitem [{\citenamefont {Kumar}\ \emph {et~al.}(2015)\citenamefont {Kumar},
  \citenamefont {Young-S.}, \citenamefont {Vudragovi\'{c}}, \citenamefont
  {Bala\v{z}}, \citenamefont {Muruganandam},\ and\ \citenamefont
  {Adhikari}}]{GP3}%
  \BibitemOpen
  \bibfield  {author} {\bibinfo {author} {\bibfnamefont {R.~K.}\ \bibnamefont
  {Kumar}}, \bibinfo {author} {\bibfnamefont {L.~E.}\ \bibnamefont {Young-S.}},
  \bibinfo {author} {\bibfnamefont {D.}~\bibnamefont {Vudragovi\'{c}}},
  \bibinfo {author} {\bibfnamefont {A.}~\bibnamefont {Bala\v{z}}}, \bibinfo
  {author} {\bibfnamefont {P.}~\bibnamefont {Muruganandam}}, \ and\ \bibinfo
  {author} {\bibfnamefont {S.~K.}\ \bibnamefont {Adhikari}},\ }\href {\doibase 10.1016/j.cpc.2015.03.024} {\bibfield  {journal} {\bibinfo  {journal}
  {Comput. Phys. Commun.}\ }\textbf {\bibinfo {volume} {195}},\ \bibinfo
  {pages} {117 } (\bibinfo {year} {2015})}\BibitemShut {NoStop}%
\bibitem [{\citenamefont {Lon\v{c}ar}\ \emph {et~al.}(2016)\citenamefont
  {Lon\v{c}ar}, \citenamefont {Bala\v{z}}, \citenamefont {Bogojevi\'{c}},
  \citenamefont {\v{S}krbi\'{c}}, \citenamefont {Muruganandam},\ and\
  \citenamefont {Adhikari}}]{GP4}%
  \BibitemOpen
  \bibfield  {author} {\bibinfo {author} {\bibfnamefont {V.}~\bibnamefont
  {Lon\v{c}ar}}, \bibinfo {author} {\bibfnamefont {A.}~\bibnamefont
  {Bala\v{z}}}, \bibinfo {author} {\bibfnamefont {A.}~\bibnamefont
  {Bogojevi\'{c}}}, \bibinfo {author} {\bibfnamefont {S.}~\bibnamefont
  {\v{S}krbi\'{c}}}, \bibinfo {author} {\bibfnamefont {P.}~\bibnamefont
  {Muruganandam}}, \ and\ \bibinfo {author} {\bibfnamefont {S.~K.}\
  \bibnamefont {Adhikari}},\ }\href {\doibase 10.1016/j.cpc.2015.11.014}
  {\bibfield  {journal} {\bibinfo  {journal} {Comput. Phys. Commun.}\ }\textbf
  {\bibinfo {volume} {200}},\ \bibinfo {pages} {406 } (\bibinfo {year}
  {2016})}\BibitemShut {NoStop}%
\bibitem [{\citenamefont {Satari\'{c}}\ \emph {et~al.}(2016)\citenamefont
  {Satari\'{c}}, \citenamefont {Slavni\'{c}}, \citenamefont {Beli\'{c}},
  \citenamefont {Bala\v{z}}, \citenamefont {Muruganandam},\ and\ \citenamefont
  {Adhikari}}]{GP5}%
  \BibitemOpen
  \bibfield  {author} {\bibinfo {author} {\bibfnamefont {B.}~\bibnamefont
  {Satari\'{c}}}, \bibinfo {author} {\bibfnamefont {V.}~\bibnamefont
  {Slavni\'{c}}}, \bibinfo {author} {\bibfnamefont {A.}~\bibnamefont
  {Beli\'{c}}}, \bibinfo {author} {\bibfnamefont {A.}~\bibnamefont
  {Bala\v{z}}}, \bibinfo {author} {\bibfnamefont {P.}~\bibnamefont
  {Muruganandam}}, \ and\ \bibinfo {author} {\bibfnamefont {S.~K.}\
  \bibnamefont {Adhikari}},\ }\href {\doibase 10.1016/j.cpc.2015.12.006}
  {\bibfield  {journal} {\bibinfo  {journal} {Comput. Phys. Commun.}\ }\textbf
  {\bibinfo {volume} {200}},\ \bibinfo {pages} {411 } (\bibinfo {year}
  {2016})}\BibitemShut {NoStop}%
\bibitem [{\citenamefont {Young-S.}\ \emph {et~al.}(2016)\citenamefont
  {Young-S.}, \citenamefont {Vudragovi\'{c}}, \citenamefont {Muruganandam},
  \citenamefont {Adhikari},\ and\ \citenamefont {Bala\v{z}}}]{GP6}%
  \BibitemOpen
  \bibfield  {author} {\bibinfo {author} {\bibfnamefont {L.~E.}\ \bibnamefont
  {Young-S.}}, \bibinfo {author} {\bibfnamefont {D.}~\bibnamefont
  {Vudragovi\'{c}}}, \bibinfo {author} {\bibfnamefont {P.}~\bibnamefont
  {Muruganandam}}, \bibinfo {author} {\bibfnamefont {S.~K.}\ \bibnamefont
  {Adhikari}}, \ and\ \bibinfo {author} {\bibfnamefont {A.}~\bibnamefont
  {Bala\v{z}}},\ }\href {\doibase 10.1016/j.cpc.2016.03.015} {\bibfield
  {journal} {\bibinfo  {journal} {Comput. Phys. Commun.}\ }\textbf {\bibinfo
  {volume} {204}},\ \bibinfo {pages} {209 } (\bibinfo {year}
  {2016})}\BibitemShut {NoStop}%
\bibitem [{\citenamefont {Lima}\ and\ \citenamefont {Pelster}(2011)}]{Lima3}%
  \BibitemOpen
  \bibfield  {author} {\bibinfo {author} {\bibfnamefont {A.~R.~P.}\
  \bibnamefont {Lima}}\ and\ \bibinfo {author} {\bibfnamefont {A.}~\bibnamefont
  {Pelster}},\ }\href {\doibase 10.1103/PhysRevA.84.041604} {\bibfield
  {journal} {\bibinfo  {journal} {Phys. Rev. A}\ }\textbf {\bibinfo {volume}
  {84}},\ \bibinfo {pages} {041604(R)} (\bibinfo {year} {2011})}\BibitemShut
  {NoStop}%
\bibitem [{\citenamefont {Lima}\ and\ \citenamefont {Pelster}(2012)}]{Lima4}%
  \BibitemOpen
  \bibfield  {author} {\bibinfo {author} {\bibfnamefont {A.~R.~P.}\
  \bibnamefont {Lima}}\ and\ \bibinfo {author} {\bibfnamefont {A.}~\bibnamefont
  {Pelster}},\ }\href {\doibase 10.1103/PhysRevA.86.063609} {\bibfield
  {journal} {\bibinfo  {journal} {Phys. Rev. A}\ }\textbf {\bibinfo {volume}
  {86}},\ \bibinfo {pages} {063609} (\bibinfo {year} {2012})}\BibitemShut
  {NoStop}%
\bibitem [{\citenamefont {Bisset}\ and\ \citenamefont
  {Blakie}(2015)}]{Blakie1}%
  \BibitemOpen
  \bibfield  {author} {\bibinfo {author} {\bibfnamefont {R.~N.}\ \bibnamefont
  {Bisset}}\ and\ \bibinfo {author} {\bibfnamefont {P.~B.}\ \bibnamefont
  {Blakie}},\ }\href {\doibase 10.1103/PhysRevA.92.061603} {\bibfield
  {journal} {\bibinfo  {journal} {Phys. Rev. A}\ }\textbf {\bibinfo {volume}
  {92}},\ \bibinfo {pages} {061603(R)} (\bibinfo {year} {2015})}\BibitemShut
  {NoStop}%
\bibitem [{\citenamefont {Ferrier-Barbut}\ \emph {et~al.}(2016)\citenamefont
  {Ferrier-Barbut}, \citenamefont {Kadau}, \citenamefont {Schmitt},
  \citenamefont {Wenzel},\ and\ \citenamefont {Pfau}}]{PfauPRL}%
  \BibitemOpen
  \bibfield  {author} {\bibinfo {author} {\bibfnamefont {I.}~\bibnamefont
  {Ferrier-Barbut}}, \bibinfo {author} {\bibfnamefont {H.}~\bibnamefont
  {Kadau}}, \bibinfo {author} {\bibfnamefont {M.}~\bibnamefont {Schmitt}},
  \bibinfo {author} {\bibfnamefont {M.}~\bibnamefont {Wenzel}}, \ and\ \bibinfo
  {author} {\bibfnamefont {T.}~\bibnamefont {Pfau}},\ }\href {\doibase 10.1103/PhysRevLett.116.215301} {\bibfield  {journal} {\bibinfo  {journal}
  {Phys. Rev. Lett.}\ }\textbf {\bibinfo {volume} {116}},\ \bibinfo {pages}
  {215301} (\bibinfo {year} {2016})}\BibitemShut {NoStop}%
\bibitem [{\citenamefont {Blakie}(2016)}]{Blakie2}%
  \BibitemOpen
  \bibfield  {author} {\bibinfo {author} {\bibfnamefont {P.~B.}\ \bibnamefont
  {Blakie}},\ }\href {\doibase 10.1103/PhysRevA.93.033644} {\bibfield
  {journal} {\bibinfo  {journal} {Phys. Rev. A}\ }\textbf {\bibinfo {volume}
  {93}},\ \bibinfo {pages} {033644} (\bibinfo {year} {2016})}\BibitemShut
  {NoStop}%
\bibitem [{\citenamefont {Xi}\ and\ \citenamefont {Saito}(2016)}]{Xi}%
  \BibitemOpen
  \bibfield  {author} {\bibinfo {author} {\bibfnamefont {K.-T.}\ \bibnamefont
  {Xi}}\ and\ \bibinfo {author} {\bibfnamefont {H.}~\bibnamefont {Saito}},\
  }\href {\doibase 10.1103/PhysRevA.93.011604} {\bibfield  {journal} {\bibinfo
  {journal} {Phys. Rev. A}\ }\textbf {\bibinfo {volume} {93}},\ \bibinfo
  {pages} {011604(R)} (\bibinfo {year} {2016})}\BibitemShut {NoStop}%
\bibitem [{\citenamefont {W\"achtler}\ and\ \citenamefont
  {Santos}(2016{\natexlab{a}})}]{Santos1}%
  \BibitemOpen
  \bibfield  {author} {\bibinfo {author} {\bibfnamefont {F.}~\bibnamefont
  {W\"achtler}}\ and\ \bibinfo {author} {\bibfnamefont {L.}~\bibnamefont
  {Santos}},\ }\href {\doibase 10.1103/PhysRevA.93.061603} {\bibfield
  {journal} {\bibinfo  {journal} {Phys. Rev. A}\ }\textbf {\bibinfo {volume}
  {93}},\ \bibinfo {pages} {061603} (\bibinfo {year}
  {2016}{\natexlab{a}})}\BibitemShut {NoStop}%
\bibitem [{\citenamefont {W\"achtler}\ and\ \citenamefont
  {Santos}(2016{\natexlab{b}})}]{Santos2}%
  \BibitemOpen
  \bibfield  {author} {\bibinfo {author} {\bibfnamefont {F.}~\bibnamefont
  {W\"achtler}}\ and\ \bibinfo {author} {\bibfnamefont {L.}~\bibnamefont
  {Santos}},\ }\href {\doibase 10.1103/PhysRevA.94.043618} {\bibfield
  {journal} {\bibinfo  {journal} {Phys. Rev. A}\ }\textbf {\bibinfo {volume}
  {94}},\ \bibinfo {pages} {043618} (\bibinfo {year}
  {2016}{\natexlab{b}})}\BibitemShut {NoStop}%
\bibitem [{\citenamefont {Bisset}\ \emph {et~al.}(2016)\citenamefont {Bisset},
  \citenamefont {Wilson}, \citenamefont {Baillie},\ and\ \citenamefont
  {Blakie}}]{Blakie3}%
  \BibitemOpen
  \bibfield  {author} {\bibinfo {author} {\bibfnamefont {R.~N.}\ \bibnamefont
  {Bisset}}, \bibinfo {author} {\bibfnamefont {R.~M.}\ \bibnamefont {Wilson}},
  \bibinfo {author} {\bibfnamefont {D.}~\bibnamefont {Baillie}}, \ and\
  \bibinfo {author} {\bibfnamefont {P.~B.}\ \bibnamefont {Blakie}},\ }\href
  {\doibase 10.1103/PhysRevA.94.033619} {\bibfield  {journal} {\bibinfo
  {journal} {Phys. Rev. A}\ }\textbf {\bibinfo {volume} {94}},\ \bibinfo
  {pages} {033619} (\bibinfo {year} {2016})}\BibitemShut {NoStop}%
\bibitem [{\citenamefont {Baillie}\ \emph {et~al.}(2016)\citenamefont
  {Baillie}, \citenamefont {Wilson}, \citenamefont {Bisset},\ and\
  \citenamefont {Blakie}}]{Blakie4}%
  \BibitemOpen
  \bibfield  {author} {\bibinfo {author} {\bibfnamefont {D.}~\bibnamefont
  {Baillie}}, \bibinfo {author} {\bibfnamefont {R.~M.}\ \bibnamefont {Wilson}},
  \bibinfo {author} {\bibfnamefont {R.~N.}\ \bibnamefont {Bisset}}, \ and\
  \bibinfo {author} {\bibfnamefont {P.~B.}\ \bibnamefont {Blakie}},\ }\href
  {\doibase 10.1103/PhysRevA.94.021602} {\bibfield  {journal} {\bibinfo
  {journal} {Phys. Rev. A}\ }\textbf {\bibinfo {volume} {94}},\ \bibinfo
  {pages} {021602} (\bibinfo {year} {2016})}\BibitemShut {NoStop}%
\bibitem [{\citenamefont {Chomaz}\ \emph {et~al.}(2016)\citenamefont {Chomaz},
  \citenamefont {Baier}, \citenamefont {Petter}, \citenamefont {Mark},
  \citenamefont {W\"achtler}, \citenamefont {Santos},\ and\ \citenamefont
  {Ferlaino}}]{Francesca2}%
  \BibitemOpen
  \bibfield  {author} {\bibinfo {author} {\bibfnamefont {L.}~\bibnamefont
  {Chomaz}}, \bibinfo {author} {\bibfnamefont {S.}~\bibnamefont {Baier}},
  \bibinfo {author} {\bibfnamefont {D.}~\bibnamefont {Petter}}, \bibinfo
  {author} {\bibfnamefont {M.~J.}\ \bibnamefont {Mark}}, \bibinfo {author}
  {\bibfnamefont {F.}~\bibnamefont {W\"achtler}}, \bibinfo {author}
  {\bibfnamefont {L.}~\bibnamefont {Santos}}, \ and\ \bibinfo {author}
  {\bibfnamefont {F.}~\bibnamefont {Ferlaino}},\ }\href {\doibase 10.1103/PhysRevX.6.041039} {\bibfield  {journal} {\bibinfo  {journal} {Phys.
  Rev. X}\ }\textbf {\bibinfo {volume} {6}},\ \bibinfo {pages} {041039}
  (\bibinfo {year} {2016})}\BibitemShut {NoStop}%
\bibitem [{\citenamefont {{Schmitt}}\ \emph {et~al.}(2016)\citenamefont
  {{Schmitt}}, \citenamefont {{Wenzel}}, \citenamefont {{B{\"o}ttcher}},
  \citenamefont {{Ferrier-Barbut}},\ and\ \citenamefont {{Pfau}}}]{Pfau2}%
  \BibitemOpen
  \bibfield  {author} {\bibinfo {author} {\bibfnamefont {M.}~\bibnamefont
  {{Schmitt}}}, \bibinfo {author} {\bibfnamefont {M.}~\bibnamefont {{Wenzel}}},
  \bibinfo {author} {\bibfnamefont {F.}~\bibnamefont {{B{\"o}ttcher}}},
  \bibinfo {author} {\bibfnamefont {I.}~\bibnamefont {{Ferrier-Barbut}}}, \
  and\ \bibinfo {author} {\bibfnamefont {T.}~\bibnamefont {{Pfau}}},\ }\href
  {\doibase 10.1038/nature20126} {\bibfield  {journal} {\bibinfo  {journal}
  {\nat}\ }\textbf {\bibinfo {volume} {539}},\ \bibinfo {pages} {259} (\bibinfo
  {year} {2016})}\BibitemShut {NoStop}%
\bibitem [{\citenamefont {Miyakawa}\ \emph {et~al.}(2008)\citenamefont
  {Miyakawa}, \citenamefont {Sogo},\ and\ \citenamefont {Pu}}]{Sogo}%
  \BibitemOpen
  \bibfield  {author} {\bibinfo {author} {\bibfnamefont {T.}~\bibnamefont
  {Miyakawa}}, \bibinfo {author} {\bibfnamefont {T.}~\bibnamefont {Sogo}}, \
  and\ \bibinfo {author} {\bibfnamefont {H.}~\bibnamefont {Pu}},\ }\href
  {\doibase 10.1103/PhysRevA.77.061603} {\bibfield  {journal} {\bibinfo
  {journal} {Phys. Rev. A}\ }\textbf {\bibinfo {volume} {77}},\ \bibinfo
  {pages} {061603} (\bibinfo {year} {2008})}\BibitemShut {NoStop}%
\bibitem [{\citenamefont {{Aikawa}}\ \emph {et~al.}(2014)\citenamefont
  {{Aikawa}}, \citenamefont {{Baier}}, \citenamefont {{Frisch}}, \citenamefont
  {{Mark}}, \citenamefont {{Ravensbergen}},\ and\ \citenamefont
  {{Ferlaino}}}]{Francesca}%
  \BibitemOpen
  \bibfield  {author} {\bibinfo {author} {\bibfnamefont {K.}~\bibnamefont
  {{Aikawa}}}, \bibinfo {author} {\bibfnamefont {S.}~\bibnamefont {{Baier}}},
  \bibinfo {author} {\bibfnamefont {A.}~\bibnamefont {{Frisch}}}, \bibinfo
  {author} {\bibfnamefont {M.}~\bibnamefont {{Mark}}}, \bibinfo {author}
  {\bibfnamefont {C.}~\bibnamefont {{Ravensbergen}}}, \ and\ \bibinfo {author}
  {\bibfnamefont {F.}~\bibnamefont {{Ferlaino}}},\ }\href {\doibase 10.1126/science.1255259} {\bibfield  {journal} {\bibinfo  {journal}
  {Science}\ }\textbf {\bibinfo {volume} {345}},\ \bibinfo {pages} {1484}
  (\bibinfo {year} {2014})}\BibitemShut {NoStop}%
\bibitem [{\citenamefont {Sogo}\ \emph {et~al.}(2009)\citenamefont {Sogo},
  \citenamefont {He}, \citenamefont {Miyakawa}, \citenamefont {Yi},
  \citenamefont {Lu},\ and\ \citenamefont {Pu}}]{Sogo2}%
  \BibitemOpen
  \bibfield  {author} {\bibinfo {author} {\bibfnamefont {T.}~\bibnamefont
  {Sogo}}, \bibinfo {author} {\bibfnamefont {L.}~\bibnamefont {He}}, \bibinfo
  {author} {\bibfnamefont {T.}~\bibnamefont {Miyakawa}}, \bibinfo {author}
  {\bibfnamefont {S.}~\bibnamefont {Yi}}, \bibinfo {author} {\bibfnamefont
  {H.}~\bibnamefont {Lu}}, \ and\ \bibinfo {author} {\bibfnamefont
  {H.}~\bibnamefont {Pu}},\ }\href
  {http://stacks.iop.org/1367-2630/11/i=5/a=055017} {\bibfield  {journal}
  {\bibinfo  {journal} {New J. Phys.}\ }\textbf {\bibinfo {volume} {11}},\
  \bibinfo {pages} {055017} (\bibinfo {year} {2009})}\BibitemShut {NoStop}%
\bibitem [{\citenamefont {Sogo}\ \emph {et~al.}(2010)\citenamefont {Sogo},
  \citenamefont {He}, \citenamefont {Miyakawa}, \citenamefont {Yi},
  \citenamefont {Lu},\ and\ \citenamefont {Pu}}]{Sogo2c}%
  \BibitemOpen
  \bibfield  {author} {\bibinfo {author} {\bibfnamefont {T.}~\bibnamefont
  {Sogo}}, \bibinfo {author} {\bibfnamefont {L.}~\bibnamefont {He}}, \bibinfo
  {author} {\bibfnamefont {T.}~\bibnamefont {Miyakawa}}, \bibinfo {author}
  {\bibfnamefont {S.}~\bibnamefont {Yi}}, \bibinfo {author} {\bibfnamefont
  {H.}~\bibnamefont {Lu}}, \ and\ \bibinfo {author} {\bibfnamefont
  {H.}~\bibnamefont {Pu}},\ }\href
  {http://stacks.iop.org/1367-2630/12/i=7/a=079801} {\bibfield  {journal}
  {\bibinfo  {journal} {New J. Phys.}\ }\textbf {\bibinfo {volume} {12}},\
  \bibinfo {pages} {079801} (\bibinfo {year} {2010})}\BibitemShut {NoStop}%
\bibitem [{\citenamefont {Zhang}\ \emph {et~al.}(2011)\citenamefont {Zhang},
  \citenamefont {Qiu}, \citenamefont {He},\ and\ \citenamefont {Yi}}]{Zhang}%
  \BibitemOpen
  \bibfield  {author} {\bibinfo {author} {\bibfnamefont {J.-N.}\ \bibnamefont
  {Zhang}}, \bibinfo {author} {\bibfnamefont {R.-Z.}\ \bibnamefont {Qiu}},
  \bibinfo {author} {\bibfnamefont {L.}~\bibnamefont {He}}, \ and\ \bibinfo
  {author} {\bibfnamefont {S.}~\bibnamefont {Yi}},\ }\href {\doibase 10.1103/PhysRevA.83.053628} {\bibfield  {journal} {\bibinfo  {journal} {Phys.
  Rev. A}\ }\textbf {\bibinfo {volume} {83}},\ \bibinfo {pages} {053628}
  (\bibinfo {year} {2011})}\BibitemShut {NoStop}%
\bibitem [{\citenamefont {Baillie}\ and\ \citenamefont
  {Blakie}(2012)}]{Baillie}%
  \BibitemOpen
  \bibfield  {author} {\bibinfo {author} {\bibfnamefont {D.}~\bibnamefont
  {Baillie}}\ and\ \bibinfo {author} {\bibfnamefont {P.~B.}\ \bibnamefont
  {Blakie}},\ }\href {\doibase 10.1103/PhysRevA.86.023605} {\bibfield
  {journal} {\bibinfo  {journal} {Phys. Rev. A}\ }\textbf {\bibinfo {volume}
  {86}},\ \bibinfo {pages} {023605} (\bibinfo {year} {2012})}\BibitemShut
  {NoStop}%
\bibitem [{\citenamefont {Lima}\ and\ \citenamefont
  {Pelster}(2010{\natexlab{a}})}]{Lima1}%
  \BibitemOpen
  \bibfield  {author} {\bibinfo {author} {\bibfnamefont {A.~R.~P.}\
  \bibnamefont {Lima}}\ and\ \bibinfo {author} {\bibfnamefont {A.}~\bibnamefont
  {Pelster}},\ }\href {\doibase 10.1103/PhysRevA.81.021606} {\bibfield
  {journal} {\bibinfo  {journal} {Phys. Rev. A}\ }\textbf {\bibinfo {volume}
  {81}},\ \bibinfo {pages} {021606(R)} (\bibinfo {year}
  {2010}{\natexlab{a}})}\BibitemShut {NoStop}%
\bibitem [{\citenamefont {Lima}\ and\ \citenamefont
  {Pelster}(2010{\natexlab{b}})}]{Lima2}%
  \BibitemOpen
  \bibfield  {author} {\bibinfo {author} {\bibfnamefont {A.~R.~P.}\
  \bibnamefont {Lima}}\ and\ \bibinfo {author} {\bibfnamefont {A.}~\bibnamefont
  {Pelster}},\ }\href {\doibase 10.1103/PhysRevA.81.063629} {\bibfield
  {journal} {\bibinfo  {journal} {Phys. Rev. A}\ }\textbf {\bibinfo {volume}
  {81}},\ \bibinfo {pages} {063629} (\bibinfo {year}
  {2010}{\natexlab{b}})}\BibitemShut {NoStop}%
\bibitem [{\citenamefont {{W\"{a}chtler}}\ \emph {et~al.}(2013)\citenamefont
  {{W\"{a}chtler}}, \citenamefont {{Lima}},\ and\ \citenamefont
  {{Pelster}}}]{Falk}%
  \BibitemOpen
  \bibfield  {author} {\bibinfo {author} {\bibfnamefont {F.}~\bibnamefont
  {{W\"{a}chtler}}}, \bibinfo {author} {\bibfnamefont {A.~R.~P.}\ \bibnamefont
  {{Lima}}}, \ and\ \bibinfo {author} {\bibfnamefont {A.}~\bibnamefont
  {{Pelster}}},\ }\href@noop {} {\bibfield  {journal} {\bibinfo  {journal}
  {ArXiv e-prints}\ } (\bibinfo {year} {2013})},\ \Eprint
  {http://arxiv.org/abs/1311.5100} {arXiv:1311.5100 [cond-mat.quant-gas]}
  \BibitemShut {NoStop}%
\bibitem [{\citenamefont {Abad}\ \emph {et~al.}(2012)\citenamefont {Abad},
  \citenamefont {Recati},\ and\ \citenamefont {Stringari}}]{Abad}%
  \BibitemOpen
  \bibfield  {author} {\bibinfo {author} {\bibfnamefont {M.}~\bibnamefont
  {Abad}}, \bibinfo {author} {\bibfnamefont {A.}~\bibnamefont {Recati}}, \ and\
  \bibinfo {author} {\bibfnamefont {S.}~\bibnamefont {Stringari}},\ }\href
  {\doibase 10.1103/PhysRevA.85.033639} {\bibfield  {journal} {\bibinfo
  {journal} {Phys. Rev. A}\ }\textbf {\bibinfo {volume} {85}},\ \bibinfo
  {pages} {033639} (\bibinfo {year} {2012})}\BibitemShut {NoStop}%
\bibitem [{\citenamefont {Liu}\ and\ \citenamefont {Yin}(2011)}]{Liu-Yin}%
  \BibitemOpen
  \bibfield  {author} {\bibinfo {author} {\bibfnamefont {B.}~\bibnamefont
  {Liu}}\ and\ \bibinfo {author} {\bibfnamefont {L.}~\bibnamefont {Yin}},\
  }\href {\doibase 10.1103/PhysRevA.84.053603} {\bibfield  {journal} {\bibinfo
  {journal} {Phys. Rev. A}\ }\textbf {\bibinfo {volume} {84}},\ \bibinfo
  {pages} {053603} (\bibinfo {year} {2011})}\BibitemShut {NoStop}%
\bibitem [{\citenamefont {Krieg}\ \emph {et~al.}(2015)\citenamefont {Krieg},
  \citenamefont {Lange}, \citenamefont {Bartosch},\ and\ \citenamefont
  {Kopietz}}]{Kopietz1}%
  \BibitemOpen
  \bibfield  {author} {\bibinfo {author} {\bibfnamefont {J.}~\bibnamefont
  {Krieg}}, \bibinfo {author} {\bibfnamefont {P.}~\bibnamefont {Lange}},
  \bibinfo {author} {\bibfnamefont {L.}~\bibnamefont {Bartosch}}, \ and\
  \bibinfo {author} {\bibfnamefont {P.}~\bibnamefont {Kopietz}},\ }\href
  {\doibase 10.1103/PhysRevA.91.023612} {\bibfield  {journal} {\bibinfo
  {journal} {Phys. Rev. A}\ }\textbf {\bibinfo {volume} {91}},\ \bibinfo
  {pages} {023612} (\bibinfo {year} {2015})}\BibitemShut {NoStop}%
\bibitem [{\citenamefont {Lange}\ \emph {et~al.}(2016)\citenamefont {Lange},
  \citenamefont {Krieg},\ and\ \citenamefont {Kopietz}}]{Kopietz2}%
  \BibitemOpen
  \bibfield  {author} {\bibinfo {author} {\bibfnamefont {P.}~\bibnamefont
  {Lange}}, \bibinfo {author} {\bibfnamefont {J.}~\bibnamefont {Krieg}}, \ and\
  \bibinfo {author} {\bibfnamefont {P.}~\bibnamefont {Kopietz}},\ }\href
  {\doibase 10.1103/PhysRevA.93.033609} {\bibfield  {journal} {\bibinfo
  {journal} {Phys. Rev. A}\ }\textbf {\bibinfo {volume} {93}},\ \bibinfo
  {pages} {033609} (\bibinfo {year} {2016})}\BibitemShut {NoStop}%
\bibitem [{\citenamefont {Naylor}\ \emph {et~al.}(2015)\citenamefont {Naylor},
  \citenamefont {Reigue}, \citenamefont {Mar\'echal}, \citenamefont {Gorceix},
  \citenamefont {Laburthe-Tolra},\ and\ \citenamefont {Vernac}}]{Cr}%
  \BibitemOpen
  \bibfield  {author} {\bibinfo {author} {\bibfnamefont {B.}~\bibnamefont
  {Naylor}}, \bibinfo {author} {\bibfnamefont {A.}~\bibnamefont {Reigue}},
  \bibinfo {author} {\bibfnamefont {E.}~\bibnamefont {Mar\'echal}}, \bibinfo
  {author} {\bibfnamefont {O.}~\bibnamefont {Gorceix}}, \bibinfo {author}
  {\bibfnamefont {B.}~\bibnamefont {Laburthe-Tolra}}, \ and\ \bibinfo {author}
  {\bibfnamefont {L.}~\bibnamefont {Vernac}},\ }\href {\doibase 10.1103/PhysRevA.91.011603} {\bibfield  {journal} {\bibinfo  {journal} {Phys.
  Rev. A}\ }\textbf {\bibinfo {volume} {91}},\ \bibinfo {pages} {011603(R)}
  (\bibinfo {year} {2015})}\BibitemShut {NoStop}%
\bibitem [{\citenamefont {Aikawa}\ \emph
  {et~al.}(2014{\natexlab{a}})\citenamefont {Aikawa}, \citenamefont {Frisch},
  \citenamefont {Mark}, \citenamefont {Baier}, \citenamefont {Grimm},\ and\
  \citenamefont {Ferlaino}}]{Er}%
  \BibitemOpen
  \bibfield  {author} {\bibinfo {author} {\bibfnamefont {K.}~\bibnamefont
  {Aikawa}}, \bibinfo {author} {\bibfnamefont {A.}~\bibnamefont {Frisch}},
  \bibinfo {author} {\bibfnamefont {M.}~\bibnamefont {Mark}}, \bibinfo {author}
  {\bibfnamefont {S.}~\bibnamefont {Baier}}, \bibinfo {author} {\bibfnamefont
  {R.}~\bibnamefont {Grimm}}, \ and\ \bibinfo {author} {\bibfnamefont
  {F.}~\bibnamefont {Ferlaino}},\ }\href {\doibase 10.1103/PhysRevLett.112.010404} {\bibfield  {journal} {\bibinfo  {journal}
  {Phys. Rev. Lett.}\ }\textbf {\bibinfo {volume} {112}},\ \bibinfo {pages}
  {010404} (\bibinfo {year} {2014}{\natexlab{a}})}\BibitemShut {NoStop}%
\bibitem [{\citenamefont {Lu}\ \emph {et~al.}(2012)\citenamefont {Lu},
  \citenamefont {Burdick},\ and\ \citenamefont {Lev}}]{Dy1}%
  \BibitemOpen
  \bibfield  {author} {\bibinfo {author} {\bibfnamefont {M.}~\bibnamefont
  {Lu}}, \bibinfo {author} {\bibfnamefont {N.~Q.}\ \bibnamefont {Burdick}}, \
  and\ \bibinfo {author} {\bibfnamefont {B.~L.}\ \bibnamefont {Lev}},\ }\href
  {\doibase 10.1103/PhysRevLett.108.215301} {\bibfield  {journal} {\bibinfo
  {journal} {Phys. Rev. Lett.}\ }\textbf {\bibinfo {volume} {108}},\ \bibinfo
  {pages} {215301} (\bibinfo {year} {2012})}\BibitemShut {NoStop}%
\bibitem [{\citenamefont {Park}\ \emph
  {et~al.}(2015{\natexlab{a}})\citenamefont {Park}, \citenamefont {Will},\ and\
  \citenamefont {Zwierlein}}]{NaK1}%
  \BibitemOpen
  \bibfield  {author} {\bibinfo {author} {\bibfnamefont {J.~W.}\ \bibnamefont
  {Park}}, \bibinfo {author} {\bibfnamefont {S.~A.}\ \bibnamefont {Will}}, \
  and\ \bibinfo {author} {\bibfnamefont {M.~W.}\ \bibnamefont {Zwierlein}},\
  }\href {\doibase 10.1103/PhysRevLett.114.205302} {\bibfield  {journal}
  {\bibinfo  {journal} {Phys. Rev. Lett.}\ }\textbf {\bibinfo {volume} {114}},\
  \bibinfo {pages} {205302} (\bibinfo {year} {2015}{\natexlab{a}})}\BibitemShut
  {NoStop}%
\bibitem [{\citenamefont {{Ni}}\ \emph {et~al.}(2008)\citenamefont {{Ni}},
  \citenamefont {{Ospelkaus}}, \citenamefont {{de Miranda}}, \citenamefont
  {{Pe'er}}, \citenamefont {{Neyenhuis}}, \citenamefont {{Zirbel}},
  \citenamefont {{Kotochigova}}, \citenamefont {{Julienne}}, \citenamefont
  {{Jin}},\ and\ \citenamefont {{Ye}}}]{KRb}%
  \BibitemOpen
  \bibfield  {author} {\bibinfo {author} {\bibfnamefont {K.-K.}\ \bibnamefont
  {{Ni}}}, \bibinfo {author} {\bibfnamefont {S.}~\bibnamefont {{Ospelkaus}}},
  \bibinfo {author} {\bibfnamefont {M.~H.~G.}\ \bibnamefont {{de Miranda}}},
  \bibinfo {author} {\bibfnamefont {A.}~\bibnamefont {{Pe'er}}}, \bibinfo
  {author} {\bibfnamefont {B.}~\bibnamefont {{Neyenhuis}}}, \bibinfo {author}
  {\bibfnamefont {J.~J.}\ \bibnamefont {{Zirbel}}}, \bibinfo {author}
  {\bibfnamefont {S.}~\bibnamefont {{Kotochigova}}}, \bibinfo {author}
  {\bibfnamefont {P.~S.}\ \bibnamefont {{Julienne}}}, \bibinfo {author}
  {\bibfnamefont {D.~S.}\ \bibnamefont {{Jin}}}, \ and\ \bibinfo {author}
  {\bibfnamefont {J.}~\bibnamefont {{Ye}}},\ }\href {\doibase 10.1126/science.1163861} {\bibfield  {journal} {\bibinfo  {journal}
  {Science}\ }\textbf {\bibinfo {volume} {322}},\ \bibinfo {pages} {231}
  (\bibinfo {year} {2008})}\BibitemShut {NoStop}%
\bibitem [{\citenamefont {Vitanov}\ \emph {et~al.}(2017)\citenamefont
  {Vitanov}, \citenamefont {Rangelov}, \citenamefont {Shore},\ and\
  \citenamefont {Bergmann}}]{Bergmann}%
  \BibitemOpen
  \bibfield  {author} {\bibinfo {author} {\bibfnamefont {N.~V.}\ \bibnamefont
  {Vitanov}}, \bibinfo {author} {\bibfnamefont {A.~A.}\ \bibnamefont
  {Rangelov}}, \bibinfo {author} {\bibfnamefont {B.~W.}\ \bibnamefont {Shore}},
  \ and\ \bibinfo {author} {\bibfnamefont {K.}~\bibnamefont {Bergmann}},\
  }\href {\doibase 10.1103/RevModPhys.89.015006} {\bibfield  {journal}
  {\bibinfo  {journal} {Rev. Mod. Phys.}\ }\textbf {\bibinfo {volume} {89}},\
  \bibinfo {pages} {015006} (\bibinfo {year} {2017})}\BibitemShut {NoStop}%
\bibitem [{\citenamefont {Park}\ \emph
  {et~al.}(2015{\natexlab{b}})\citenamefont {Park}, \citenamefont {Will},\ and\
  \citenamefont {Zwierlein}}]{NaK2}%
  \BibitemOpen
  \bibfield  {author} {\bibinfo {author} {\bibfnamefont {J.~W.}\ \bibnamefont
  {Park}}, \bibinfo {author} {\bibfnamefont {S.~A.}\ \bibnamefont {Will}}, \
  and\ \bibinfo {author} {\bibfnamefont {M.~W.}\ \bibnamefont {Zwierlein}},\
  }\href {http://stacks.iop.org/1367-2630/17/i=7/a=075016} {\bibfield
  {journal} {\bibinfo  {journal} {New J. Phys.}\ }\textbf {\bibinfo {volume}
  {17}},\ \bibinfo {pages} {075016} (\bibinfo {year}
  {2015}{\natexlab{b}})}\BibitemShut {NoStop}%
\bibitem [{\citenamefont {Kuznetsova}\ \emph {et~al.}(2009)\citenamefont
  {Kuznetsova}, \citenamefont {Gacesa}, \citenamefont {Pellegrini},
  \citenamefont {Yelin},\ and\ \citenamefont {Côté}}]{Kuznetsova}%
  \BibitemOpen
  \bibfield  {author} {\bibinfo {author} {\bibfnamefont {E.}~\bibnamefont
  {Kuznetsova}}, \bibinfo {author} {\bibfnamefont {M.}~\bibnamefont {Gacesa}},
  \bibinfo {author} {\bibfnamefont {P.}~\bibnamefont {Pellegrini}}, \bibinfo
  {author} {\bibfnamefont {S.~F.}\ \bibnamefont {Yelin}}, \ and\ \bibinfo
  {author} {\bibfnamefont {R.}~\bibnamefont {Côté}},\ }\href
  {http://stacks.iop.org/1367-2630/11/i=5/a=055028} {\bibfield  {journal}
  {\bibinfo  {journal} {New J. Phys.}\ }\textbf {\bibinfo {volume} {11}},\
  \bibinfo {pages} {055028} (\bibinfo {year} {2009})}\BibitemShut {NoStop}%
\bibitem [{\citenamefont {Gregory}\ \emph {et~al.}(2015)\citenamefont
  {Gregory}, \citenamefont {Molony}, \citenamefont {Köppinger}, \citenamefont
  {Kumar}, \citenamefont {Ji}, \citenamefont {Lu}, \citenamefont {Marchant},\
  and\ \citenamefont {Cornish}}]{Gregory}%
  \BibitemOpen
  \bibfield  {author} {\bibinfo {author} {\bibfnamefont {P.~D.}\ \bibnamefont
  {Gregory}}, \bibinfo {author} {\bibfnamefont {P.~K.}\ \bibnamefont {Molony}},
  \bibinfo {author} {\bibfnamefont {M.~P.}\ \bibnamefont {Köppinger}},
  \bibinfo {author} {\bibfnamefont {A.}~\bibnamefont {Kumar}}, \bibinfo
  {author} {\bibfnamefont {Z.}~\bibnamefont {Ji}}, \bibinfo {author}
  {\bibfnamefont {B.}~\bibnamefont {Lu}}, \bibinfo {author} {\bibfnamefont
  {A.~L.}\ \bibnamefont {Marchant}}, \ and\ \bibinfo {author} {\bibfnamefont
  {S.~L.}\ \bibnamefont {Cornish}},\ }\href
  {http://stacks.iop.org/1367-2630/17/i=5/a=055006} {\bibfield  {journal}
  {\bibinfo  {journal} {New J. Phys.}\ }\textbf {\bibinfo {volume} {17}},\
  \bibinfo {pages} {055006} (\bibinfo {year} {2015})}\BibitemShut {NoStop}%
\bibitem [{\citenamefont {Schleich}(2005)}]{SchleichBook}%
  \BibitemOpen
  \bibfield  {author} {\bibinfo {author} {\bibfnamefont {W.~P.}\ \bibnamefont
  {Schleich}},\ }\href@noop {} {\emph {\bibinfo {title} {Quantum Optics in
  Phase Space}}}\ (\bibinfo  {publisher} {Wiley-VCH},\ \bibinfo {address}
  {Berlin},\ \bibinfo {year} {2005})\BibitemShut {NoStop}%
\bibitem [{\citenamefont {Bogojevi\'{c}}\ \emph {et~al.}(2005)\citenamefont
  {Bogojevi\'{c}}, \citenamefont {Bala\v{z}},\ and\ \citenamefont
  {Beli\'{c}}}]{STE1}%
  \BibitemOpen
  \bibfield  {author} {\bibinfo {author} {\bibfnamefont {A.}~\bibnamefont
  {Bogojevi\'{c}}}, \bibinfo {author} {\bibfnamefont {A.}~\bibnamefont
  {Bala\v{z}}}, \ and\ \bibinfo {author} {\bibfnamefont {A.}~\bibnamefont
  {Beli\'{c}}},\ }\href {\doibase 10.1103/PhysRevE.72.036128} {\bibfield
  {journal} {\bibinfo  {journal} {Phys. Rev. E}\ }\textbf {\bibinfo {volume}
  {72}},\ \bibinfo {pages} {036128} (\bibinfo {year} {2005})}\BibitemShut
  {NoStop}%
\bibitem [{\citenamefont {Bogojevi\'{c}}\ \emph {et~al.}(2008)\citenamefont
  {Bogojevi\'{c}}, \citenamefont {Vidanovi\'{c}}, \citenamefont {Bala\v{z}},\
  and\ \citenamefont {Beli\'{c}}}]{STE2}%
  \BibitemOpen
  \bibfield  {author} {\bibinfo {author} {\bibfnamefont {A.}~\bibnamefont
  {Bogojevi\'{c}}}, \bibinfo {author} {\bibfnamefont {I.}~\bibnamefont
  {Vidanovi\'{c}}}, \bibinfo {author} {\bibfnamefont {A.}~\bibnamefont
  {Bala\v{z}}}, \ and\ \bibinfo {author} {\bibfnamefont {A.}~\bibnamefont
  {Beli\'{c}}},\ }\href {\doibase 10.1016/j.physleta.2008.01.079} {\bibfield
  {journal} {\bibinfo  {journal} {Phys. Lett. A}\ }\textbf {\bibinfo {volume}
  {372}},\ \bibinfo {pages} {3341 } (\bibinfo {year} {2008})}\BibitemShut
  {NoStop}%
\bibitem [{\citenamefont {Vidanovi\'{c}}\ \emph {et~al.}(2009)\citenamefont
  {Vidanovi\'{c}}, \citenamefont {Bogojevi\'{c}}, \citenamefont {Bala\v{z}},\
  and\ \citenamefont {Beli\'{c}}}]{STE3}%
  \BibitemOpen
  \bibfield  {author} {\bibinfo {author} {\bibfnamefont {I.}~\bibnamefont
  {Vidanovi\'{c}}}, \bibinfo {author} {\bibfnamefont {A.}~\bibnamefont
  {Bogojevi\'{c}}}, \bibinfo {author} {\bibfnamefont {A.}~\bibnamefont
  {Bala\v{z}}}, \ and\ \bibinfo {author} {\bibfnamefont {A.}~\bibnamefont
  {Beli\'{c}}},\ }\href {\doibase 10.1103/PhysRevE.80.066706} {\bibfield
  {journal} {\bibinfo  {journal} {Phys. Rev. E}\ }\textbf {\bibinfo {volume}
  {80}},\ \bibinfo {pages} {066706} (\bibinfo {year} {2009})}\BibitemShut
  {NoStop}%
\bibitem [{\citenamefont {{Bala\v{z}}}\ \emph
  {et~al.}(2011{\natexlab{a}})\citenamefont {{Bala\v{z}}}, \citenamefont
  {{Vidanovi\'{c}}}, \citenamefont {{Bogojevi\'{c}}}, \citenamefont
  {{Beli{\'c}}},\ and\ \citenamefont {{Pelster}}}]{STE4}%
  \BibitemOpen
  \bibfield  {author} {\bibinfo {author} {\bibfnamefont {A.}~\bibnamefont
  {{Bala\v{z}}}}, \bibinfo {author} {\bibfnamefont {I.}~\bibnamefont
  {{Vidanovi\'{c}}}}, \bibinfo {author} {\bibfnamefont {A.}~\bibnamefont
  {{Bogojevi\'{c}}}}, \bibinfo {author} {\bibfnamefont {A.}~\bibnamefont
  {{Beli{\'c}}}}, \ and\ \bibinfo {author} {\bibfnamefont {A.}~\bibnamefont
  {{Pelster}}},\ }\href {\doibase 10.1088/1742-5468/2011/03/P03004} {\bibfield
  {journal} {\bibinfo  {journal} {J. Stat. Mech.-Theory Exp.}\ }\textbf
  {\bibinfo {volume} {2011}},\ \bibinfo {pages} {P03004} (\bibinfo {year}
  {2011}{\natexlab{a}})}\BibitemShut {NoStop}%
\bibitem [{\citenamefont {{Bala\v{z}}}\ \emph
  {et~al.}(2011{\natexlab{b}})\citenamefont {{Bala\v{z}}}, \citenamefont
  {{Vidanovi\'{c}}}, \citenamefont {{Bogojevi\'{c}}}, \citenamefont
  {{Beli{\'c}}},\ and\ \citenamefont {{Pelster}}}]{STE5}%
  \BibitemOpen
  \bibfield  {author} {\bibinfo {author} {\bibfnamefont {A.}~\bibnamefont
  {{Bala\v{z}}}}, \bibinfo {author} {\bibfnamefont {I.}~\bibnamefont
  {{Vidanovi\'{c}}}}, \bibinfo {author} {\bibfnamefont {A.}~\bibnamefont
  {{Bogojevi\'{c}}}}, \bibinfo {author} {\bibfnamefont {A.}~\bibnamefont
  {{Beli{\'c}}}}, \ and\ \bibinfo {author} {\bibfnamefont {A.}~\bibnamefont
  {{Pelster}}},\ }\href {\doibase 10.1088/1742-5468/2011/03/P03005} {\bibfield
  {journal} {\bibinfo  {journal} {J. Stat. Mech.-Theory Exp.}\ }\textbf
  {\bibinfo {volume} {2011}},\ \bibinfo {pages} {P03005} (\bibinfo {year}
  {2011}{\natexlab{b}})}\BibitemShut {NoStop}%
\bibitem [{\citenamefont {Baillie}\ \emph {et~al.}(2015)\citenamefont
  {Baillie}, \citenamefont {Bisset},\ and\ \citenamefont {Blakie}}]{Blakie}%
  \BibitemOpen
  \bibfield  {author} {\bibinfo {author} {\bibfnamefont {D.}~\bibnamefont
  {Baillie}}, \bibinfo {author} {\bibfnamefont {R.~N.}\ \bibnamefont {Bisset}},
  \ and\ \bibinfo {author} {\bibfnamefont {P.~B.}\ \bibnamefont {Blakie}},\
  }\href {\doibase 10.1103/PhysRevA.91.013613} {\bibfield  {journal} {\bibinfo
  {journal} {Phys. Rev. A}\ }\textbf {\bibinfo {volume} {91}},\ \bibinfo
  {pages} {013613} (\bibinfo {year} {2015})}\BibitemShut {NoStop}%
\bibitem [{\citenamefont {Bohn}\ \emph {et~al.}(2009)\citenamefont {Bohn},
  \citenamefont {Cavagnero},\ and\ \citenamefont {Ticknor}}]{sigma}%
  \BibitemOpen
  \bibfield  {author} {\bibinfo {author} {\bibfnamefont {J.~L.}\ \bibnamefont
  {Bohn}}, \bibinfo {author} {\bibfnamefont {M.}~\bibnamefont {Cavagnero}}, \
  and\ \bibinfo {author} {\bibfnamefont {C.}~\bibnamefont {Ticknor}},\ }\href
  {http://stacks.iop.org/1367-2630/11/i=5/a=055039} {\bibfield  {journal}
  {\bibinfo  {journal} {New J. Phys.}\ }\textbf {\bibinfo {volume} {11}},\
  \bibinfo {pages} {055039} (\bibinfo {year} {2009})}\BibitemShut {NoStop}%
\bibitem [{\citenamefont {Baran}\ \emph {et~al.}(2005)\citenamefont {Baran},
  \citenamefont {Colonna}, \citenamefont {Greco},\ and\ \citenamefont
  {Toro}}]{Baran}%
  \BibitemOpen
  \bibfield  {author} {\bibinfo {author} {\bibfnamefont {V.}~\bibnamefont
  {Baran}}, \bibinfo {author} {\bibfnamefont {M.}~\bibnamefont {Colonna}},
  \bibinfo {author} {\bibfnamefont {V.}~\bibnamefont {Greco}}, \ and\ \bibinfo
  {author} {\bibfnamefont {M.~D.}\ \bibnamefont {Toro}},\ }\href {\doibase 10.1016/j.physrep.2004.12.004} {\bibfield  {journal} {\bibinfo  {journal}
  {Phys. Rep.}\ }\textbf {\bibinfo {volume} {410}},\ \bibinfo {pages} {335 }
  (\bibinfo {year} {2005})}\BibitemShut {NoStop}%
\bibitem [{\citenamefont {Tabacu}\ \emph {et~al.}(2015)\citenamefont {Tabacu},
  \citenamefont {Raportaru}, \citenamefont {Slusanschi}, \citenamefont
  {Baran},\ and\ \citenamefont {Nicolin}}]{Alex2}%
  \BibitemOpen
  \bibfield  {author} {\bibinfo {author} {\bibfnamefont {R.}~\bibnamefont
  {Tabacu}}, \bibinfo {author} {\bibfnamefont {M.~R.}\ \bibnamefont
  {Raportaru}}, \bibinfo {author} {\bibfnamefont {E.}~\bibnamefont
  {Slusanschi}}, \bibinfo {author} {\bibfnamefont {V.}~\bibnamefont {Baran}}, \
  and\ \bibinfo {author} {\bibfnamefont {A.~I.}\ \bibnamefont {Nicolin}},\
  }\href {https://www.nipne.ro/rjp/2015_60_9-10/1441_1449.pdf} {\bibfield
  {journal} {\bibinfo  {journal} {Rom. J. Phys.}\ }\textbf {\bibinfo {volume}
  {60}},\ \bibinfo {pages} {1441 } (\bibinfo {year} {2015})}\BibitemShut
  {NoStop}%
\bibitem [{\citenamefont {Petrovi\'{c}}\ \emph {et~al.}(2002)\citenamefont
  {Petrovi\'{c}}, \citenamefont {Raspopovi\'{c}}, \citenamefont {Dujko},\ and\
  \citenamefont {Makabe}}]{ZLjP1}%
  \BibitemOpen
  \bibfield  {author} {\bibinfo {author} {\bibfnamefont {Z.~L.}\ \bibnamefont
  {Petrovi\'{c}}}, \bibinfo {author} {\bibfnamefont {Z.~M.}\ \bibnamefont
  {Raspopovi\'{c}}}, \bibinfo {author} {\bibfnamefont {S.}~\bibnamefont
  {Dujko}}, \ and\ \bibinfo {author} {\bibfnamefont {T.}~\bibnamefont
  {Makabe}},\ }\href {\doibase 10.1016/S0169-4332(02)00018-1} {\bibfield
  {journal} {\bibinfo  {journal} {Appl. Surf. Sci.}\ }\textbf {\bibinfo
  {volume} {192}},\ \bibinfo {pages} {1 } (\bibinfo {year} {2002})},\ \bibinfo
  {note} {advance in Low Temperature \{RF\} Plasmas}\BibitemShut {NoStop}%
\bibitem [{\citenamefont {Dujko}\ \emph {et~al.}(2011)\citenamefont {Dujko},
  \citenamefont {Ebert}, \citenamefont {White},\ and\ \citenamefont
  {Petrovi\'{c}}}]{ZLjP2}%
  \BibitemOpen
  \bibfield  {author} {\bibinfo {author} {\bibfnamefont {S.}~\bibnamefont
  {Dujko}}, \bibinfo {author} {\bibfnamefont {U.}~\bibnamefont {Ebert}},
  \bibinfo {author} {\bibfnamefont {R.~D.}\ \bibnamefont {White}}, \ and\
  \bibinfo {author} {\bibfnamefont {Z.~L.}\ \bibnamefont {Petrovi\'{c}}},\
  }\href {http://stacks.iop.org/1347-4065/50/i=8S1/a=08JC01} {\bibfield
  {journal} {\bibinfo  {journal} {Jpn. J. Appl. Phys.}\ }\textbf {\bibinfo
  {volume} {50}},\ \bibinfo {pages} {08JC01} (\bibinfo {year}
  {2011})}\BibitemShut {NoStop}%
\bibitem [{\citenamefont {Dusling}\ and\ \citenamefont
  {Sch\"afer}(2011)}]{Dusling}%
  \BibitemOpen
  \bibfield  {author} {\bibinfo {author} {\bibfnamefont {K.}~\bibnamefont
  {Dusling}}\ and\ \bibinfo {author} {\bibfnamefont {T.}~\bibnamefont
  {Sch\"afer}},\ }\href {\doibase 10.1103/PhysRevA.84.013622} {\bibfield
  {journal} {\bibinfo  {journal} {Phys. Rev. A}\ }\textbf {\bibinfo {volume}
  {84}},\ \bibinfo {pages} {013622} (\bibinfo {year} {2011})}\BibitemShut
  {NoStop}%
\bibitem [{\citenamefont {Pantel}\ \emph {et~al.}(2015)\citenamefont {Pantel},
  \citenamefont {Davesne},\ and\ \citenamefont {Urban}}]{Urban}%
  \BibitemOpen
  \bibfield  {author} {\bibinfo {author} {\bibfnamefont {P.-A.}\ \bibnamefont
  {Pantel}}, \bibinfo {author} {\bibfnamefont {D.}~\bibnamefont {Davesne}}, \
  and\ \bibinfo {author} {\bibfnamefont {M.}~\bibnamefont {Urban}},\ }\href
  {\doibase 10.1103/PhysRevA.91.013627} {\bibfield  {journal} {\bibinfo
  {journal} {Phys. Rev. A}\ }\textbf {\bibinfo {volume} {91}},\ \bibinfo
  {pages} {013627} (\bibinfo {year} {2015})}\BibitemShut {NoStop}%
\bibitem [{\citenamefont {Chiacchiera}\ \emph {et~al.}(2009)\citenamefont
  {Chiacchiera}, \citenamefont {Lepers}, \citenamefont {Davesne},\ and\
  \citenamefont {Urban}}]{Urban2}%
  \BibitemOpen
  \bibfield  {author} {\bibinfo {author} {\bibfnamefont {S.}~\bibnamefont
  {Chiacchiera}}, \bibinfo {author} {\bibfnamefont {T.}~\bibnamefont {Lepers}},
  \bibinfo {author} {\bibfnamefont {D.}~\bibnamefont {Davesne}}, \ and\
  \bibinfo {author} {\bibfnamefont {M.}~\bibnamefont {Urban}},\ }\href
  {\doibase 10.1103/PhysRevA.79.033613} {\bibfield  {journal} {\bibinfo
  {journal} {Phys. Rev. A}\ }\textbf {\bibinfo {volume} {79}},\ \bibinfo
  {pages} {033613} (\bibinfo {year} {2009})}\BibitemShut {NoStop}%
\bibitem [{\citenamefont {Chiacchiera}\ \emph {et~al.}(2011)\citenamefont
  {Chiacchiera}, \citenamefont {Lepers}, \citenamefont {Davesne},\ and\
  \citenamefont {Urban}}]{Urban3}%
  \BibitemOpen
  \bibfield  {author} {\bibinfo {author} {\bibfnamefont {S.}~\bibnamefont
  {Chiacchiera}}, \bibinfo {author} {\bibfnamefont {T.}~\bibnamefont {Lepers}},
  \bibinfo {author} {\bibfnamefont {D.}~\bibnamefont {Davesne}}, \ and\
  \bibinfo {author} {\bibfnamefont {M.}~\bibnamefont {Urban}},\ }\href
  {\doibase 10.1103/PhysRevA.84.043634} {\bibfield  {journal} {\bibinfo
  {journal} {Phys. Rev. A}\ }\textbf {\bibinfo {volume} {84}},\ \bibinfo
  {pages} {043634} (\bibinfo {year} {2011})}\BibitemShut {NoStop}%
\bibitem [{\citenamefont {Kadanoff}\ and\ \citenamefont {Baym}(1962)}]{Baym}%
  \BibitemOpen
  \bibfield  {author} {\bibinfo {author} {\bibfnamefont {L.~P.}\ \bibnamefont
  {Kadanoff}}\ and\ \bibinfo {author} {\bibfnamefont {G.}~\bibnamefont
  {Baym}},\ }\href@noop {} {\emph {\bibinfo {title} {Quantum statistical
  mechanics}}}\ (\bibinfo  {publisher} {Benjamin},\ \bibinfo {address} {New
  York},\ \bibinfo {year} {1962})\BibitemShut {NoStop}%
\bibitem [{\citenamefont {Pedri}\ \emph {et~al.}(2003)\citenamefont {Pedri},
  \citenamefont {Gu\'ery-Odelin},\ and\ \citenamefont
  {Stringari}}]{Stringari-ansatz}%
  \BibitemOpen
  \bibfield  {author} {\bibinfo {author} {\bibfnamefont {P.}~\bibnamefont
  {Pedri}}, \bibinfo {author} {\bibfnamefont {D.}~\bibnamefont
  {Gu\'ery-Odelin}}, \ and\ \bibinfo {author} {\bibfnamefont {S.}~\bibnamefont
  {Stringari}},\ }\href {\doibase 10.1103/PhysRevA.68.043608} {\bibfield
  {journal} {\bibinfo  {journal} {Phys. Rev. A}\ }\textbf {\bibinfo {volume}
  {68}},\ \bibinfo {pages} {043608} (\bibinfo {year} {2003})}\BibitemShut
  {NoStop}%
\bibitem [{\citenamefont {Castin}\ and\ \citenamefont {Dum}(1996)}]{Castin}%
  \BibitemOpen
  \bibfield  {author} {\bibinfo {author} {\bibfnamefont {Y.}~\bibnamefont
  {Castin}}\ and\ \bibinfo {author} {\bibfnamefont {R.}~\bibnamefont {Dum}},\
  }\href {\doibase 10.1103/PhysRevLett.77.5315} {\bibfield  {journal} {\bibinfo
   {journal} {Phys. Rev. Lett.}\ }\textbf {\bibinfo {volume} {77}},\ \bibinfo
  {pages} {5315} (\bibinfo {year} {1996})}\BibitemShut {NoStop}%
\bibitem [{\citenamefont {Giorgini}\ \emph {et~al.}(2008)\citenamefont
  {Giorgini}, \citenamefont {Pitaevskii},\ and\ \citenamefont
  {Stringari}}]{Stringari2}%
  \BibitemOpen
  \bibfield  {author} {\bibinfo {author} {\bibfnamefont {S.}~\bibnamefont
  {Giorgini}}, \bibinfo {author} {\bibfnamefont {L.~P.}\ \bibnamefont
  {Pitaevskii}}, \ and\ \bibinfo {author} {\bibfnamefont {S.}~\bibnamefont
  {Stringari}},\ }\href {\doibase 10.1103/RevModPhys.80.1215} {\bibfield
  {journal} {\bibinfo  {journal} {Rev. Mod. Phys.}\ }\textbf {\bibinfo {volume}
  {80}},\ \bibinfo {pages} {1215} (\bibinfo {year} {2008})}\BibitemShut
  {NoStop}%
\bibitem [{\citenamefont {Bohn}\ and\ \citenamefont {Jin}(2014)}]{Bohn-Jin}%
  \BibitemOpen
  \bibfield  {author} {\bibinfo {author} {\bibfnamefont {J.~L.}\ \bibnamefont
  {Bohn}}\ and\ \bibinfo {author} {\bibfnamefont {D.~S.}\ \bibnamefont {Jin}},\
  }\href {\doibase 10.1103/PhysRevA.89.022702} {\bibfield  {journal} {\bibinfo
  {journal} {Phys. Rev. A}\ }\textbf {\bibinfo {volume} {89}},\ \bibinfo
  {pages} {022702} (\bibinfo {year} {2014})}\BibitemShut {NoStop}%
\bibitem [{\citenamefont {Aikawa}\ \emph
  {et~al.}(2014{\natexlab{b}})\citenamefont {Aikawa}, \citenamefont {Frisch},
  \citenamefont {Mark}, \citenamefont {Baier}, \citenamefont {Grimm},
  \citenamefont {Bohn}, \citenamefont {Jin}, \citenamefont {Bruun},\ and\
  \citenamefont {Ferlaino}}]{tau}%
  \BibitemOpen
  \bibfield  {author} {\bibinfo {author} {\bibfnamefont {K.}~\bibnamefont
  {Aikawa}}, \bibinfo {author} {\bibfnamefont {A.}~\bibnamefont {Frisch}},
  \bibinfo {author} {\bibfnamefont {M.}~\bibnamefont {Mark}}, \bibinfo {author}
  {\bibfnamefont {S.}~\bibnamefont {Baier}}, \bibinfo {author} {\bibfnamefont
  {R.}~\bibnamefont {Grimm}}, \bibinfo {author} {\bibfnamefont {J.~L.}\
  \bibnamefont {Bohn}}, \bibinfo {author} {\bibfnamefont {D.~S.}\ \bibnamefont
  {Jin}}, \bibinfo {author} {\bibfnamefont {G.~M.}\ \bibnamefont {Bruun}}, \
  and\ \bibinfo {author} {\bibfnamefont {F.}~\bibnamefont {Ferlaino}},\ }\href
  {\doibase 10.1103/PhysRevLett.113.263201} {\bibfield  {journal} {\bibinfo
  {journal} {Phys. Rev. Lett.}\ }\textbf {\bibinfo {volume} {113}},\ \bibinfo
  {pages} {263201} (\bibinfo {year} {2014}{\natexlab{b}})}\BibitemShut
  {NoStop}%
\bibitem [{\citenamefont {Bluhm}\ \emph {et~al.}(2011)\citenamefont {Bluhm},
  \citenamefont {K\"ampfer},\ and\ \citenamefont {Redlich}}]{qgp1}%
  \BibitemOpen
  \bibfield  {author} {\bibinfo {author} {\bibfnamefont {M.}~\bibnamefont
  {Bluhm}}, \bibinfo {author} {\bibfnamefont {B.}~\bibnamefont {K\"ampfer}}, \
  and\ \bibinfo {author} {\bibfnamefont {K.}~\bibnamefont {Redlich}},\ }\href
  {http://stacks.iop.org/1742-6596/270/i=1/a=012062} {\bibfield  {journal}
  {\bibinfo  {journal} {J. Phys.: Conf. Ser.}\ }\textbf {\bibinfo {volume}
  {270}},\ \bibinfo {pages} {012062} (\bibinfo {year} {2011})}\BibitemShut
  {NoStop}%
\bibitem [{\citenamefont {Florkowski}\ \emph {et~al.}(2012)\citenamefont
  {Florkowski}, \citenamefont {Ryblewski},\ and\ \citenamefont
  {Strickland}}]{qgp2}%
  \BibitemOpen
  \bibfield  {author} {\bibinfo {author} {\bibfnamefont {W.}~\bibnamefont
  {Florkowski}}, \bibinfo {author} {\bibfnamefont {R.}~\bibnamefont
  {Ryblewski}}, \ and\ \bibinfo {author} {\bibfnamefont {M.}~\bibnamefont
  {Strickland}},\ }\href {\doibase 10.1103/PhysRevD.86.085023} {\bibfield
  {journal} {\bibinfo  {journal} {Phys. Rev. D}\ }\textbf {\bibinfo {volume}
  {86}},\ \bibinfo {pages} {085023} (\bibinfo {year} {2012})}\BibitemShut
  {NoStop}%
\bibitem [{\citenamefont {Florkowski}(2010)}]{urhic}%
  \BibitemOpen
  \bibfield  {author} {\bibinfo {author} {\bibfnamefont {W.}~\bibnamefont
  {Florkowski}},\ }\href@noop {} {\emph {\bibinfo {title} {Phenomenology of
  ultra-relativistic heavy-ion collisions}}}\ (\bibinfo  {publisher} {World
  Scientific},\ \bibinfo {address} {Singapore},\ \bibinfo {year}
  {2010})\BibitemShut {NoStop}%
\bibitem [{\citenamefont {Bittencourt}(2004)}]{Bittencourt}%
  \BibitemOpen
  \bibfield  {author} {\bibinfo {author} {\bibfnamefont {J.~A.}\ \bibnamefont
  {Bittencourt}},\ }\href@noop {} {\emph {\bibinfo {title} {Fundamentals of
  plasma physics}}},\ \bibinfo {edition} {3rd}\ ed.\ (\bibinfo  {publisher}
  {Springer},\ \bibinfo {address} {New York},\ \bibinfo {year}
  {2004})\BibitemShut {NoStop}%
\bibitem [{\citenamefont {Brennan}(1999)}]{Brennan}%
  \BibitemOpen
  \bibfield  {author} {\bibinfo {author} {\bibfnamefont {K.~F.}\ \bibnamefont
  {Brennan}},\ }\href@noop {} {\emph {\bibinfo {title} {The physics of
  semiconductors with applications to optoelectronic devices}}}\ (\bibinfo
  {publisher} {Cambridge University Press},\ \bibinfo {address} {Cambridge},\
  \bibinfo {year} {1999})\BibitemShut {NoStop}%
\end{thebibliography}
\end{document}